\pdfoutput=1

\PassOptionsToPackage{
pdfencoding=auto,
pdfnewwindow=true,
pdfusetitle=false,
bookmarks=true,
bookmarksnumbered=true,
bookmarksopen=true,
pdfpagemode=UseThumbs, 
bookmarksopenlevel=1, 
pdfpagelabels=false,   
breaklinks=true,  
backref=false, 
colorlinks=true,
}{hyperref}
\RequirePackage{etex}
\PassOptionsToPackage{skip=0pt, font=small,labelfont=bf}{caption} 
\documentclass[10pt,showpacs,letterpaper,aps,pra,twocolumn,nofootinbib,superscriptaddress,floatfix]{revtex4-2} 
\usepackage{hyperref,makecell}
\usepackage{csquotes}

\usepackage{amsthm,mathtools,amssymb,changepage} 
\usepackage{centernot}
\usepackage[none]{hyphenat}
\usepackage{fancyhdr}
\newcommand{\ket}[1]{\left| #1 \right\rangle}
\newcommand{\bra}[1]{\left\langle #1 \right|}

\newcommand{\beq}{\begin{equation}}
\newcommand{\eeq}{\end{equation}}
\newcommand{\bea}{\begin{align}}
\newcommand{\eea}{\end{align}}

\usepackage{graphicx}
\usepackage[usenames,dvipsnames,table]{xcolor} 
\input{preamble}

\definecolor{googleblue}{RGB}{34, 0, 204}
\definecolor{panblue}{RGB}{0,24,150}
\definecolor{carmine}{RGB}{150, 0, 24}
\hypersetup{
unicode=true,
bookmarksnumbered=false,
bookmarksopen=false,
breaklinks=false,
colorlinks=true,
linkcolor=carmine,
citecolor=googleblue,
urlcolor=panblue,
anchorcolor=OliveGreen}

\newcommand{\Hilb}[1][]{\ensuremath{\mathcal{H}_{#1}}}
\newcommand{\Ket}[1]{\ensuremath{\left \vert #1 \right \rangle}}
\newcommand{\Bra}[1]{\ensuremath{\left \langle #1 \right \vert}}
\newcommand{\BraKet}[2]{\ensuremath{\left \langle #1 \middle \vert #2 \right \rangle}}
\newcommand{\KetBra}[2]{\ensuremath{\left \vert #1 \middle \rangle \middle \langle #2 \right \vert}}
\newcommand{\Proj}[1]{\ensuremath{\KetBra{#1}{#1}}}

\allowdisplaybreaks

\usepackage{subcaption}
\usepackage[labelformat=simple]{subcaption}

\usepackage{microtype}
\microtypecontext{spacing=nonfrench}
\microtypesetup{
expansion={true,nocompatibility},
protrusion={true,nocompatibility},
activate={true,nocompatibility},
tracking=true,
kerning=true,
spacing={true}
}

\usepackage[all=normal,floats=tight,mathspacing=tight,wordspacing=tight,paragraphs=normal,tracking=tight,charwidths=tight,mathdisplays=normal,sections=normal,margins=normal]{savetrees}

\everypar=\expandafter{\the\everypar\loosness=-1 }
\newcommand{\nocontentsline}[3]{}
\let\oldaddcontentsline\addcontentsline
\newcommand{\tocless}[2]{%
  \let\addcontentsline=\nocontentsline#1{#2}
  \let\addcontentsline\oldaddcontentsline}

\begin{document}
\title{A review and analysis of six extended Wigner's friend arguments}
\author{David Schmid}
\email{davidschmid10@gmail.com}
\affiliation{International Centre for Theory of Quantum Technologies, University of Gda\'nsk, 80-308 Gda\'nsk, Poland}
\author{Y\`{i}l\`{e} Y{\=\i}ng}
\email{yying@perimeterinstitute.ca}
\affiliation{Perimeter Institute for Theoretical Physics, Waterloo, Ontario, Canada, N2L 2Y5}
\affiliation{Department of Physics and Astronomy, University of Waterloo, Waterloo, Ontario, Canada, N2L 3G1}
\author{Matthew S. Leifer}
\email{leifer@chapman.edu}
\affiliation{Institute for Quantum Studies \& Schmid College of Science and Technology, Chapman University, One University Drive, Orange, CA, 92866, USA}

\begin{abstract}
The Wigner's friend thought experiment was intended to illustrate the difficulty one has in describing an agent as a quantum system when that agent performs a measurement. While it does pose a challenge to the orthodox interpretation of quantum theory, most modern interpretations have no trouble in resolving the difficulty. Recently, a number of extensions of Wigner's ideas have been proposed. We provide a gentle introduction to six such arguments, modifying the specifics of many of them so that they are as simple and unified as possible. In particular, we show that all of the arguments hinge on assumptions about correlations between measurement outcomes that are not accessible to any observer, even in principle. We then provide a critical analysis of each argument, focusing especially on how well one can motivate the required assumptions regarding these inaccessible correlations. Although we argue that some of these assumptions are not entirely well-motivated, all of the arguments do shed light on the nature of quantum theory, especially when concerning the description of agents and their measurements. 
\end{abstract}

\maketitle
\tableofcontents 

\section{Introduction}

A key subject of contention in quantum foundations is the measurement problem~\cite{maudlin1995three}. In his famous thought experiment, Wigner introduced a dramatization of the measurement problem that arises when an observer, Wigner's friend, performs a measurement on a quantum system, and when one attempts to view this situation from two different perspectives. In one perspective, the Friend is viewed as an abstract agent, and the Friend's measurement induces a measurement outcome together with a collapse of the wavefunction. In the second perspective, the Friend is viewed as a dynamical quantum system in its own right, in which case the Friend's measurement is viewed as a unitary interaction with the system being measured, resulting in an entangled state between the system and Friend. While both of these perspectives are supposedly valid within the context of textbook quantum theory, arguments like Wigner's raise the question of whether they are in fact consistent with each other.

As we will discuss in detail below, Wigner's thought experiment does pose a challenge for some interpretations of quantum theory, most notably the orthodox interpretation, but most interpretations sidestep the argument easily. In recent years, however, various extensions of the Wigner's friend argument have been proposed. These arguments aim to be more rigorous than Wigner's original argument and to have notable metaphysical (or epistemological) implications for a broader class of interpretations. 

In this work, we will provide a gentle introduction to a number of these thought experiments, followed by a critical analysis of each. We focus primarily on Brukner's argument~\cite{brukner2015quantum,bruknerNogo2018}, the Frauchiger-Renner argument~\cite{Frauchiger2018,Rio2023,Nurgalieva_2020}, an argument due to Pusey-Masanes~\cite{PuseyYoutube,LeiferYoutube} and related works~\cite{Healey_2018,leegwaterWhen2022,ormrod2022no}, the Local Friendliness argument~\cite{bongStrong2020,Cavalcanti2021,haddaraPossibilistic2022,Wiseman2023thoughtfullocal,utrerasalarcon2023allowing}, an argument by Gao~\cite{Gao2019}, and an argument by Gu{\'e}rin et. al.~\cite{allard2021no}. (We have chosen to focus on these arguments for pedagogical reasons and because we believe that together they represent all the essential features of the different arguments found in the current literature.)

Our presentation of these arguments aims to draw out the essential features of each argument, and our critical analysis of each argument focuses on the particular assumptions and features of the argument that we believe are most open to criticism. 
We will not attempt to analyze the consequences of these arguments for every possible interpretation, nor to summarize or evaluate the many attempted responses to these no-go theorems~\cite{lazaroviciHow2018,vilasini2022general,Sudbery_2017,debrota2020respecting,stacey2019qbism,brukner2015quantum,kastner2020unitary,cavalcanti2023consistency,bub2017bohr,renes2021consistency,polychronakos2022quantum,aaronson2018,okon2022reassessing,Okon2022,Lawrence2023relativefactsof,baumann2016measurement,sudbery2019hidden,Elouard2021quantumerasing,Zukowski2021,Baumann2018formalisms,Healey_2018,pittphilsci16292,drezet2018wigner,Nurgalieva_2020,Wiseman2023thoughtfullocal,Cavalcanti2021,dibiagioStable2021}. Rather, we will comment on specific interpretative viewpoints only when they serve to illustrate some relevant point.

Moreover, we modify and simplify somewhat the presentation of (most, if not all, of) the extended Wigner's friend (EWF) arguments in order to clarify the key features of each and better demonstrate how similar many of them actually are. This also makes it easier to relate the different arguments, and to understand how
their consequences coincide or differ.

Roughly speaking, one can situate the different EWF arguments in this work as follows. Brukner's argument constitutes the first work to recognize the conceptual significance of combining components of Wigner's friend scenarios with those of Bell scenarios. The Local Friendliness argument can be viewed as taking Brukner's experimental setup and intuitive argument and turning them into a rigorous no-go theorem (based on weaker assumptions). The Pusey-Masanes no-go theorem simplifies the Local Friendliness setup by having each agent performing a fixed measurement rather than choosing from some set of possible measurements. The Frauchiger-Renner no-go theorem is constructed in exactly the same experimental setup as Pusey-Masanes's, but appeals primarily to epistemological assumptions rather than metaphysical ones. All four of these arguments, unlike Gao's and Gu{\'e}rin et. al.'s, can be constructed around Bell-inequality violating correlations. Gao's argument introduces the question of what quantum theory predicts for correlations between sequential measurements (but also involves two parties measuring their halves of a shared entangled state). Gu{\'e}rin et. al.'s argument, meanwhile, involves sequential measurements on a single system alone. (See also Section~\ref{comparisonsec} for a more extensive comparison of the Local Friendliness, Pusey-Masanes, and Frauchiger-Renner no-go theorems.)

An important feature of all six EWF arguments is that they do not invoke any notable assumptions of (classical) realism~\cite{schmid2021unscrambling}, such as the existence of hidden variables. Most no-go theorems about quantum theory, such as Bell's theorem, begin by making assumptions that each system has some objective physical properties that exist independently of the observer and can be described by a classical variable. Since many interpretations deny this very first assumption, they survive these theorems.\footnote{As one example, many Copenhagenish interpretations~\cite{forthcomingCopenhagenish} (which we take to include Bub and Pitowsky's \enquote{information} interpretation~\cite{bub2008dogmas}, Healey's Quantum Pragmatism~\cite{Healey2020}, Rovelli's Relational Quantum Mechanics~\cite{rovelli1996relational}, the QBism of Fuchs, Schack, et. al.~\cite{fuchs2009quantumbayesian,fuchs2010qbism,FuchsMerminSchack,fuchs2019qbism}, and the views of various others, such as Peres~\cite{peresQuantum1995}, Brukner and Zeilinger~\cite{Brukner2003}) think that very little, if anything, in the quantum formalism actually represents objective physical reality and there is no deeper description of a quantum system to be had than that provided by quantum theory. As another example, the research program of quantum causal models rejects the idea that common causes are described by classical variables like $\lambda$.}  The remarkable achievement of recent extended Wigner's friend arguments is that they manage to constrain interpretations of this sort.  Such arguments are difficult to come by, and much work remains to be done to uncover what they have to teach us about the nature of reality.

Loosely, each EWF argument assumes that observers can be treated quantum mechanically, that measurement processes in which these observers perform measurements can be described as unitary interactions, and that an agent with sufficiently advanced technology could undo such a unitary interaction, erasing the observed outcome while reverting the observer and the system in question to their initial state before the measurement. It is also assumed that observed measurement outcomes are single and absolute (that is, objective---not perspectival or relative). 
 These `background' assumptions are common to all EWF arguments.
In addition, each EWF argument we will discuss makes an additional assumption about the joint frequency with which certain pairs of measurement outcomes will be seen, including in physical circumstances where this joint frequency cannot be observed by any observer, even in principle. A key question for each argument, then, is how compelling this last assumption is, since it is primarily this last assumption that distinguishes the different arguments from each other.

In our view, the most compelling EWF argument is the Local Friendliness argument. 
 The additional assumption made in this argument is that any freely chosen measurement setting is uncorrelated with any set of relevant \emph{observed} events outside its future light cone. 
Provided that one grants the background assumptions of the theorem, then, one is either forced to give up on this weak notion of locality, or one is forced to the conclusion that measurement outcomes are not single and absolute, but rather are only well-defined, for example, relative to the agent who observed them. Either conclusion is quite radical, even more radical than the positions forced on us by previous no-go theorems (including Bell's theorem and proofs of the failure of noncontextuality). 

For the other EWF arguments we will discuss, it is less clear that the extra assumption (beyond the background ones listed above) can be given solid motivation. In some cases, the assumptions are contentious but interesting to consider, while in some other cases, we argue that the assumptions are unmotivated or even flawed.

A very coarse-grained comparison of the six arguments is given in the following table.
\begin{table}[h!]
    \centering
    \setlength\tabcolsep{0.3mm}
     \begin{adjustwidth}{-3.5mm}{}
\begin{tabular}{|c|c|c|}
    \hline
    \rowcolor{gray!15}
        \textbf{EWF Argument}  & \textbf{Type} & \textbf{Key assumption} \\ \hline
    Brukner~\cite{bruknerNogo2018} & Parallel & Local Causality (essentially)  \\ \hline
    Local Friendliness~\cite{bongStrong2020} & Parallel & Local Agency  \\ \hline
    Pusey-Masanes~\cite{PuseyYoutube,LeiferYoutube} & Parallel & Born Inaccessible Correlations \\ \hline
    Frauchiger-Renner~\cite{Frauchiger2018} & Parallel & \makecell{Born Inaccessible Correlations  \vspace{-1mm} \\ Consistency}  \\ \hline
    Gao~\cite{Gao2019} & Mixed & unclear  \\ \hline
    Gueren et. al.~\cite{allard2021no} & Sequential & (a kind of) Linearity  \\ \hline
\end{tabular}
      \end{adjustwidth}
      \vspace{2mm}
\caption{ A rough comparison between the six arguments. Note that the descriptors of `parallel' and `sequential' do not necessarily refer to the arrangement of measurements in the scenario, but rather refer to whether or not the assumptions being used in the argument constrain inaccessible correlations between measurements that are done in parallel or in sequence. Assumptions with an asterisk are those we will argue below are problematic.} 
\label{tab_comp}
\end{table} 

\section[Building up to EWF arguments]{Building up to extended Wigner's friend arguments}
\label{sec_Build}

We begin with a warm-up discussion of the measurement problem, Wigner's original thought experiment, and minor extensions of the latter. This will serve the purpose of introducing the basic sort of reasoning and notation we will use, while also allowing us to explain how most modern interpretations are fully capable of resolving these apparent problems. This is precisely why the extended Wigner's friend arguments we will discuss in later sections are so important: they provide significantly stronger constraints on these interpretations.

\subsection{The measurement problem}
\label{sec_MMP}

Broadly speaking, the measurement problem arises because the postulates of quantum theory refer to a notion of \enquote{measurement}, but do not define precisely what sorts of processes count as measurements. Moreover, these postulates entail that there are two different ways of describing a measurement process, and 
there is no precise specification of when one or the other description is required.

Suppose a qubit ${\rm S}$ with an associated Hilbert space $\Hilb[S]$ is prepared in the state $\Ket{+}_{\rm S} = \frac{1}{\sqrt{2}} \left ( \Ket{0}_{\rm S} + \Ket{1}_{\rm S} \right )$, and is measured in the $\left \{ \Ket{0}_{\rm S}, \Ket{1}_{\rm S} \right \}$ basis.  According to the textbook postulates of quantum mechanics, the $\Ket{0}_{\rm S}$ outcome will be obtained with probability $1/2$, and if it is obtained, the state is updated to $\Ket{0}_{\rm S}$. Similarly, the $\Ket{1}_{\rm S}$ outcome will be obtained with probability $1/2$, in which case the state is updated to $\Ket{1}_{\rm S}$.  The net result is that after the measurement, the state will be either $\Ket{0}_{\rm S}$ or $\Ket{1}_{\rm S}$, with probability $1/2$ each. 

However, a measurement device is just a physical system, made of atoms like everything else, so we should be able to associate it with a Hilbert space $\Hilb[M]$, so that the interaction between the qubit and the measuring device can be described by a unitary on $\Hilb[S]\otimes\Hilb[M]$ according to the Schr{\"o}dinger equation. Note that $\Hilb[M]$ should be taken to be a large enough subsystem of the universe such that the interaction between the system and the measuring device can always be modeled by a unitary.  If one thinks that environmental decoherence is necessary for the device to register a definite measurement outcome, then a significant portion of the environment beyond the measurement device itself should also be included in $\Hilb[M]$.

Suppose that the measurement device starts in the ready state $\Ket{R}_{\rm M}$, a state in which the measurement device is ready to start a new measurement, e.g., it is calibrated and the pointer needle points to zero.  Note that there will typically be many possible $\Ket{R}_{\rm M}$ states, e.g., because parts of the apparatus may be in thermal contact with some environment.  All that matters is that the measurement device is in one such state, as they will all behave similarly in the measurement interaction.

Assuming that the measurement device works correctly when the system is prepared in either the $\Ket{0}_{\rm S}$ state or the $\Ket{1}_{\rm S}$ state, the unitary evolution $U_{\rm SM}$ describing the measurement must satisfy
\begin{equation}
\label{eq_Usm1}
    \begin{split}
	U_{\rm SM} \Ket{0}_{\rm S}  \Ket{R}_{\rm M} & = \Ket{0}_{\rm S}  \Ket{0}_{\rm M}, \\
	 U_{\rm SM} \Ket{1}_{\rm S}  \Ket{R}_{\rm M} & = \Ket{1}_{\rm S}  \Ket{1}_{\rm M}, 
    \end{split}
\end{equation}
where $\Ket{0}_{\rm M}$ and $\Ket{1}_{\rm M}$ are states in which the measurement device is registering the outcome $0$ and $1$ respectively.  Note that, as with $\Ket{R}_{\rm M}$, there will typically be many possible $\Ket{0}_{\rm M}$ and $\Ket{1}_{\rm M}$ states, but all that matters is that the evolution takes $\Ket{0}_{\rm S}  \Ket{R}_{\rm M}$ and $ \Ket{1}_{\rm S}  \Ket{R}_{\rm M}$ to one such pair of states.  Typically we want to have $\BraKet{0}{1}_{\rm M} \approx 0$, so that the outcome can be determined with good accuracy by measuring the measurement device, but this is actually immaterial for the argument.

Without loss of generality, we can take $\Ket{R}_{\rm M}:=\Ket{0}_{\rm M}$, in which case the unitary evolution for the measurement satisfying Eq.~\eqref{eq_Usm1} is the CNOT gate. From now on, we assume this convention, so Eq.~\eqref{eq_Usm1} becomes
\begin{equation}
    \label{eq_Usm}
        \begin{split}
        U_{\rm SM} \Ket{0}_{\rm S}  \Ket{0}_{\rm M} & = \Ket{0}_{\rm S}  \Ket{0}_{\rm M}, \\
         U_{\rm SM} \Ket{1}_{\rm S}  \Ket{0}_{\rm M} & = \Ket{1}_{\rm S}  \Ket{1}_{\rm M}, 
        \end{split}
    \end{equation}

Now consider the case where the qubit is prepared in the $\Ket{+}_{\rm S}$ state, and thus the joint state of the qubit and the measuring device is $\Ket{+}_{\rm S} \Ket{0}_{\rm M}$. We apply $U_{\rm SM}$ to $\Ket{+}_{\rm S} \Ket{0}_{\rm M}$ and obtain, by linearity, the joint state after the measurement
\begin{align*}
   \Ket{\phi^{+}}_{\rm SM} & = U_{\rm SM} \Ket{+}_{\rm S} \Ket{0}_{\rm M} \\
   &= \frac{1}{\sqrt{2}} \left ( U_{\rm SM} \Ket{0}_{\rm S}  \Ket{0}_{\rm M} + U_{\rm SM} \Ket{1}_{\rm S}  \Ket{0}_{\rm M} \right ) \\
                              & = \frac{1}{\sqrt{2}} \left ( \Ket{0}_{\rm S}  \Ket{0}_{\rm M} + \Ket{1}_{\rm M}  \Ket{1}_{\rm M}\right ).
\end{align*}

This is clearly not the same quantum state as one obtains in the first description of the measurement process (either $\Ket{0}_{\rm S}$ or $\Ket{1}_{\rm S}$).  From the point of view of the orthodox interpretation, these two descriptions are in flat-out contradiction. A central feature of this interpretation (sometimes called the textbook interpretation or Dirac-von-Neumann interpretation), is the eigenstate-eigenvalue link, which states that the set of possible properties a system can have is in one-to-one correspondence with the set of projectors on the Hilbert space associated to that system.
Then, if the state of the system is either $\Ket{0}_{\rm S}$ or $\Ket{1}_{\rm S}$, then the system has a property corresponding to the projector $\Proj{0}_{\rm S}$, which takes either value $1$ or $0$ depending on whether the state is $\Ket{0}_{\rm S}$ or $\Ket{1}_{\rm S}$.  If the state of the system and the measurement device is $\Ket{\phi^{+}}_{\rm SM}$, then no such property exists because $\Ket{\phi^{+}}_{\rm SM}$ is not an eigenstate of $\Proj{0}_{\rm S} \otimes  I_{\rm M}$.  From the orthodox view, then, one of these descriptions must be right and the other wrong.  There must be a fact of the matter about whether a measurement, and hence a state collapse, has occurred. So the orthodox interpretation is simply wrong (provided one grants that the measurement can be given a unitary description).

However, this \enquote{measurement problem} is not problematic from the perspective of most other interpretations. To start with, if an interpretation gives a clear specification for when an objective collapse has happened, then there is no ambiguity about whether a measurement should be modeled by a unitary or the projection postulate. This is the case in collapse theories~\cite{GRW,alloriWave2020,Penrose94,ChaMcQ21}. In addition, many interpretations reject the eigenstate-eigenvalue link, without which the contradiction is straightforwardly blocked. Many worlds~\cite{Everett1957} is such an example. In some other interpretations, systems have properties beyond those associated to projectors for which the system is in an eigenstate; for instance, in Bohmian mechanics~\cite{Bohm1952,Bohm1952II,Bohm}, the properties of the system are determined not just by the wavefunction, but also by the positions of Bohmian particles. Moreover, some interpretations (most notably Copenhagenish interpretations such as QBism) deny that any properties of a system can be read off of the wavefunction at all. 
If one rejects the idea that state assignments imply anything about the properties of the system in question, then having two different quantum state assignments that are both correct does not imply a contradiction. More broadly, interpretations which take an epistemic view of the quantum state\footnote{Such interpretations hold that the quantum state is not a direct representation of all and only the properties of the system, rather, it is a representation of an agent's knowledge about a system.} resolve the measurement problem easily: the two different descriptions of the system in question are both correct---they merely correspond to the descriptions of two agents with differing knowledge of the state.

It is worth noting that the measurement problem (as well as the Wigner's friend, enemy, and stalkee arguments in Section~\ref{sec_enemy}-\ref{sec_OGwigner}) ultimately arises as a consequence of the tension between unitary evolution and the scope of possible state-update rules in quantum theory (e.g., the projection postulate). In contrast, the extensions of these Wigner's friend arguments we will discuss in the rest of the paper focus on a different tension, namely that between unitary evolution and the Born rule. (A priori, it is not particularly surprising that there is a tension between unitary evolution and state update rules for quantum measurements, since these two might naturally be taken to constitute different dynamical laws for the system in question. It seems less obvious to us that there would be any tension between unitary evolution and the Born rule, since the latter merely stipulates the frequency at which outcomes will be observed in a measurement, and says nothing about the state of the system after the measurement.)

\subsection{Wigner's friend} 
\label{sec_OGwigner}

The original Wigner's friend thought experiment~\cite{Wigner1995} is essentially a dramatization of the measurement problem where one imagines an observer (the Friend) who performs a computational basis measurement on a system ${\rm S}$ prepared in the initial state $\Ket{+}_{\rm S} = \frac{1}{\sqrt{2}} \left ( \Ket{0}_{\rm S} + \Ket{1}_{\rm S} \right )$, and models this procedure from two different perspectives. 

From the Friend's perspective, an outcome is observed, based on which the state is updated using the projection postulate. If outcome $0$ is observed, then the state of the system ${\rm S}$ after the measurement will be $\Ket{0}_{\rm S}$; if outcome $1$ is observed, then the state of the system ${\rm S}$ after the measurement will be $\Ket{1}_{\rm S}$.  Each possibility is equally likely.

From Wigner's perspective, in contrast, the Friend is just like another quantum system, and he models the Friend in the same way as the measurement device described in Section~\ref{sec_MMP} and thus, the measurement process is a unitary interaction between the Friend and system ${\rm S}$. 

Denote the system representing the Friend (and potentially the environment necessary for them to observe a definite outcome according to decoherence) as $\Hilb[F]$, and denote their ready state as $\Ket{0}_{\rm F}$. Just as in the analysis of Section~\ref{sec_MMP}, the unitary evolution $U_{\rm SF}$ describing their measurement must satisfy
\begin{align*}
	U_{\rm SF} \Ket{0}_{\rm S}  \Ket{0}_{\rm F} & = \Ket{0}_{\rm S}  \Ket{0}_{\rm F}, & U_{\rm SF} \Ket{1}_{\rm S}  \Ket{0}_{\rm F} & = \Ket{1}_{\rm S}  \Ket{1}_{\rm F},
\end{align*}
where $\Ket{0}_{\rm F}$ and $\Ket{1}_{\rm F}$ are coarse-grained quantum states in which the Friend has observed the outcome $0$ or $1$, respectively. 

Wigner models the Friend's measurement by applying  $U_{\rm SF}$ to the initial state $\Ket{+}_{\rm S} \Ket{0}_{\rm F}$, obtaining one of the Bell states
\begin{align}
   \Ket{\phi^{+}}_{\rm SF} & = U_{\rm SF} \Ket{+}_{\rm S} \Ket{0}_{\rm F} \\ \nonumber
    & = \frac{1}{\sqrt{2}} \left ( \Ket{0}_{\rm S}  \Ket{0}_{\rm F} + \Ket{1}_{\rm M}  \Ket{1}_{\rm F}\right ).
\end{align}
Since Wigner is describing the state of another observer, he is often referred to as superobserver. This term will be even more appropriate in later sections on Wigner's enemy, stalkee and their extensions, where we imagine that some agents moreover have the extreme technological capability to apply coherent quantum operations to macroscopic systems including other observers (like the Friend here).

From the point of view of the orthodox interpretation, this implies a contradiction in exactly the same way as the standard measurement problem of the previous section.
Also as in the previous section, there is no contradiction within most modern interpretations. For example, in epistemic interpretations,  the different state assignments made by Wigner and by the Friend do not reflect different states of reality, just different states of knowledge about the system.

\subsection{Wigner's enemy}
\label{sec_enemy}

The thought experiments in this and the next subsection are similar to those introduced by Deutsch~\cite{deutschQuantum1985}. 

Since Wigner and his Friend's perspectives are not contradictory, despite their disagreeing quantum state assignments, one might hope that in the long run (e.g., after the experiment is over), agreement can be reached between the Friend and Wigner. This will happen, for example, if the Friend tells Wigner their measurement outcome, in which case Wigner will agree with the Friend's quantum state assignment.

However, this is not always possible if Wigner is a superobserver with the technological capabilities to apply unitary operations to the Friend and the system. For example, he could apply $U_{\rm SF}^{\dagger}$, the inverse of the unitary $U_{\rm SF}$, on the joint system of the Friend and the system ${\rm S}$, effectively reversing (or \enquote{undoing}) the Friend's measurement. According to Wigner, undoing the measurement returns the state of the Friend and system ${\rm S}$ to their initial state, as
\begin{align} \label{eq_Wign}
    U_{\rm SF}^{\dagger}\Ket{\phi^{+}}_{\rm SF}=\Ket{+}_{\rm S}\Ket{0}_{\rm F}.
\end{align}
In so doing, Wigner completely erases all traces of the measurement including the Friend's memory of it. (Hence why we call this the Wigner's enemy scenario: friends do not recohere friends!)

What is the Friend's perspective on this experiment?
The Friend can argue that before the undoing of the measurement, they and the system were either in state $\Ket{0}_{\rm S}\Ket{0}_{\rm F}$ or the state $\Ket{1}_{\rm S}\Ket{1}_{\rm F}$, depending on whether the Friend observes outcome $0$ or $1$. 
Consequently, the Friend might predict that after Wigner's unitary is applied, the state will be
\begin{align}\label{eq_EnemyF1}
    U_{\rm SF}^{\dagger}\Ket{0}_{\rm S}\Ket{0}_{\rm F}=\Ket{0}_{\rm S}\Ket{0}_{\rm F}
\end{align}
or
\begin{align}\label{eq_EnemyF2}
    U_{\rm SF}^{\dagger}\Ket{1}_{\rm S}\Ket{1}_{\rm F}=\Ket{1}_{\rm S}\Ket{0}_{\rm F}. 
\end{align}

After the undoing, then, both Wigner and the Friend agree that the Friend is uncorrelated with the system, but they disagree on the state of the spin. (Furthermore, the Friend will not agree that the $U_{\rm SF}^{\dagger}$ unitary constitutes a reversal of their measurement, since they do not describe their measurement by a unitary process, much less by $U_{\rm SF}$.) Just as in the last sections, these different state assignments lead to a contradiction within the orthodox interpretation, but do not lead to any problems in most other interpretations.

\subsection{Wigner's stalkee}
\label{sec_stalkee}

The preceding arguments focused on the different quantum state assignments made by the Friend and Wigner. The following thought experiment turns on the fact that different state assignments lead to different operational predictions for measurement outcomes.

Assuming again that Wigner has the extreme technological capabilities of a superobserver, consider the case where he performs a measurement on the joint system comprised of the Friend together with system ${\rm S}$ in a basis containing the projector onto the entangled state $\Ket{\phi^{+}}_{\rm SF}$. As Wigner's description of the state of ${\rm SF}$ is $\Ket{\phi^{+}}_{\rm SF}$, he expects to always get the outcome corresponding to this projector. 
(Moreover, note that Wigner can do this measurement without disturbing the quantum state of the Friend and system when the state is indeed an eigenstate of the measurement. Hence, one might say that Wigner is stalking his Friend!)

However, if the Friend's quantum state assignment is $\ket{0}_{\rm S}\ket{0}_{\rm F}$ or $\ket{1}_{\rm S}\ket{1}_{\rm F}$ (as it would be, e.g., in the orthodox interpretation), then the Friend predicts that Wigner's measurement outcome only has probability $\frac{1}{2}$ to be the $\ket{\phi^{+}}_{\rm SF}$ outcome. Thus, Wigner and his Friend make different predictions for the outcomes of Wigner's measurement.

In the orthodox interpretation, this is a bona-fide contradiction, and one finds that one cannot make an unambiguous prediction for the setup in question.

If one believes there is an objective fact of the matter about the frequency with which Wigner's $\Ket{\phi^{+}}_{\rm SF}$ outcome occurs, then a fundamental physical theory should predict this frequency. Thus, it seems natural to assert that at least one of the Friend's and Wigner's use of quantum theory to make the prediction is wrong. 
In the Bohmian interpretation, for instance, the Friend's prediction is incorrect (owing to the Friend assuming an effective collapse has happened in a situation where this is inapplicable), while in a collapse theory, Wigner's prediction can be incorrect (owing to Wigner assuming no collapse has happened, while in reality one did happen).
 
In some Copenhagenish interpretations, one could instead take the position that the Friend is not allowed to treat themself as a quantum system. In this latter case, the Friend's prediction that the outcome occurs with probability $\frac{1}{2}$ can be rejected as invalid, leaving Wigner's prediction as the correct one. (This is also a sufficient response to resolve the Wigner's enemy contradiction.) 
However, this position implies that the Friend is {\em unable to make any predictions whatsoever} about what Wigner will observe. The Friend is not even able to take on board Wigner's perspective and Wigner's prediction, since these rely on assigning a quantum state to themself. 
If quantum theory is meant to be a fundamental theory, it is quite awkward to believe that there are some experiments for which an observer \emph{cannot use it} to make predictions. 
We will see that this matter arises again in further extended Wigner's friend arguments, such as the Frauchiger-Renner argument of Section~\ref{sec_evaCons}. 

Another alternative would be to take a perspectival (or relational) view, wherein when one reasons about what {\em Wigner} would observe, one should not take into account measurement outcomes observed by other observers, unless these can be (and are) transmitted to Wigner. But the Friend's measurement outcome is (by assumption) not transmitted to Wigner. Consequently, the Friend should not use the state assignments in Eq.~\eqref{eq_EnemyF1} or Eq.~\eqref{eq_EnemyF2} to make predictions about what Wigner will observe, and rather should assign the state Eq.~\eqref{eq_Wign} to herself and system ${\rm S}$. Consequently, in such relational views, the Friend agrees with Wigner's prediction. 
This might seem odd, insofar as the Friend neglects her own information---the specific outcomes they perceived---in making their prediction. But this is exactly what one expects in a perspectival picture where the Friend's outcome is only meaningful relative to the Friend, and is not relevant to Wigner or what he will observe in any sense.\footnote{By contrast, it is unclear in non-perspectival interpretations why the Friend would neglect their own information, as this information presumably would be relevant to making predictions for Wigner's outcome.} There simply is no global picture that incorporates both Wigner's outcome and Friend's outcome in this scenario, since both outcomes are not accessible (even in principle) to any single observer. 

\subsection[A useful fact]{A useful fact regarding Wigner's measurement}\label{usefulfact}

Finally, let us note a mathematical fact that we will make use of a few times in this manuscript, which is that the quantum predictions when Wigner measures the joint Friend-plus-system state in the entangled basis containing the projector 
\begin{equation}
    \Pi^{\phi^+}_{\rm SF}:=\ket{\phi^{+}}_{\rm SF}\bra{\phi^{+}}=\frac{1}{2}(\ket{00}+\ket{11})_{\rm SF}(\bra{00}+\bra{11})
\end{equation}
are the same as the quantum predictions when Wigner instead applies $U_{\rm SF}^{\dagger}$, the inverse of the unitary describing the Friend's measurement, and then directly measures (only) the system in the basis $\big\{\ket{+}_{\rm S},\ket{-}_{\rm S}\big\}$. This fact is well-known (see e.g. Ref.~\cite{aaronson2018}).

To see this explicitly, note that tensoring the state $\rho_{\rm S}$ of the system ${\rm S}$ with the ready state $\Ket{0}_{\rm F}$ of the Friend and then applying the unitary $U_{\rm SF}$ defines an isometry $V$, which maps a state of the system ${\rm S}$ to a state of the joint system ${\rm SF}$, i.e.,
\begin{align}
    V\Ket{0}_{\rm S}=\Ket{0}_{\rm S}\Ket{0}_{\rm F}, \ V\Ket{1}_{\rm S}=\Ket{1}_{\rm S}\Ket{1}_{\rm F}.
\end{align}

Using this notation, $\Pi^{\phi^+}_{\rm SF}$, the projector onto the entangled state can be expressed as 
\begin{align}\label{eq_piSF}
    \Pi^{\phi^+}_{\rm SF}=V\ket{+}_{\rm S}\Bra{+}V^{\dagger},
\end{align}
while the action of Wigner undoing the Friend's measurement can be expressed as applying the adjoint of the isometry on $\rho_{\rm SF}$, the density operator for the joint state of the Friend and the system. This takes $\rho_{\rm SF}$ to a state of the system ${\rm S}$, namely
\begin{align}\label{eq_V}
    \rho_{\rm S}:=V^{\dagger}\rho_{\rm SF}V.
\end{align}

The quantum prediction for the probability of obtaining the outcome corresponding to the projector $\Pi^{\phi^+}_{\rm SF}$ on the joint system $\rho_{\rm SF}$ is given by the Born rule as
\begin{align}
    {\rm Tr}\left(\Pi^{\phi^+}_{\rm SF}\rho_{\rm SF}\right).
\end{align}
By Eqs.~\eqref{eq_piSF} and~\eqref{eq_V}, the term in the trace can be expanded as
\begin{equation}
    {\rm Tr}\left[\left(V\ket{+}_{\rm S}\Bra{+}V^{\dagger}\right)\left(V\rho_{\rm S}V^{\dagger}\right)\right]
\end{equation}
Since $V^{\dagger}V=I$, this expression can be simplified to
\begin{align}
    {\rm Tr}\left(V\ket{+}_{\rm S}\Bra{+}\rho_{\rm S}V^{\dagger}\right)
\end{align}
Since the trace of a set of linear operators is invariant under cyclic permutations, the above expression can be further simplified to
\begin{align}
    {\rm Tr}\left(\ket{+}_{\rm S}\Bra{+}\rho_{\rm S}V^{\dagger}V\right)
    ={\rm Tr}\left(\ket{+}_{\rm S}\Bra{+}\rho_{\rm S}\right).
\end{align}
Thus, 
\begin{align}
    {\rm Tr}\left(\Pi^{\phi^+}_{\rm SF}\rho_{\rm SF}\right)={\rm Tr}\left(\ket{+}_{\rm S}\Bra{+}\rho_{\rm S}\right).
\end{align}
Therefore, applying the undoing operation $V^{\dagger}$ followed by measuring the $\ket{+}_{\rm S}\Bra{+}$ observable on the system ${\rm S}$ gives the same observable statistics as instead measuring the $\Pi^{\phi^+}_{\rm SF}$ observable on the joint ${\rm SF}$ system.

We will use this to simplify the presentation of various EWF arguments: rather than referring to a superobserver measuring an observer-plus-system in an entangled basis, we will describe them as undoing the observer's measurement (which is possible since the observer's measurements are modeled as unitary interactions in the EWF arguments we present; see Section~\ref{sec_backass}) and then just measuring the observer's system in the $\pm$ basis.

One should note that depending on one's interpretation of quantum theory, this operational equivalence may not correspond to ontological equivalence, so it is conceivable that some interpretation might have a different response to an EWF argument depending on which of these protocols Wigner follows. However, we are not aware of any cases where this occurs.

\section{Background assumptions going into EWF arguments}
\label{sec_backass}

There are a number of background assumptions that are common to all EWF arguments. These assumptions were already used, whether implicitly or explicitly, in the previous sections. Note that some of the assumptions we list here are slightly stronger than what is strictly necessary, but we choose to focus on the essential feature of each assumption rather than the weakest possible version that is required.

The assumptions in question relate to agents who perform measurements and to what quantum theory says about such situations. As is standard, we will not bother to rigorously define what is meant by an \enquote{observer}, but we consider it a sufficient condition that the entity in question is similarly constituted to ourselves---e.g., a human. Similarly, the definition of what it means to \enquote{perform a measurement} is difficult to pin down precisely, but what is meant by this is clear enough for the purpose of constructing thought experiments.

First, it is assumed\footnote{In the Frauchiger-Renner argument, it has been argued that the assumption of Absoluteness of Observed events is not needed, but we challenge this in Section~\ref{sec_shiftyAOE}.} that the outcome obtained when an observer carries out a measurement process is single and absolute (even if the observation is part of a unitary process). This assumption is often called the {\em Absoluteness of Observed Events}~\cite{bongStrong2020,ormrod2022no}. Note that one does not require that any record of the outcome persists at later times.
Two notable classes of interpretations that reject this assumption are many-worlds interpretations~\cite{Everett1957} (in which there are generally multiple outcomes in multiple worlds) and perspectival (or \enquote{relational}) interpretations (in which each outcome generally has meaning only relative to the observer that observed it). Note that Absoluteness of Observed Events is also necessary~\cite{wisemanCausarum2017} for most no-go results like Bell's theorem, although this is often left implicit in that context.

Second, it is assumed that any system can be described by a Hilbert space, that the evolution of any closed system can be described by unitary dynamics, and that when a measurement is performed, the frequency with which its outcome arises always obeys the Born rule. In particular, these assumptions apply even to macroscopic systems, such as observers.  One notable class of interpretations that rejects unitarity for macroscopic objects is collapse theories~\cite{GRW,alloriWave2020,Penrose94,ChaMcQ21}, in which macroscopic enough systems do not evolve unitarily. Meanwhile, the claim that measurement outcome frequencies must obey the Born rule is rejected, for example, by many-worlds interpretations and QBism.

These assumptions are sometimes called the \enquote{universality of quantum theory}. 
However, note that here we are not making any assumptions about the projection postulate from textbook quantum theory. This is because (unlike the arguments in Section~\ref{sec_Build}) the EWF arguments we will cover in this paper (at least as we have presented them\footnote{In the original~\cite{Frauchiger2018} (measure-prepare) presentation of the Frauchiger-Renner argument one actually does appeal to the state-update postulate, a feature of the argument that has been criticized~\cite{Sudbery_2017,lazaroviciHow2018,baumann2016measurement,leegwaterWhen2022}.}) do not involve combining reasoning which uses both the projection postulate and the unitary evolution rule. Rather, they rely only on the Born rule and unitarity.

This is interesting because (as we mentioned at the end of Section~\ref{sec_MMP}) it seems intuitive that the two distinct descriptions of how a state is updated upon measurement (the projection postulate on the one hand, and unitarity on the other) are in tension, but it is much less clear why the Born rule itself (which has nothing to do with how one makes predictions for {\em future} measurements on the same system) would be in tension with the unitary evolution rule.\footnote{Interestingly, the Local Friendliness argument can be given in a manner that sidesteps the issue of whether the Born rule can be applied to outcomes of measurements that must be modeled unitarily to arrive at a contradiction. That is, inequality constraints on empirical correlations can be derived from the Local Friendliness assumptions, which do not refer to quantum theory at all; moreover, the predictions for these empirical correlations using quantum theory do not require one to apply Born rule on those measurements that must be modeled as unitary dynamics for violating the inequalities.}

Third, it is assumed that at least in principle, it is possible for a sufficiently advanced agent (often called a superobserver~\cite{Wheeler1957}) to apply an arbitrary unitary to any system (possibly excluding themself). In particular, even if one considers a macroscopic system containing another agent, such a superobserver could apply an arbitrary unitary to that system. Given (by the universal applicability of unitary dynamics) that the process of an observer carrying out a measurement process on some system can be treated as a unitary interaction $U_{\rm SF}$ between that observer (${\rm F}$) and the system (${\rm S}$), it follows that it is possible at least in principle that a superobserver could later apply the inverse unitary $U^\dagger_{\rm SF}$, completely reversing the measurement process. Such a reversal erases all records of the measurement outcome and returns the state of the observer, the measured system, and everything else in the joint system ${\rm SF}$ to the initial quantum state prior to the measurement.

These are the key background assumptions required for all EWF arguments. The specific EWF arguments we will cover in this paper each make an additional one or two assumptions beyond these. We will present these extra assumptions as they are needed. The degree to which each EWF argument is compelling is largely determined by how compelling the extra assumptions it requires are. 

As mentioned in the introduction, EWF arguments do \emph{not} invoke any notable assumptions of classical realism (such as ontic states). This makes them robust against many of the usual ways out of no-go theorems, such as interpretations that reject or aim to generalize such classical realist concepts.

\subsection{Motivating the background assumptions}
\label{sec_motivate}

Let us now mention and respond to a common objection to these background assumptions. According to the assumptions, an observer may carry out a measurement, followed by which a superobserver can completely reverse that measurement by applying the appropriate inverse unitary. But one sometimes hears it asserted (see e.g. Ref.~\cite{Lawrence2023relativefactsof}) that measurement processes are necessarily irreversible, and consequently that either 1) the reversal is not possible, or 2) the process in question was \enquote{not really a measurement}. 

This is a valid position to take---a position which rejects the universal applicability of unitary dynamics discussed in the previous section.\footnote{Alternatively, one could reject the idea that superobservers could possibly exist, as done by Peres~\cite{peresQuantum1995}. However, we are not aware of any actual arguments for why this would be the case; certainly Peres does not properly justify it in Ref.~\cite{peresQuantum1995}. Moreover, Ref.~\cite{Wiseman2023thoughtfullocal} has given arguments for why it may be possible to have a superobserver.} 
However, for this to be a sensible position, one must provide a specification of exactly how and under what circumstances a system evolves irreversibly (and moreover, a specification of exactly what kind of process constitutes a measurement). It is {\em not} sufficient to appeal to standard decoherence theory, since standard decoherence (as with all unitary interactions) is  reversible {\em in principle}, even if it is not reversible {\em in practice}. 
So, taking this view seriously requires that quantum theory is {\em incorrect} at some scales, and presumably leads one to adopt a collapse theory.

But one sometimes hears it claimed that the background assumptions going into EWF arguments are \enquote{trivially} wrong---they are dismissed {\em out-of-hand} due to the belief that it is an \enquote{incorrect application of quantum theory} to give a unitary description of a measurement process where there is a single and absolute measurement outcome. (Similarly, it is sometimes argued that the Born rule will never be relevant to any unitary process, and the fact that many EWF arguments apply it to processes that are treated unitarily is seen as problematic.)
Such objections miss the point. 
The validity or falsehood of the conjunction of the background assumptions in Section~\ref{sec_backass} {\em cannot} be deduced by operational quantum theory---it is a matter of interpretation. 
And there do exist serious interpretations that hold that all systems evolve unitarily---even systems containing agents who observe measurement outcomes. Certain Copenhagenish interpretations are of this sort, and as another example, in Bohmian mechanics, it is possible to describe a measurement process as a unitary while having an absolute measurement outcome for it.  So the background assumptions are not in immediate logical contradiction, and many physicists (both experts and non-experts) do subscribe to all of them. This is why EWF no-go theorems are useful---because they are a tool for studying and formalizing any apparent tension between such assumptions.

For those inclined to reject these background assumptions, EWF arguments provide fuel for the fire.
If one {\em grants} the background assumptions, then a given EWF argument constitutes an argument against the additional assumptions (which we have not yet introduced, but will introduce as we go along) going into that specific argument.
Either way, these arguments clearly demonstrate the fact that textbook quantum theory is not sufficiently precise as a physical theory so that everyone agrees on how to apply it in certain experiments to make predictions. 

\section[Preliminary: Hardy's argument]{Preliminary: Hardy's proof of Bell nonclassicality}
\label{sec_Hardy}

We now recount a simple proof of Bell nonclassicality, since the mathematics of this proof form a basis around which many extended Wigner's friend arguments can be built (namely the Local Friendliness argument and those by Brukner, Frauchiger-Renner, Pusey-Masanes, as well as a few related works~\cite{Healey_2018,leegwaterWhen2022,ormrod2022no}). Furthermore, familiarizing oneself with various possible responses to Bell's theorem helps one better understand the value of these EWF arguments (in particular, because one sees that some kinds of responses to the former are no longer adequate responses to the latter).  The proof we give is a simple case of an argument due to Hardy~\cite{hardyNonlocality1993}. 

\begin{figure}[htb!]
    \centering
    \begin{subfigure}[t]{0.23\textwidth}
        \centering
        \includegraphics[width=\textwidth]{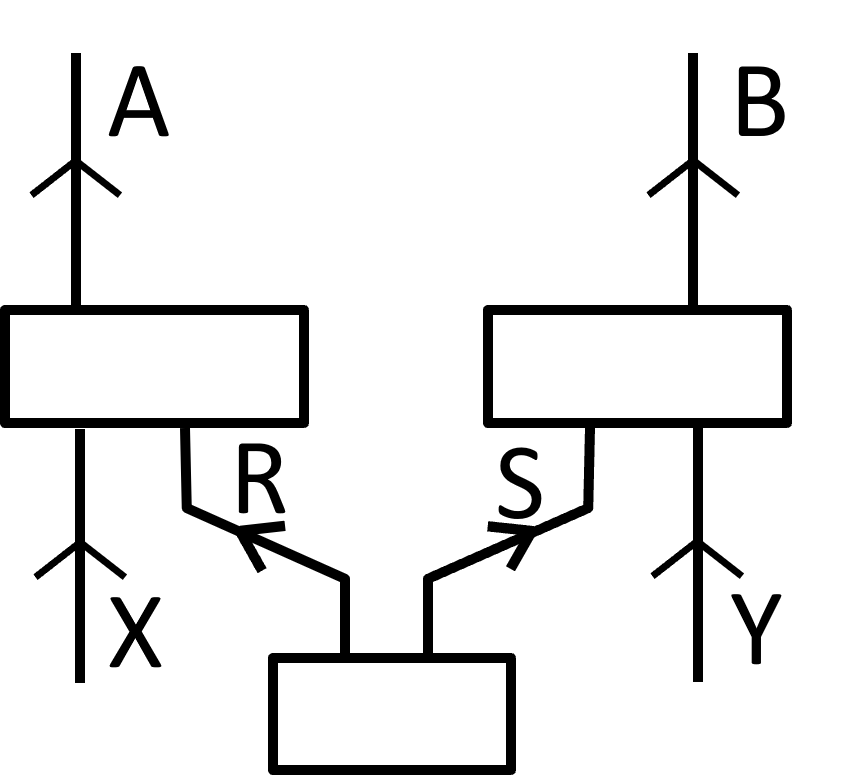}
        \caption{}
        \label{HD1}
    \end{subfigure}
    \hfill
    \begin{subfigure}[t]{0.23\textwidth}
        \centering
        \includegraphics[width=\textwidth]{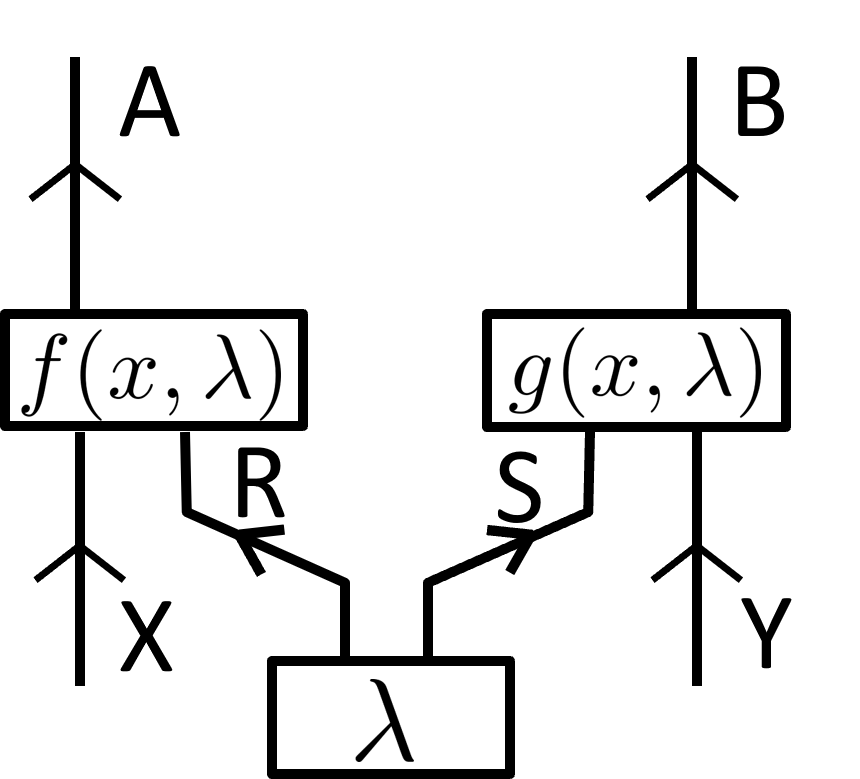}
        \caption{}
        \label{HD2}
    \end{subfigure}
    \caption{a) The quantum circuit for Hardy's proof of Bell nonclassicality. Here, ${\rm R}$ and ${\rm S}$ are quantum systems. b) A classical circuit of the same form as in (a), but where systems ${\rm R}$ and ${\rm S}$ are associated with classical variables rather than Hilbert spaces. Hardy's argument establishes that the set of correlations $p(ab|xy)$ which can be generated by the circuit in (a) is strictly larger than that which can be generated by the circuit in (b).  }
    \label{fig_HardyNL}
\end{figure}

Consider a scenario wherein local measurements are performed on each half of a bipartite quantum system ${\rm RS}$, as shown in the circuit of Figure~\ref{fig_HardyNL}. These measurements have binary settings $X(Y)$ and outcomes $A(B)$. Take the state in question to be 
\begin{equation}
\ket{\psi_{\rm Hardy}}_{\rm RS} := \frac{1}{\sqrt{3}}(\ket{00}_{\rm RS}+\ket{01}_{\rm RS}+\ket{10}_{\rm RS}),
\end{equation}
and take the measurements to be in the computational basis for $x=0$ ($y=0$), and to be in the complementary $\pm$ basis for $x=1$ ($y=1$). That is, for $x=0$, system ${\rm R}$ is measured in the computational basis
\begin{equation}
\big\{\ket{0}\bra{0},\ket{1}\bra{1}\big\}, 
\end{equation}
while for $x=1$, it is measured in the complementary basis
\begin{equation}
\big\{\ket{+}\bra{+},\ket{-}\bra{-}\big\}.
\end{equation}
System ${\rm S}$ is measured in these same two bases when $y=0$ and $y=1$, respectively.

One can immediately verify a specific set of predictions~\cite{Santos2021} made by quantum theory, summarized in Table~\ref{tab_HardyNL}. When the settings are $(x,y) = (0,0)$, one will never observe the joint outcome $(a,b) = (1,1)$; when the settings are $(0,1)$, the joint outcome $(0,-)$ is never observed; when the settings are $(1,0)$, the joint outcome $(-,0)$ is never observed; finally, when the settings are $(1,1)$, the joint outcome {\em is sometimes} $(-,-)$. That is,
\begin{align}
    p(a=1,b=1|x=0,y=0)&=0, \label{QMpredHD1} \\
    p(a=0,b=-|x=0,y=1)&=0, \label{QMpredHD2}\\
    \label{QMpredHD3}
    p(a=-,b=0|x=1,y=0)&=0, \\ 
    \label{QMpredHD4} p(a=-,b=-|x=1,y=1) &\neq 0.
\end{align}

Rephrased in modern terms~\cite{wood2015lesson}, the question Bell asked was whether the quantum predictions for the circuit in Figure~\ref{fig_HardyNL}(a) can be reproduced by a common-cause classical causal model~\cite{pearl2009causality} of the same form, as shown in Figure~\ref{fig_HardyNL}(b).

In this causal model, the correlations between $a$ and $b$ are generated by a classical variable $\lambda$ which acts as a common cause. The outcome $a$ depends causally on $x$ and $\lambda$ according to the function\footnote{Note that there is no loss of generality in taking these dependences to be functional rather than stochastic, as first proved by Fine~\cite{Fine1982}.}
$f(x,\lambda)$, while $b$ depends causally on $y$ and $\lambda$ according to the function $g(y,\lambda)$.
For any given $\lambda$, the function ${\rm f}$ specifies the value of $a$ that would be observed for every possible $x$. So for each $\lambda$, we can define the value assignments to the outcomes of the observables corresponding to $x=0$ and $x=1$, namely, the value assignments $A_0 := f(\lambda,0)$ and ${A_1 := f(\lambda,1)}$. Similarly, one can define the value assignments ${B_0 := g(\lambda,0)}$ and $B_1 := g(\lambda,1)$. Note that at most one of the value assignments $A_0$ and $A_1$ are observable in a given run of an experiment, since the measurements which would reveal these value assignments are incompatible, and similarly for $B_0$ and $B_1$. 
Thus, the other two value assignments are counterfactual quantities in that given run.

To be consistent with the quantum prediction that outcome $(a,b)=(1,1)$ never occurs for settings $(x,y)=(0,0)$, it must be that the value assignments satisfy
\begin{equation}
    B_0 = 1 \ \implies \ A_0 = 0 \label{eq_HD1}
\end{equation}
for every possible value of $\lambda$ that could arise given a preparation of the Hardy state. Similarly, the other three quantum predictions for settings $(x,y)=(0,1)$, $(1,0)$, and $(1,1)$ as shown in the first two columns of Table~\ref{tab_HardyNL} imply that all such $\lambda$'s must make value assignments respectively satisfy 
\begin{align}
    A_0 = 0 \ \implies \ B_1 = +, \label{eq_HD2}\\
    A_1 =- \ \implies \ B_0 = 1, \label{eq_HD3}\\
    A_1 = - \ \centernot\implies \ B_1 = +.\label{eq_HD4}
\end{align}
These implications are summarized in the third column of Table~\ref{tab_HardyNL}.

\begin{table}[h]
    \centering
    \setlength\tabcolsep{1mm}
\begin{tabular}{|c|c|c|}
    \hline
    \rowcolor{gray!15}
    \textbf{settings $\mathbf{(x,y)}$} & \textbf{$\mathbf{p(a,b|x,y)}$} & \textbf{implication} \\ \hline
    0,0 & $p(1,1|0,0)=0$ & $B_0 = 1 \ \implies \ A_0 = 0$  \\ \hline
    0,1 & $p(0,-|0,1)=0$ & $A_0 = 0 \ \implies \ B_1 = +$  \\ \hline
    1,0 & $p(-,0|1,0)=0$ & $A_1 =- \ \implies \ B_0 = 1$  \\ \hline
    1,1 & $p(-,-|1,1) \ > \ 0$ & $A_1 = - \ \centernot\implies \ B_1 = +$  \\ \hline
\end{tabular}
\caption{The quantum predictions for the correlations between pairs of outcomes in the Hardy argument and their implications for the value assignments made by $\lambda$.} 
\label{tab_HardyNL}
\end{table}

However, it is not hard to see that no set of value assignments can satisfy these constraints. Since logical implications are transitive, chaining the implication in Eq.~\eqref{eq_HD3} with that in Eq.~\eqref{eq_HD1} and then that in Eq.~\eqref{eq_HD2} yields 
\begin{equation}
\label{eq_chain}
A_1 =- \ \implies \ B_0 = 1 \ \implies \ A_0 = 0 \ \implies \ B_1 = +,
\end{equation}
in direct contradiction with the implication in Eq.~\eqref{eq_HD4}.

Consequently, it follows that no classical common-cause circuit like that shown in Fig.~\ref{fig_HardyNL}(b) can reproduce the quantum predictions for these measurements. This is known as Bell's theorem.

The classical causal modeling framework is essentially equivalent to the framework of ontological models~\cite{Harrigan}, so Bell's theorem appeals to the usual notion of classical realism in physics. Meanwhile, the assumption that the causal structure is common-cause is motivated by locality: by imagining that the measurements performed by Alice and Bob are at space-like separation and appealing to relativity theory. 

\subsection{Several common responses to Bell's theorem} 
\label{sec_responseBell}

One common type of response to Bell's theorem is to reject the idea that the underlying causal structure of a Bell experiment is the common-cause structure represented in Fig.~\ref{fig_HardyNL}.  For example, one could allow for an additional cause-effect relation between a setting at one wing and an outcome at another wing. For spacelike separated measurements, this implies superluminal causation\footnote{The common term \enquote{Bell nonlocality} strongly connotes this response, which is one of the many possible responses.  Hence, we prefer the more neutral term Bell nonclassicality~\cite{Wolfe2020quantifyingbell}.}, so such strategies are difficult to reconcile with relativity theory. However, they have the advantage that they allow one to retain the classical causal modeling framework (or equivalently, the framework of ontological models). 
If one allows the outcome $a$ to depend also on the setting $y$ and the outcome $b$ to depend also on the setting $x$, then the relevant value assignments can depend on the nonlocal context, and one must in general denote them ${A_{x,y}:=g(\lambda,x,y)}$ and $B_{x,y}:=g(\lambda,x,y)$. Now, the third and the first implications in Table~\ref{tab_HardyNL} would become ${A_{x=1,y=0} =-} \ \implies \ {B_{x=1,y=0} = 1}$ and $B_{x=0,y=0} = 1 \ \implies \ A_{x=0,y=0} = 0$, respectively. These value assignments cannot be chained together using transitivity, and so no contradiction arises. (Retrocausation and superdeterminism account for Bell nonclassicality in a mathematically similar way, where the value assignments for the outcomes would be denoted by other quantities than $A_x$ and $B_y$.) The most well-known interpretation which involves such superluminal causal influences is Bohmian mechanics.

However, many physicists reject this type of response to Bell's theorem on the grounds that it introduces tension with relativity, with noncontextuality~\cite{Spe05,SpekLeibniz19}, and with an assumption of no-fine-tuning~\cite{wood2015lesson}. Motivated by these, one can seek alternative sorts of causal explanations wherein one preserves a kind of locality by modifying the background assumptions going into Bell's theorem---for instance, assumptions embedded in the classical causal modeling framework. 

One research program which aims to do this is that of quantum causal modeling~\cite{costa2016quantum,allen2017quantum,Barrett2019}, or more generally, of nonclassical causal modeling~\cite{henson2014theory,schmid2021unscrambling}.  A key background assumption in the framework of classical causal models is Reichenbach's principle, which can be refined into a qualitative assumption and a quantitative assumption~\cite{cavalcantiModifications2014,wisemanCausarum2017}. The qualitative part of Reichenbach's principle demands the existence of a common cause between any two correlated variables that do not causally influence one another, while the quantitative part demands that this common cause is a classical variable with the property that when one conditions on it, the correlations are eliminated.
A guiding idea of the research agenda of nonclassical causal modeling is to generalize the quantitative Reichenbach's principle, while holding on to the qualitative Reichenbach's principle. That is, one demands that the causal structure of a Bell scenario is indeed common-cause, but allows for the possibility that the common cause in question is described by something more general than a classical variable. A proposed generalization of the quantitative Reichenbach's principle can be found, for example, in Ref.~\cite{allen2017quantum}. 

Another key component of the framework of classical causal modeling that can be challenged is the calculus it provides for doing counterfactual reasoning~\cite{pearl2009causality,spirtes2000causation}. Some interpretations---most notably operationalism and Copenhagenism---are cautious about or outright dismissive of counterfactual reasoning.
In this view, one rejects the idea that one can combine the data from different experimental arrangements which implement incompatible measurements. As Peres famously put it, \enquote{\emph{unperformed experiments have no results}}~\cite{Peres1978}.
Taken seriously, this view leads one to reject the idea that the shared bipartite quantum system can be associated with some underlying classical common cause $\lambda$ which gives meaning to all four value assignments in any single run of the experiment. 
Rather, in a given run where $x=x_0$ and $y=y_0$ are the setting choices, the only two values about which one is allowed to reason are $A_{x_0}$ and $B_{y_0}$, by virtue of the fact that the associated measurements are actually performed. So one can evade Bell's theorem by rejecting notions such as ontic states and the value assignments they make.  (Although in such an approach, it is not clear to us that there is any nontrivial notion of locality which might be preserved.)

A final strategy for avoiding Bell's theorem is to reject the idea that measurement outcomes are single and absolute. Rather, the outcomes might only have meaning relative to some particular kind of context. Depending on the specific perspectival interpretation in question, measurement outcomes might be defined relative to
\begin{itemize}
\item the observer~\cite{fuchs2009quantumbayesian,fuchs2010qbism,FuchsMerminSchack,fuchs2019qbism}
\item the systems involved in the interaction~\cite{rovelli1996relational,dibiagioStable2021}
\item the world-branch~\cite{Everett1957}  (in some Many Worlds interpretations)
\item what future measurements are done~\cite{renes2021consistency,polychronakos2022quantum},
\end{itemize}
and so on.
If measurement outcomes only have meaning relative to one of these, then it will not always be possible to combine statements about outcomes defined relative to different entities into a single global picture. For example, if Alice's outcome is only be meaningful to Alice, and Bob's outcome is only meaningful to Bob, then it is meaningless to try to make inferences about one based on the other (at least if the two do not interact or communicate to each other.)

\section{Brukner's argument}
\label{sec_brukner}

Brukner~\cite{brukner2015quantum,bruknerNogo2018} was the first\footnote{The argument presented in Ref.~\cite{myrvoldModal2002} by Myrvold in 2002 resembles a number of EWF arguments~\cite{Healey_2018,leegwaterWhen2022,ormrod2022no} developed later, and precedes Brukner's argument in Ref.~\cite{brukner2015quantum}. However, Myrvold did not make any connection to Wigner's friend in Ref.~\cite{myrvoldModal2002}.} to recognize the importance of combining elements of the Wigner's friend thought experiment with elements of a Bell setup. He provided an argument using such a mashup and aimed to undermine the objectivity of measurement outcomes. 

Unfortunately, as we will see, Brukner's argument uses assumptions that are unnecessarily strong and so do not support the intended conclusion. Nevertheless, the ideas which motivate Brukner's work (at least as we see it) are quite illustrative, and this work inspired the closely related Local Friendliness argument of Ref.~\cite{bongStrong2020,Cavalcanti2021} (covered in Section~\ref{sec_LF}), which uses the same experimental setup as Brukner's, and which {\em does} succeed in  highlighting the tension between the objectivity of outcomes and other a priori reasonable assumptions.

Recall from the previous section that one possible response to proofs of Bell nonclassicality is to reject the assumption of classical realism, and in particular, the notion of value assignments it entails. That is, one can reject arguments for Bell nonclassicality on the grounds that only two of the four value assignments $A_{0},A_{1},B_{0},B_{1}$---namely, the two that are actually observed---are meaningful in any given run of the experiment. 

Brukner's argument can be motivated as an attempt to challenge such responses. To do this, one introduces superobservers into the Hardy argument in such a way that all of the incompatible measurements are {\em actually} measured in a single run of the experiment---albeit by different observers, and in a way that no single observer can ever know both outcomes of any pair of incompatible measurements. In so doing, the argument shifts focus from value assignments $A_{0},A_{1},B_{0},B_{1}$ (whose existence is contentious) to {\em actually observed measurement outcomes}, whose existence and objectivity are much harder to question. Such an argument need not appeal to the strong assumption of classical realism which underpins Bell-type arguments, and so it is clear that rejecting the notions of ontic states or classical causal explanations will not help one to evade the conclusion. Rather, one is forced to a more radical conclusion---for example, the view that outcomes of measurements are observer-dependent. 

\begin{figure}[htb!]
    \centering
    \includegraphics[width=0.49\textwidth]{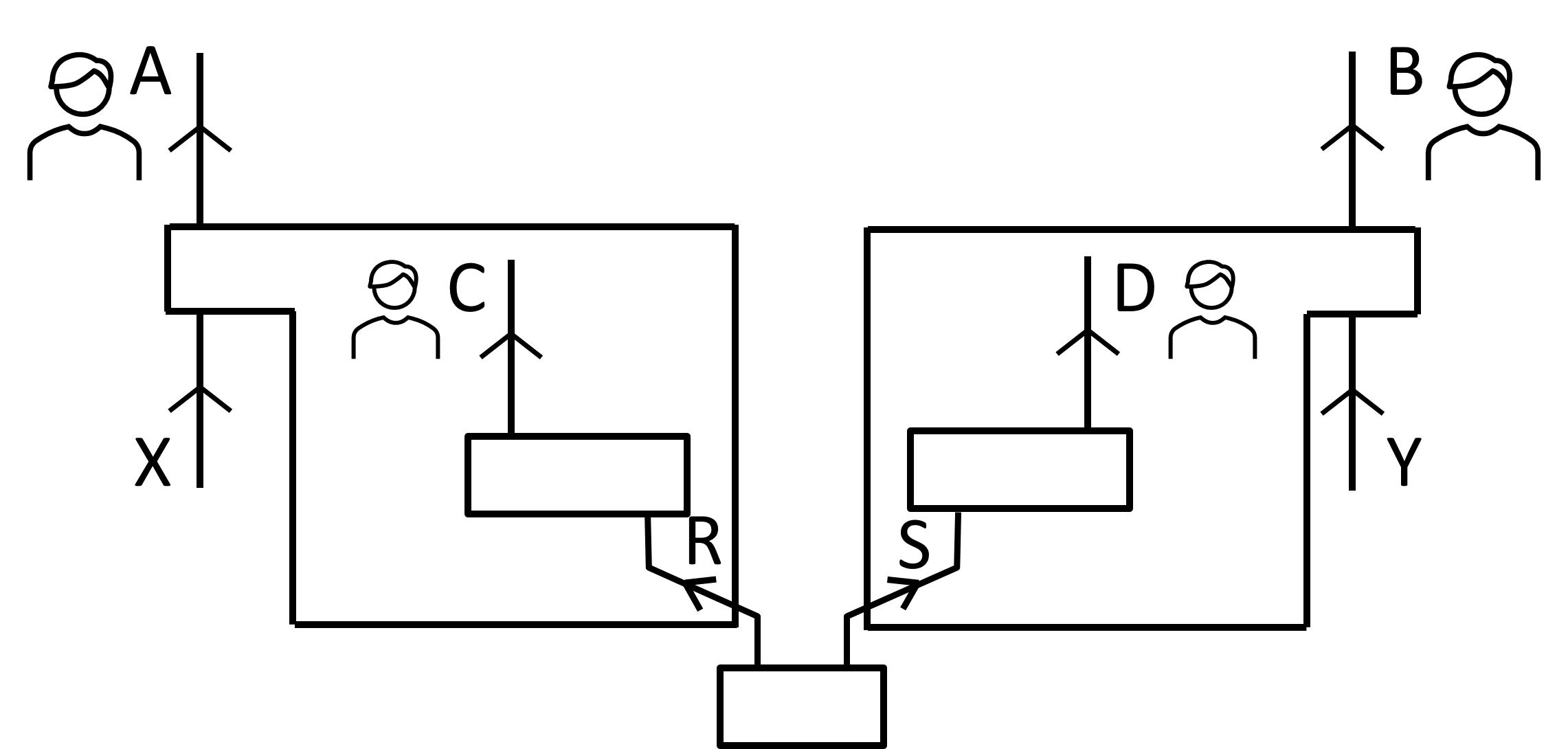}
    \caption{Brukner's extended Wigner's friend setup. Here, Alice and Bob are two superobservers who have control over Charlie's lab and Debbie's lab, respectively, and each superobserver has a binary measurement setting. Charlie and Debbie share an entangled state on ${\rm RS}$.}
    \label{fig_BruknerScenario}
\end{figure}

Brukner's setup considers two friends and two superobservers, as depicted in Fig.~\ref{fig_BruknerScenario}. The friends, Charlie and Debbie, share halves of an entangled system in the Hardy state
\begin{equation}
    \ket{\psi_{\rm Hardy}}_{\rm RS} = \frac{1}{\sqrt{3}}(\ket{00}_{\rm RS}+\ket{01}_{\rm RS}+\ket{10}_{\rm RS}).
    \end{equation}
Each friend performs a fixed measurement on their half of the system (${\rm R}$ and ${\rm S}$, respectively). This fixed measurement is in the computational basis
    \begin{equation}
    \big\{\ket{0}\bra{0},\ket{1}\bra{1}\big\}.
    \end{equation}    

Following the assumption from Section~\ref{sec_backass} that all processes can be given a unitary description, Charlie and Debbie's measurements can be modeled as unitary interactions. Thus, after Charlie and Debbie performed their measurements, the resulting joint state of Charlie (${\rm C}$), Debbie  (${\rm D}$) and the bipartite system ${\rm RS}$ is
\begin{equation}
    \frac{1}{\sqrt{3}}(\ket{00}_{\rm RS}\ket{0}_{\rm C}\ket{0}_{\rm D}+\ket{01}_{\rm RS}\ket{0}_{\rm C}\ket{1}_{\rm D}+\ket{10}_{\rm RS}\ket{1}_{\rm C}\ket{0}_{\rm D}).
\end{equation}

The superobservers, Alice and Bob, are assumed to be so technologically advanced that they can reverse Charlie's and Debbie's  measurements, respectively.
After Charlie does his measurement, Alice has two choices: when her measurement choice $x$ is 0, she asks Charlie what he observed; when her measurement choice $x$ is 1, she undoes Charlie's measurement by performing the inverse unitary, and then measures his system ${\rm R}$ in the complementary $\pm$ basis    
\begin{equation}
    \big\{\ket{+}\bra{+},\ket{-}\bra{-}\big\}.
\end{equation}
Similarly, after Debbie does her measurement, Bob has two choices: when his measurement choice $y$ is 0, he asks Debbie what she observed; when his measurement choice $y$ is 1, he undoes Debbie's measurement and then measures her system ${\rm S}$ in the same complementary basis. In this way, Alice and Bob each choose between two incompatible measurements; it is just that in the first case, the measurement in question has already been performed by the Friend. (Since no further quantum operations are performed on Alice and Bob in this experiment procedure, there is no reason to treat Alice and Bob's measurements unitarily.)

Note that Brukner's presentation in Ref.~\cite{brukner2015quantum} has each superobserver measuring their respective friend and the system in a particular entangled basis, rather than undoing the friend's measurement and measuring only the system in a complementary basis. These are equivalent in terms of operational statistics, as we proved in Section~\ref{usefulfact}, but we believe our presentation is simpler. 

Next, Brukner makes an assumption which he terms Observer-Independent Facts. It states that {\em  \enquote{The truth values of the propositions of all observers form a Boolean algebra. Moreover, the algebra is equipped with a (countably additive) positive measure for all statements in the Boolean algebra, which gives the probability for a given statement to be true}}. Here, Brukner does not specify what the set of propositions in question is. From the context, there are two likely candidates: i) the set corresponding to outcomes for every possible measurement that an observer {\em might do}, and ii) the set corresponding to the outcomes for every measurement that some observer {\em actually did}.

The assumption of Observer-Independent Facts seems meant to be akin to Absoluteness of Observed Events. However, if one takes the first option, then this assumption is in fact a much stronger assumption. Rather than merely claiming that there is an objective fact of the matter about what the outcome of a particular {\em realized} measurement was, the assumption would also demand that there is a fact of the matter about the outcomes of all possible-but-unperformed measurements.

To best see this, let us first denote Alice's measurement outcome when $x=0$ by $a_0$, and her outcome when $x=1$ by $a_1$. In particular, $a_0$ is the outcome when system ${\rm R}$ is measured in the computational basis,  and $a_1$ is the outcome when it is measured in the $\pm$ basis. Define $b_0$ and $b_1$ similarly, so $b_0$ is the outcome when system ${\rm S}$ is measured in the computational bases and $b_1$ is the outcome when it is measured in the $\pm$ basis. These quantities are useful for comparison with the value assignments $A_{0},A_{1},B_{0},B_{1}$ from the previous section.

Brukner rightly points out that if observed events are absolute, then $a_{0}$ and $b_{0}$ are well-defined, definite quantities in every run of the experiment, since they are {\em actually measured} by Charlie and Debbie in every run. 
However, one cannot similarly argue that $a_{1}$ and $b_{1}$ are necessarily well-defined, definite quantities in every run of the experiment, since $a_{1}$ is {\em not} measured in runs when $x=0$, and  $b_{1}$ is not measured when $y=0$. 

So when Brukner states that \emph{\enquote{The assumption of \enquote{observer-independent facts}... [requires] an assignment of truth values to statements $a_0$ and $a_1$ independently of which measurement Wigner [the superobserver] performs}} (notation modified to fit our conventions), he is stating that $a_1$ is well-defined even in runs where it is not measured. But this requires assuming that even unperformed measurements have objective outcomes. Consequently, Brukner's notion of observer-independent facts is much stronger than the assumption of Absoluteness of Observed Events. Together with Brukner's additional assumption that such outcomes are independent of its distant context (for example, $a_0$ is assumed to be independent of Bob's choice), 
one basically recovers the same assumptions going into Bell's theorem. (This seems to be an oversight rather than an intentional choice.)  Thus, Brukner's theorem does not provide any new constraints on nature beyond those provided by prior proofs of Bell nonclassicality, like the one we gave in Section~\ref{sec_Hardy}. 

However, one \emph{can} obtain a more useful argument if one takes the second reading of Brukner's assumption of Observer-Independent Facts (that it applies only to propositions concerning outcomes of measurements that were actually performed). 
To do so, one focuses on the particular case when the superobservers measure in the $\pm$ basis, since in this case the systems on both wings are also being measured in the computational basis (by Charlie and Debbie, respectively); consequently, both measurements must have truth values in a single run of the experiment, despite the fact that they are incompatible. By following this line of reasoning, one arrives at the Local Friendliness argument~\cite{bongStrong2020,Cavalcanti2021}, which we will present now.

\section{The Local Friendliness no-go theorem}
\label{sec_LF}

The Local Friendliness no-go theorem was first proven in Ref.~\cite{bongStrong2020}, and then proven in a scenario using Hardy correlations in Ref.~\cite{haddaraPossibilistic2022}. The argument can be viewed as a completion and formalization of the ideas behind Brukner's argument from the previous section (i.e., the idea of transforming value assignments in question into observed outcomes by introducing an appropriate superobserver), extending them into a precise, theory-independent and empirically testable no-go theorem. This argument is built on assumptions that are strictly weaker than Brukner's (and Bell's), and it succeeds in providing an argument against the Absoluteness of Observed Events (modulo other assumptions in the theorem).

The Local Friendliness argument has been given in many different set-ups~\cite{bongStrong2020,Wiseman2023thoughtfullocal,haddaraPossibilistic2022,utrerasalarcon2023allowing,yile2023}. For simplicity and ease of comparison with the other EWF arguments, we will present the argument in a setup that is identical to Brukner's (where all value assignments in question are transformed into observed outcomes), and we will derive the no-go result in the context of quantum theory (instead of presenting it in a theory-independent form). Our presentation of the argument is inspired by Ref.~\cite{haddaraPossibilistic2022}. 

Consider again two friends, Charlie and Debbie, who share a Hardy state, $\ket{\psi_{\rm Hardy}}_{\rm RS}$, and measure it in the computational basis, and imagine two superobservers, Alice and Bob, with perfect quantum control over Charlie's lab and Debbie's lab, respectively. Alice and Bob each choose between two possible measurements. When Alice's measurement setting is $0$, she asks Charlie what he observed and takes Charlie's answer as her measurement outcome.\footnote{Actually, when $x=0$, Alice's measurement outcomes play no role in our particular presentation of the argument, so in this case Alice actually does not need to copy Charlie's outcome, but could instead do nothing at all. Similarly for Bob. This simplifies the argument and 
removes one possible source of contention, as we will discuss in Section~\ref{sec_track}.}
 When Alice's measurement setting is $1$, she undoes Charlie's measurement and then measures system ${\rm R}$ in the $\pm$ basis. Bob does the same: when his setting is $0$, he asks Debbie what she observed and copies Debbie's outcome; otherwise, he undoes Debbie's measurement and then measures system ${\rm S}$ in the $\pm$ basis. 

 We imagine that Alice and Charlie's operations are at spacelike separation from Debbie and Bob's operations. (In fact, all that is really needed for the argument is that $c$, $d$, $x$ and $a$ are not in the future lightcone of $y$, and that $c$, $d$, $y$, and $b$ are not in the future lightcone of $x$.)

\begin{figure}[htb!]
    \centering
    \includegraphics[width=0.49\textwidth]{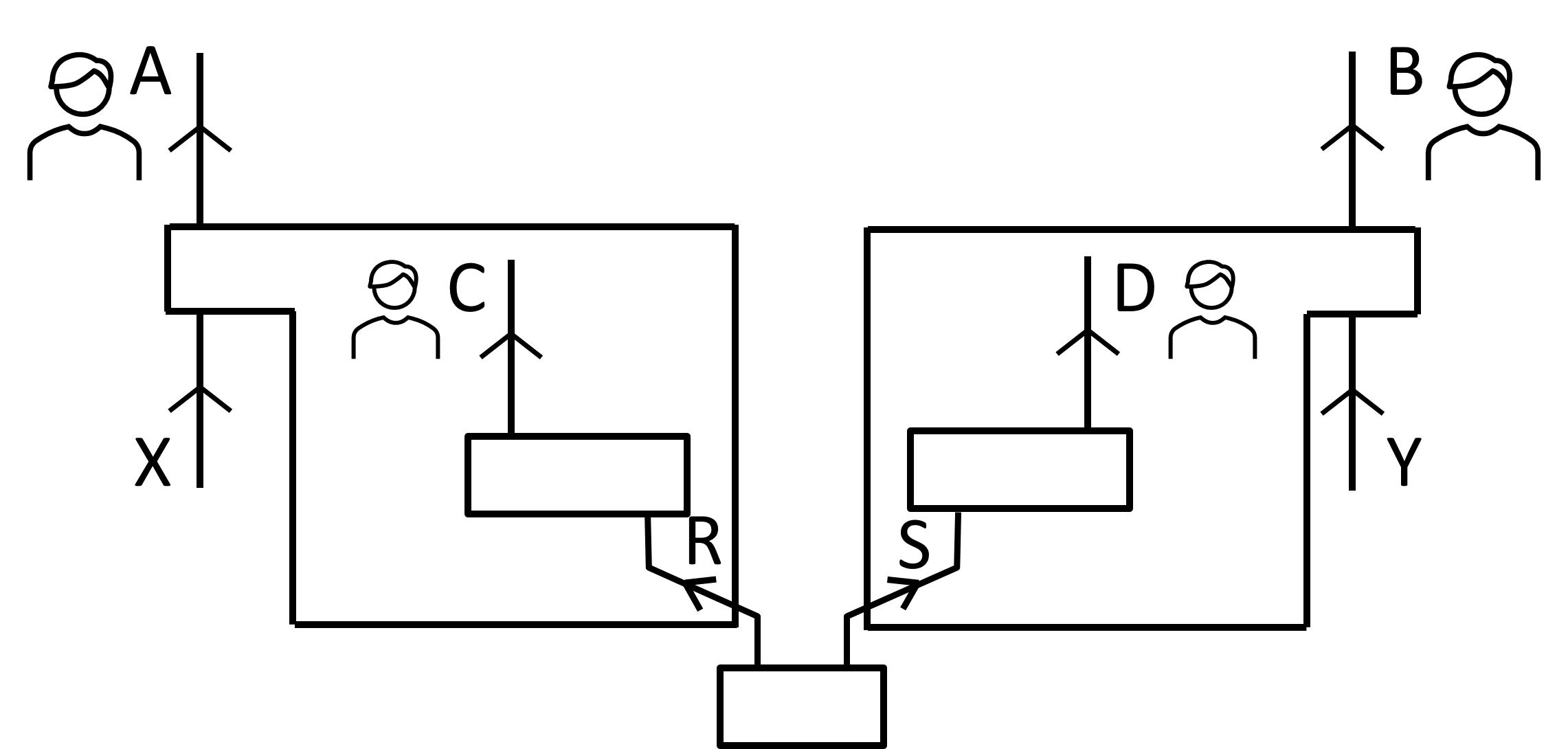}
    \caption{The Local Friendliness setup. (Exactly the same as Brukner's setup.) Here, Alice and Bob are two superobservers who have control over Charlie's lab and Debbie's lab, respectively, and each superobserver has a binary measurement setting. Charlie and Debbie share an entangled state on ${\rm RS}$.  }
    \label{fig_LF}
\end{figure}

The argument makes an assumption called {\em Local Friendliness}, which is the conjunction of two metaphysical assumptions. 

First, it assumes \emph{Absoluteness of Observed Events}, stating that an observed event is an absolute single event, and not relative to anything or anyone. (This was one of the background assumptions we mentioned in Section~\ref{sec_backass}.) Because Alice, Bob, Charlie, and Debbie are all agents who observe outcomes, which we denote by $a,b,c,d$, respectively, there is a single joint outcome for $(a,b,c,d,x,y)$ and consequently, a well-defined joint probability $p(a,b,c,d,x,y)$ in any run of the experiment. 

The second assumption is a weak notion of locality which we call \emph{Local Agency for Observed Events}.\footnote{This is weaker than the assumption of Local Agency used in Refs.~\cite{bongStrong2020,Cavalcanti2021,haddaraPossibilistic2022}, which demands that any freely chosen measurement setting is uncorrelated with any set of relevant events---observed or not---outside its future light cone.} This assumption demands that any freely chosen measurement setting is uncorrelated with any set of relevant \emph{observed} events outside its future light cone.\footnote{Even though this assumption reduces to operational no-signalling when applied to a Bell experiment, it is a stronger assumption, since it also constrains quantities that are not operationally accessible, such as the joint frequency with which Alice's setting and Debbie's outcome arise. It is a weaker assumption than parameter independence, since parameter independence requires the framework of classical causal models or ontological models as a prerequisite.} This is motivated by the idea that a free and randomly chosen variable can only be correlated with variables that it causally influences, and that it can only causally influence things in its future lightcone. Thus, this assumption encompasses a kind\footnote{Really, it is a weaker version of these assumptions than is needed for Bell's theorem, since here we need only constrain the relation between settings and observed events.} of no-superdeterminism, no-retrocausation, and no-superluminality.

To prove the Local Friendliness theorem, first note the following quantum predictions: 
\begin{align}
    p(c=1,d=1|x=0,y=0)&=0,\label{eq_AOELF1}\\
    p(c=0,b=-|x=0,y=1)&=0, \label{eq_AOELF2}\\
    p(a=-,d=0|x=1,y=0)&=0,  \label{eq_AOELF3}\\
    \label{QMpredLF4} p(a=-,b=-|x=1,y=1) &\neq 0,
\end{align}
which, up to labeling, are the same as those we saw in the previous section. These predictions can be directly verified by experiment.

Next, we will convert all the predictions in Eqs.~\eqref{eq_AOELF1}-\eqref{eq_AOELF3} about different pairs of measurement settings into predictions about a single pair of measurement settings, namely, $x=1$ and $y=1$.
This is critical because it is {\em only} when $x=1$ and $y=1$ that {\em both} the computational basis measurements and the $\pm$ basis measurements are performed on each half of the bipartite system (by some observer), and consequently, it is only in these runs that Absoluteness of Observed Events requires a joint outcome for all these measurements. (Recall that this was the primary intuition behind Brukner's argument; now we will see it formalized.)
To achieve this, we apply Local Agency for Observed Events to Eq.~\eqref{eq_AOELF1}-\eqref{eq_AOELF3}.

Since neither $c$ nor $d$ is in the future light cone of the setting $x$ or $y$, according to Local Agency for Observed Events, $c$ and $d$ should be uncorrelated with $x$ and $y$, meaning that $p(c,d|x,y)=p(c,d)$. Equivalently, $p(c,d|x,y) = p(c,d|x',y')$ for all $x,y,x',y'$, and so from Eq.~\eqref{eq_AOELF1}, we have
\begin{equation}\label{eq_LALF1}
    p(c=1,d=1|x=1,y=1)=0.
\end{equation}
Intuitively: if $x$ and $y$ had been different from what they were in Eq.~\eqref{eq_AOELF1} (namely 0), then the probability distribution over $(c,d)$  would have remained unchanged, because neither $c$ nor $d$ is in the future light cone of $x$ or $y$.\footnote{This intuitive argument is framed as a counterfactual statement, but one could maintain that the local agency for Observed Events assumption itself is not. So although one cannot avoid reasoning about different incompatible measurement contexts in this argument, one can avoid appealing to counterfactuals, if one cares to.}

Similarly, since neither $c$ nor $b$ is in the future light cone of the free choice $x$, we have $p(c,b|x,y)=p(c,b|y)$, and so Eq.~\eqref{eq_AOELF2} implies
\begin{equation}\label{eq_LALF2}
    p(c=0,b=-|x=1,y=1)=0.
\end{equation}
Finally, since neither $a$ nor $d$ is in the future light cone of the free choice $y$, we have $p(a,d|x,y)=p(a,d|x)$, and so Eq.~\eqref{eq_AOELF3} implies
\begin{equation}\label{eq_LALF3}
    p(a=-,d=0|x=1,y=1)=0.
\end{equation}

Absoluteness of Observed Events demands the existence of a joint outcome $(a,b,c,d)$ when $x=1$ and $y=1$, since all four of these outcomes are actually observed by some observer in a single run. From Eqs.~\eqref{eq_LALF1}-\eqref{eq_LALF3} and Eq.~\eqref{QMpredLF4}, we have that in such runs of the experiment, it must be that
\begin{align}    
    d = 1 \ &\implies \ c = 0, \label{eq_LF1}\\
    c = 0 \ &\implies \ b = +, \label{eq_LF2}\\
    a =- \ &\implies \ d = 1, \\
    a = - \ &\centernot\implies \ b = +.
\end{align}
We summarize these implications in Table~\ref{tab_LF}. This table is analogous to that for Hardy nonlocality (see Table~\ref{tab_HardyNL}).

 By transitivity, the first three of these imply that
\begin{equation}
    a=- \implies d=1 \implies c=0 \implies b=+,
\end{equation}
which contradicts the forth. 

Therefore, the quantum predictions cannot satisfy both Absoluteness of Observed Events and Local Agency for Observed Events. 

\begin{table}[h]
    \setlength\tabcolsep{0.5mm}
    \begin{adjustwidth}{-5mm}{}
    \begin{tabular}{|c|c|c|}
        \hline
        \rowcolor{gray!15}
        \textbf{quant. pred. } & \textbf{applying Local Agency} & \textbf{implication} \\ \hline
        Eq.~\eqref{eq_AOELF1} & $p(c=1,d=1|x=1,y=1)=0$ & $d = 1 \ \implies \ c = 0$  \\ \hline
        Eq.~\eqref{eq_AOELF2} & $p(c=0,b=-|x=1,y=1)=0$ & $c = 0 \ \implies \ b = +$  \\ \hline
        Eq.~\eqref{eq_AOELF3} & $p(a=-,d=0|x=1,y=1)=0$ & $a =- \ \implies \ d = 1$  \\ \hline
        Eq.~\eqref{QMpredLF4} &  {\rm (not needed)}  & $a = - \ \centernot\implies \ b = +$  \\ \hline
    \end{tabular}
    \end{adjustwidth}
    \caption{The quantum predictions, their transformation (using Local Agency for Observed Outcomes) into constraints on the runs where $x=1$ and $y=1$, and their implications for each pair of observed outcomes in the Local Friendliness argument.}
    \label{tab_LF}
\end{table}

At a technical level, Ref.~\cite{bongStrong2020} develops this no-go result considerably further. It characterizes the entire polytope of correlations consistent with the theory-independent Local Friendliness assumptions (for several choices of cardinalities for Alice and Bob's measurement settings\footnote{\label{foot_LA}In cases where these cardinalities are greater than two, not all measurements involved in the argument on system ${\rm R}$ and ${\rm S}$ can be performed in a single run of the experiment. Thus, it is still susceptible to the criticism that \enquote{unperformed measurements have no results} mentioned in Section~\ref{sec_responseBell}. Nevertheless, even in these cases, Local Agency for Observed Events is still enough for deriving a contradiction despite being a weaker assumption than Bell's local causality, as will be discussed in Section~\ref{sec_LAOE}.} and  outcomes). As such, the argument is not merely a no-go theorem against quantum theory, but against any theory or experimental data which violates the Local Friendliness inequalities defining the polytope. These further results make the Local Friendliness argument the most technically well-characterized EWF argument to date, and the Local Friendliness no-go theorem is the only empirically testable EWF argument so far. 

\subsection[Evaluating Local Agency for Observed Events]{Evaluating the Local Agency for Observed Events assumption}  
\label{sec_LAOE}

 Unlike most of the EWF arguments in this paper, the Local Friendliness argument involves an explicit assumption relating to locality. 
This locality assumption can be motivated by relativity theory.  One might naively think that this implies it to be on a similar footing to Bell's theorem, and that the standard responses to Bell's theorem will also be valid responses to the Local Friendliness theorem. But this is not the case, because the assumption of Local Agency for Observed Events needed in this argument is significantly weaker than Bell's assumption of local causality.

The fact that the former is weaker than the latter is explained thoroughly in Ref.~\cite{Cavalcanti2021}, which shows (as mentioned in Section~\ref{sec_responseBell}) how Bell's notion of locality (namely local causality) can be viewed as a consequence of the conjunction of the quantitative {\em and} qualitative Reichenbach's principles~\cite{reichenbachDirection1991,cavalcantiModifications2014}, while Local Agency for Observed Events can follow from the qualitative version alone. In particular, Local Agency for Observed Events (unlike Bell's local causality) does not imply that one can explain the correlations arising between any two spacelike separated measurements in terms of some classical variable (the complete common cause) that eliminates the correlations when conditioned upon.

Another way to see the difference between the two is to note that Bell's theorem can be understood as assuming a common-cause explanation {\em within the framework of classical causal models} (where the common cause must be a classical variable). Meanwhile, the Local Friendliness theorem can be understood as assuming a common-cause explanation, {\em but for any arbitrary nonclassical causal model} as captured within the framework of Ref.~\cite{henson2014theory} (where common causes may be associated with systems in any generalized probabilistic theory~\cite{Hardy,barrett2007,chiribella2010probabilistic}). In other words, just as Bell inequalities are causal compatibility inequalities within the classical framework for causal modeling, Local Friendliness inequalities are causal compatibility inequalities within the more general framework of nonclassical causal modeling~\cite{yile2023}.

A final way to see that the assumptions of the Local Friendliness theorem are weaker than those of Bell's theorem is that the polytope of correlations defined by the assumptions of the former is strictly larger (in some scenarios) than the polytope of locally causal correlations~\cite{bongStrong2020}. (However, one requires the arguments in the previous two paragraphs to show that the assumptions are weakened {\em in a physically motivated sense}.)

These facts imply that the Local Friendliness no-go theorem has important implications for interpretations that aim to give causal accounts of correlations using a nonclassical generalization of the framework of classical causal modeling (as first recognized in Ref.~\cite{Cavalcanti2021}). 

First of all, even if the program of quantum causal models succeeds in providing a satisfactory causal explanation of Bell correlations by introducing a notion of common-cause explanation that goes beyond the framework of classical causal models (e.g., by allowing the common causes in question to be generalized probabilistic systems and generalizing the quantitative Reichenbach's principle accordingly), such a strategy is insufficient on its own to evade the Local Friendliness no-go theorem.

Moreover, and as we foreshadowed in Section~\ref{sec_brukner}, the Local Friendliness no-go theorem also implies that it is not sufficient to merely reject the notion of value assignments (e.g., by dropping the ontological models framework, or the notion of classical causal models). Interpretations (such as some Copenhagenish interpretations) that take this basic strategy may need to go further; for instance, by rejecting Absoluteness of Observed Events and so embracing perspectivalism. Alternatively, they might try to provide arguments for why one should reject Local Agency for Observed Events in this context (even though they would maintain it in the context of Bell experiments, where it reduces to operational no-signaling).
For example, they might object that Local Agency for Observed Events allows one to reasoning about ``unperformed experiments'', especially in Local Friendliness set-ups where not all measurements involved in the argument on system ${\rm R}$ and ${\rm S}$ can be performed in a single run of the experiment, as mentioned in Footnote~\ref{foot_LA}.

Interpretations that stipulate that measurement outcomes are perspectival evade the Local Friendliness no-go theorem in much the same way as they evade Bell's no-go theorem. Strategies that change the causal structure (by appealing to superluminal influences, for example) also work similarly in both cases. 

\subsection{The Tracking assumption} 
\label{sec_track}

The argument presented in Ref.~\cite{haddaraPossibilistic2022} does not begin directly from the quantum predictions in Eqs.~\eqref{eq_AOELF1}-\eqref{QMpredLF4}; 
rather, it starts from the quantum predictions for $p(ab|xy)$, namely
\begin{align}
    p(a=1,b=1|x=0,y=0)&=0, \label{QMpredLF1p} \\
    p(a=0,b=-|x=0,y=1)&=0, \label{QMpredLF2p}\\
    \label{QMpredLF3p}
    p(a=-,b=0|x=1,y=0)&=0, \\ 
    \label{QMpredLF4p} p(a=-,b=-|x=1,y=1) &\neq 0.
\end{align}
The predictions for these correlations, like those of Eqs.~\eqref{eq_AOELF1}-\eqref{QMpredLF4}, are directly observable. 

One can then derive Eqs.~\eqref{eq_AOELF1}-\eqref{QMpredLF4} from the fact that $a=c$ whenever $x=0$ and $b=d$ whenever $y=0$, or equivalently
\begin{equation}\label{eq_track}
    \begin{split} 
    p(a | c, x = 0) = \delta_{a,c}, \\
    p(b | d, y = 0) = \delta_{b,d}.
    \end{split}
\end{equation}
Eq.~\eqref{eq_track} is an instance of an assumption that we will call the {\em Tracking} assumption, since it ensures that Alice's outcome tracks Charlie's perfectly, and that Bob's outcome tracks Debbie's perfectly, in the relevant runs of the experiment. Eq.~\eqref{eq_track} is simply meant to formalize the fact that when Alice's measurement setting is $x=0$, she asks Charlie what he observed and takes Charlie's answer as her measurement outcome, and similarly for Bob and Debbie. 

However, one sometimes hears the instance of the Tracking assumption in Eq.~\eqref{eq_track} challenged. 
For example, Ref.~\cite{okon2022reassessing} points out that there are alternative ways in which Alice can carry out her $x=1$ measurement yielding the same relative frequency for $a$ but that make this assumption seem much less innocent.  Specifically, when $x=1$, if Alice obtains the outcome for the computational basis measurement on the system ${\rm R}$ not by asking Charlie, but by reversing Charlie's measurement and then herself measures the system ${\rm R}$ in the computational basis, then there is no guarantee that she will get the same outcome as Charlie. Even though the effective dynamics of Charlie's measurement on ${\rm R}$ followed by its reversing is represented as an identity channel on ${\rm R}$ in quantum theory, in interpretations that violate transformation noncontextuality (such as Bohmian mechanics) or that have inherent stochasticity, it is entirely possible that Alice's outcome will not be the same as Charlie's outcome.

Still, as long as Alice {\em actually} implements her measurement by asking Charlie what he observed (as in the original Local Friendliness proposal), then Eq.~\eqref{eq_track} as an instance of Tracking is hard to deny. 
Indeed, one might even say that this Tracking assumption is not a metaphysical assumption, but rather a methodological one~\cite{Wiseman2023thoughtfullocal} whose validity is guaranteed by the very choice of experimental arrangement. 

Moreover, while it is not often recognized explicitly, the Tracking assumption more generally is also required for Bell's theorem (and arguably for {\em all} scientific experiments where different observers share their observations, or even where a single observer collects and uses data). 
Tracking posits that an observer can faithfully transmit their observations to other observers. To deny this is quite a radical position. If this assumption fails in a Bell scenario, for instance, one could imagine that when Alice and Bob get together to collect their joint statistics, the outcomes they share with each other are not the same as the ones they actually observed, but rather are somehow generated at the moment when the two meet. (This is possible even if all outcomes obey Absoluteness of Observed Events.) In this case, Bell-violating correlations (and even signaling correlations) can be generated in a local manner. And it is not clear to us why Eq.~\eqref{eq_track} is any more suspicious than the use of the Tracking assumption in Bell's theorem and other scientific experiments.

In Ref.~\cite{bongStrong2020}, Eq.~\eqref{eq_track} is presented as an implication of Absoluteness of Observed Events in the Local Friendliness setup. In our view, Eq.~\eqref{eq_track} is a consequence of the combination of both Absoluteness of Observed Events and Tracking. Specifically, Absoluteness of Observed Events allows one to represent the outcomes as tuples of classical random variables $(a,b,c,d)$, while the Tracking assumption gives the equivalence between the value of $a$ and that of $c$ in the runs where Alice copies Charlie's outcome, and similarly for $b$ and $d$. 

So really, the Tracking assumption is independent from Absoluteness of Observed Events. An example that illustrates this is the Parallel Lives view~\cite{brassard2019parallel,raymond2021local} of Many Worlds. In such an interpretation, when one of the parallel Alices and one of the parallel Bobs meet, the outcome transmitted by that Alice to that Bob does correspond to the actual outcome observed by that Alice, so the Tracking assumption is satisfied even though Absoluteness of Observed Events is violated.

But in any case, our presentation of the Local Friendliness argument makes it clear that the specific instance of the Tracking assumption in Eq.~\eqref{eq_track} is {\em not necessary}, so one cannot avoid the no-go argument by challenging it. To see this, note that this condition was only required to get from Eqs.~\eqref{QMpredLF1p}-\eqref{QMpredLF3p} to Eqs.~\eqref{eq_AOELF1}-\eqref{eq_AOELF3}, but these latter equations {\em are themselves directly observable predictions of quantum theory}. So one can begin the argument from this step, after which the argument makes no reference to Eq.~\eqref{eq_track} (which is how we presented the argument).

As far as we can tell, Eq.~\eqref{eq_track} was used in Ref.~\cite{bongStrong2020} only so that there is a single probability distribution, namely $p(ab|xy)$, to focus the argument on, which makes it easier to think about objects like the polytope of correlations consistent with the Local Friendliness assumptions. In contrast, our argument also refers to probability distributions such as $p(cd|xy)$. This latter distribution may not be always observable for all choices of $x$ and $y$---although, critically, the only elements of it that appear in the argument are those which \emph{are} actually observable (i.e., those wherein $x=0$ and $y=0$).

\section{The Pusey-Masanes no-go theorem} \label{PMsec}

 The main idea in Brukner's and the Local Friendliness arguments presented in Section~\ref{sec_brukner} and Section~\ref{sec_LF} (respectively) was to ensure that the outcomes of both the computational basis and $\pm$ basis measurements both on system ${\rm R}$ and on system ${\rm S}$ are {\em actually observed} in a single run of experiment.
Then, one can demand a joint distribution over them via Absoluteness of Observed Events without appealing to value assignments for measurements that might have been but were not performed in a given run of experiment (as one must do in proofs of Bell's theorem). 
Again employing this strategy, we will now see that one can further construct extended Wigner's friend arguments wherein each agent performs a fixed measurement, so there are no measurement choices.

Two such arguments are the Pusey-Masanes and Frauchiger-Renner arguments. The Pusey-Masanes argument was first presented in a talk by Pusey~\cite{PuseyYoutube} as a simplified presentation of the Frauchiger-Renner argument. Because the latter is (in our view) considerably more nuanced and contentious despite having the similar core features and metaphysical consequences regarding interpretations of quantum theory, we postpone discussing it and its relationship to the Pusey-Masanes argument to Section~\ref{sec_FR}. 

Note that Pusey attributes this argument to Luis Masanes. Although neither claims ownership of it, we will refer to this theorem as the Pusey-Masanes theorem. It was subsequently expanded on in a talk by one of the present authors (Leifer~\cite{LeiferYoutube}), and it is further formalized here. These presentations of the argument were built around the probabilistic violation of a CHSH inequality. For ease of comparison with the other no-go theorems herein, we will make the argument using the deterministic logic of the Hardy argument. One can also find a number of related arguments in the literature~\cite{myrvoldModal2002,Healey_2018,leegwaterWhen2022,ormrod2022no,ormrod2023theories}, and we will expand on these in Section~\ref{sec_relatedPM}.

The setup is similar to that of the Local Friendliness argument, but now the superobservers do not choose among different possible measurement settings: they each implement the same measurement across all runs. Charlie and Debbie always measure their half of the Hardy state $\ket{\psi_{\rm Hardy}}_{\rm RS}$ in the computational basis. Two superobservers, Alice and Bob (respectively), undo these measurements unitarily and then measure systems ${\rm R}$ and ${\rm S}$ (respectively) in the complementary $\pm$ basis.

\begin{figure}[htb!]
    \centering
    \includegraphics[width=0.49\textwidth]{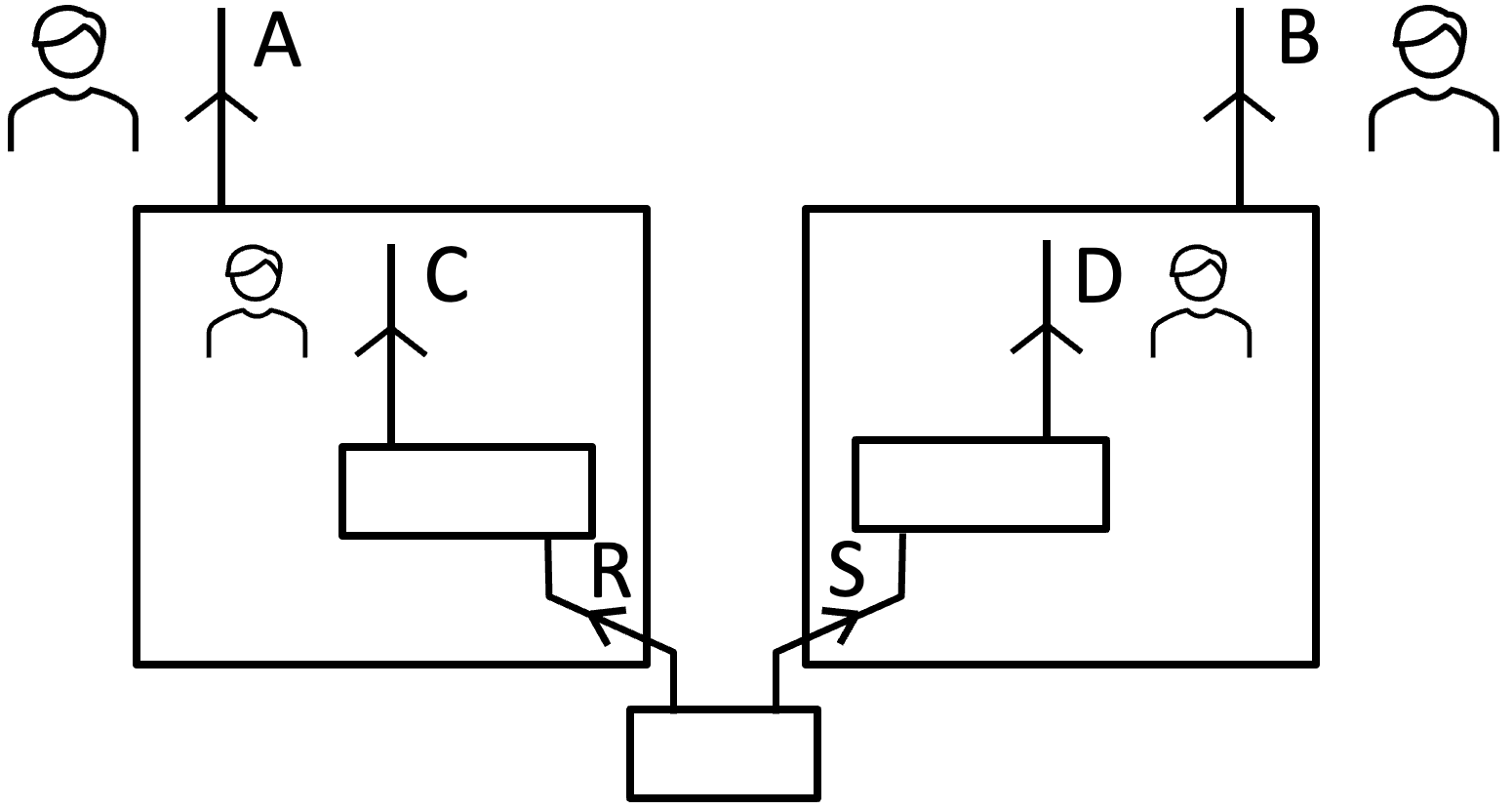}
    \caption{The Pusey-Masanes extended Wigner's friend setup.   Charlie and Debbie share an Hardy state, while Alice and Bob are two superobservers who have control over Charlie's and Debbie's lab. Each agent performs a fixed measurement.}
    \label{PM}
\end{figure}

Assuming Absoluteness of Observed Events, there is a fact of the matter about what the outcome of each measurement is in every run of the experiment. As before, we denote these outcomes $a,b,c$ and $d$. If the experiment is repeated many times, then the relative frequency for $p(a,b,c,d)$ exists. 
Of course, this relative frequency cannot be observed by any observer, since all records of Charlie's outcome $c$ are necessarily erased by Alice's unitary undoing operation prior to Alice's outcome $a$ being generated, and similarly, all records of Debbie's outcome $d$ are erased before Bob's outcome $b$ can be generated. Nonetheless, quantum theory seems to place constraints on this relative frequency.

In particular, if one evolves the initial state unitarily until the relevant measurements occur, and then applies the Born rule to this evolved state together with the relevant measurement effects, one computes that
\begin{align}
p(c=1,d=1)&=0 \label{QMpredPM1} \\
p(c=0,b=-)&=0 \\
\label{QMpredPM3}
p(a=-,d=0)&=0 \\ 
\label{QMpredPM4} p(a=-,b=-) &\neq 0.
\end{align}

Much as in the previous two arguments, it follows from these that
\begin{align}    
d = 1 \ &\implies \ c = 0, \label{eq_PM1}\\
c = 0 \ &\implies \ b = +, \label{eq_PM2}\\
a =- \ &\implies \ d = 1, \label{eq_PM3}\\
a = - \ &\centernot\implies \ b = +.
\end{align}
We summarize the quantum predictions and these implications in Table~\ref{tab_PM}. This table, once again, is analogous to that for Hardy nonlocality (see Table~\ref{tab_HardyNL}).

Again, these four implications are not consistent, as the first three chain together by transitivity to imply that $a = - \ \implies \ b = +$, contradicting the fourth. As in the other EWF arguments, this chain of reasoning does not assume classical realism---it refers only to actually observed outcomes $a$, $b$, $c$, and $d$.

\begin{table}[h]
\setlength\tabcolsep{0.5mm}
    \begin{tabular}{|c|c|c|}
        \hline
        \rowcolor{gray!15}
        \textbf{agents} & \textbf{quantum prediction} & \textbf{implication} \\ \hline
        Charlie, Debbie & $p(c=1,d=1)=0$ & $d = 1 \ \implies \ c = 0$  \\ \hline
        Charlie, Bob & $p(c=0,b=-)=0$ & $c = 0 \ \implies \ b = +$  \\ \hline
        Alice, Debbie & $p(a=-,d=0)=0$ & $a =- \ \implies \ d = 1$  \\ \hline
        Alice, Bob & $p(a=-,b=-) \ > \ 0$ & $a = - \ \centernot\implies \ b = +$  \\ \hline
    \end{tabular}
\caption{The quantum predictions and implications about each pair of observed outcomes in the Pusey-Masanes argument.}
\label{tab_PM}
\end{table}

However, there is a critical feature of this argument that could be questioned. Although the correlations in Eqs.~\eqref{QMpredPM1}-\eqref{QMpredPM4} are those which one computes directly from the quantum circuit using unitarity and the Born rule, it is not the case that the actual relative frequencies in the experiment {\em must} match these probabilities {\em on pain of falsifying quantum theory}. This is because only one of the four (namely Eq.~\eqref{QMpredPM4}) is actually observed in the experiment (when Alice and Bob get together to collect their results). Because Charlie's and Debbie's results are erased during the undoing operations, the joint probabilities involving them, namely $p(c,d)$, $p(c,b)$, and $p(a,d)$, cannot be observed by collecting results in a similar fashion.

Not only are the correlations in Eqs.~\eqref{QMpredPM1}-\eqref{QMpredPM4} not all {\em observed} in the experiment, some of them are also {\em inaccessible even in principle}, even though they \emph{do exist} according to Absoluteness of Observed Events.  
This is most obvious if we arrange the experiment such that Charlie and Alice's operations are all spacelike separated from Bob and Debbie's, in which case $p(c,d)$, $p(c,b)$, and $p(a,d)$ are all inaccessible to all observers, even in principle, since Charlie's outcome is erased before it could possibly be communicated to any observer with access to Debbie's or Bob's outcomes, and similarly, Debbie's outcome is erased before it could possibly be communicated to any observer with access to Alice's or Charlie's outcomes. This problem persists for other spatiotemporal arrangements, as we discuss in Section~\ref{sec_BIC}. 

Consequently, the Pusey-Masanes argument relies on an assumption we will term \emph{Born Inaccessible Correlations}. It requires that when two measurements are made in parallel, the two outcomes always arise with a joint frequency given by the Born rule, even if no single observer could access both outcomes, even in principle. 

To be clear, we are not saying that this is problematic on the grounds that unobservable quantities are necessarily meaningless---such an extreme position would only be held by a strict operationalist. Rather, we are saying that what precisely can be said about unobservable quantities is an interpretation-laden question, and consequently one that is often contentious~\cite{baumanncomments2019,healeyreply2019}.
For example, it is possible for an interpretation to explicitly violate the assumption in question, even though they exactly reproduce the Born rule predictions for all observable correlations, as we will discuss further in Section~\ref{sec_BIC}.

\subsection[Evaluating Born Inaccessible Correlations]{Evaluating the assumption of Born Inaccessible Correlations}
\label{sec_BIC}

In the Local Friendliness scenario presented in Sec~\ref{sec_LF}, Born Inaccessible Correlations is implied by Local Agency for Observed Events (together with the Born rule for accessible correlations), since one could arrive at (for example) Eq.~\eqref{eq_LALF1} from Eq.~\eqref{eq_AOELF1}. (Although this implication does not hold in general, such as in the Pusey-Masanes argument, where there are no measurement settings, and so Local Agency for Observed Events has no consequence.) However, Born Inaccessible Correlations does not imply Local Agency for Observed Events, since Born Inaccessible Correlations is only about parallel measurement outcomes without reference to measurement settings (or, consequently, any \enquote{unperformed experiments}), while Local Agency for Observed Events constrains the relation between observed events and any distant settings (and consequently, unperformed experiments).\footnote{This fact is especially important for Local Friendliness setups of the sort mentioned in Footnote~\ref{foot_LA}, where not all relevant measurements on ${\rm R}$ and ${\rm S}$ can be realized in a single run of experiment. Such experiments involve not only existing-but-inaccessible correlations, but also unrealized (hence, of course, also inaccessible) correlations. Consequently, in some of these cases, such as the so-called minimal Local Friendliness scenario~\cite{Wiseman2023thoughtfullocal,yile2023}, an interpretation of quantum theory, such as Bohmian mechanics, \emph{can} resolve the no-go result without violating the instance of Born Inaccessible Correlations implied by Local Agency for Observed Events while holding onto Absoluteness of Observed Events.  } 
Consequently, Born Inaccessible Correlations is a weaker assumption when applied to the correlations in the Local Friendliness argument. 

However, what matters the most is not the strength of the assumption, but how well the assumption can be motivated.
So let us now try to see how one could motivate Born Inaccessible Correlations (without introducing different measurement settings such as by using Local Agency for Observed Events) in the Pusey-Masanes argument. We will begin by considering variants of the experiment that differ only by how long each agent waits to perform their respective actions.

\begin{figure}[htb!]
    \centering
    \begin{subfigure}[t]{0.213\textwidth}
        \centering
        \includegraphics[width=\textwidth]{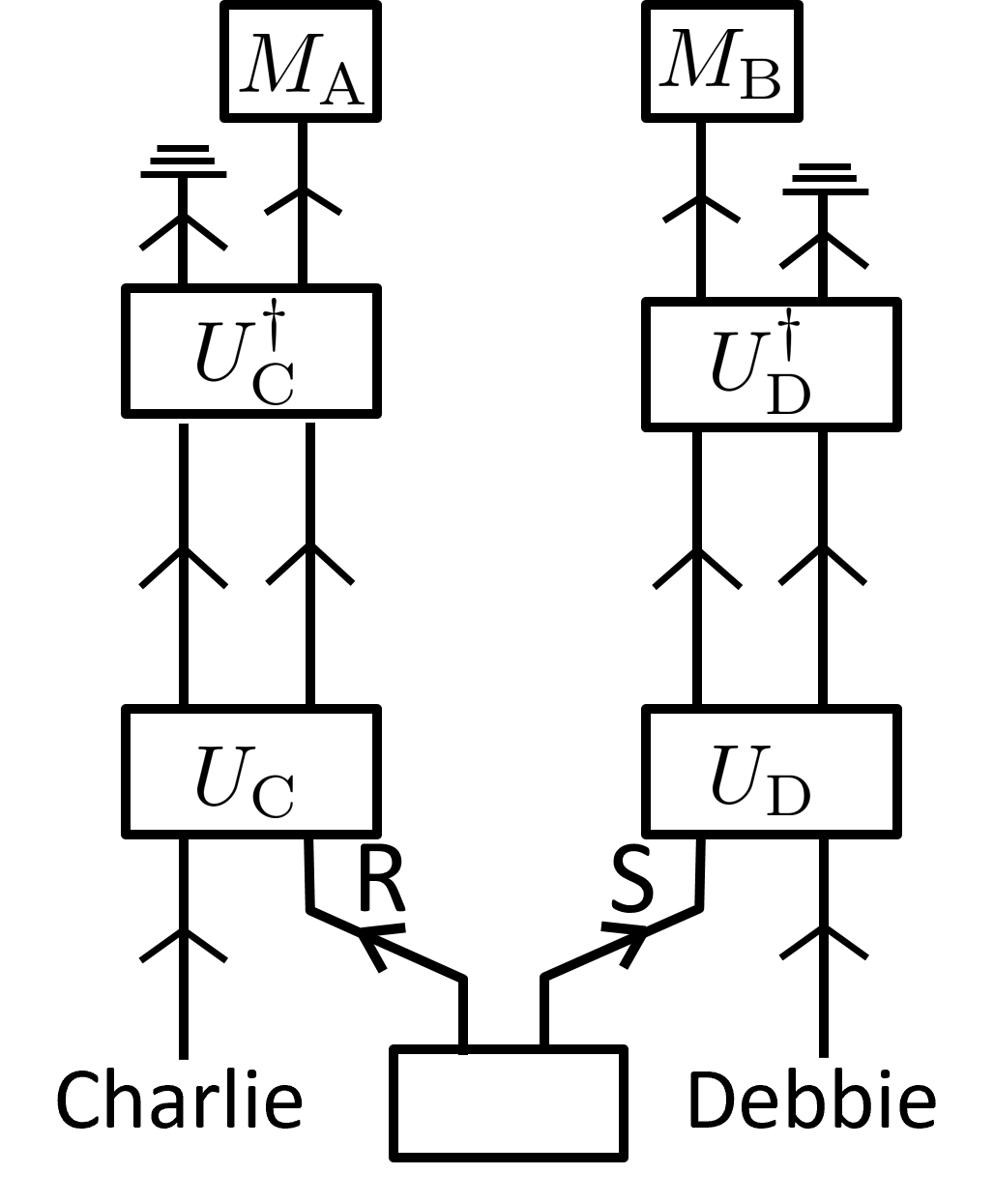}
        \caption{}
        \label{PM1}
    \end{subfigure}
    \hfill
    \begin{subfigure}[t]{0.23\textwidth}
        \centering
        \includegraphics[width=\textwidth]{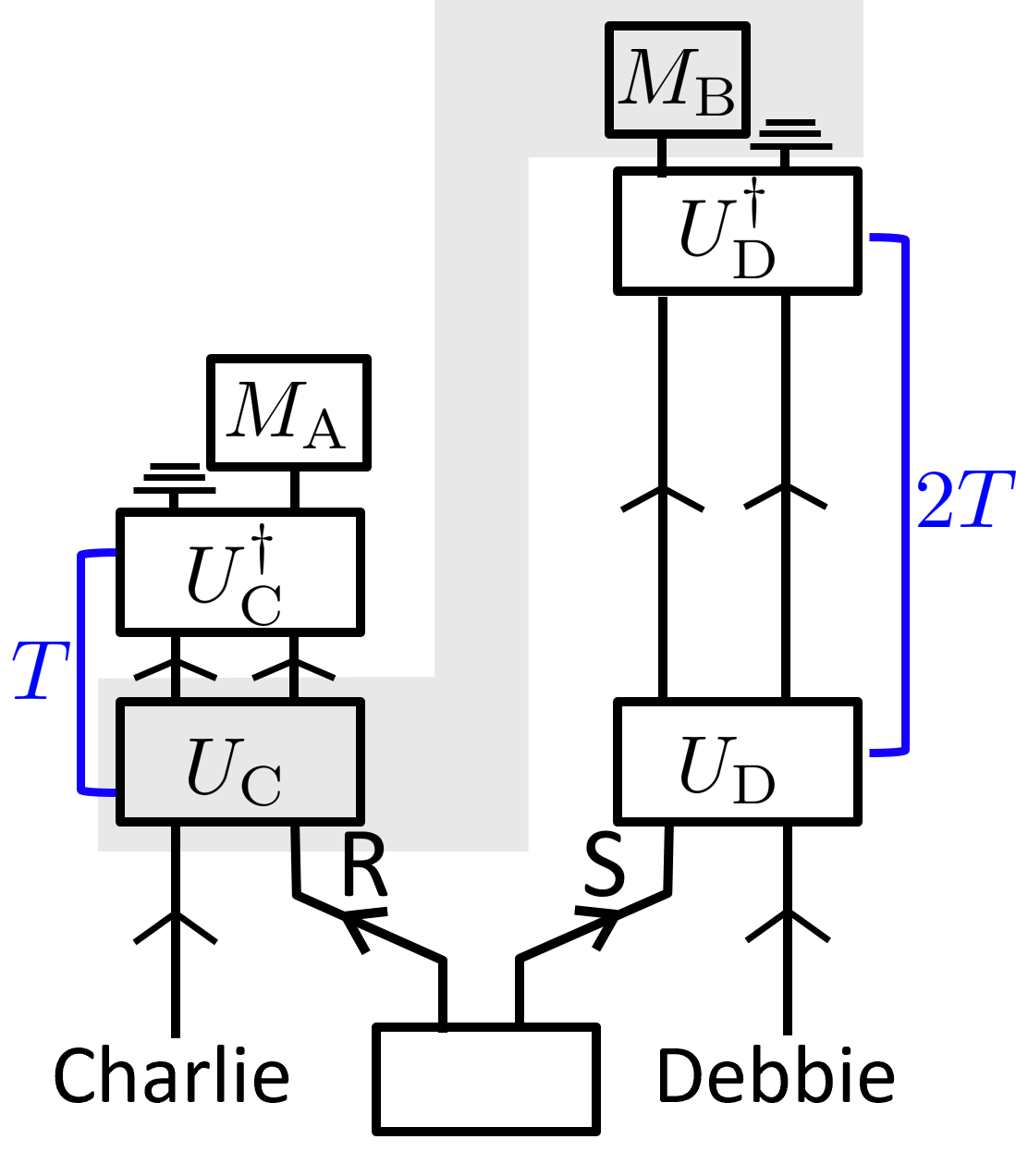}
        \caption{}
        \label{PM2}
    \end{subfigure}
    \caption{a) Circuit representation of an external view of the experiment. b) Depending on the timing, at most three of the joint frequencies in Eqs.~\eqref{QMpredPM1}-\eqref{QMpredPM4} can be made observable to any agent. Here, the Charlie-Bob joint frequency (shaded grey) is not observable, because all records of Charlie's outcome are unitarily erased before Bob's outcome is generated. }
    \label{PM_Loophole}
\end{figure}

First, consider the circuit in Figure~\ref{PM_Loophole}(a), which is just an alternative circuit representation of the Pusey-Masanes experiment. The advantage of this representation over that in Figure~\ref{PM} is that it depicts Charlie and Debbie as quantum systems in the quantum circuit, and so one can explicitly depict the undoing unitaries; the disadvantage is that in this representation, Charlie and Debbie's measurements are represented directly as unitary channels, and so one cannot naturally depict the outcomes that Charlie and Debbie are assumed to observe. 

In the previous section, we discussed how $p(c,d)$, $p(c,b)$, and $p(a,d)$ are in principle inaccessible if Alice and Charlie's operations are at spacelike separation from Bob and Debbie's. But even if one considers other spatiotemporal arrangements for the experiment, at least one of the probability distributions will be inaccessible. 
(Here and henceforth, the terms accessible and inaccessible will always refer to the question of in-principle accessibility.) 
Consider for example Figure~\ref{PM_Loophole}(b), where we depict this same experiment but with a different choice of timings. Namely, we imagine that Charlie and Debbie perform their measurements at the same time, that Alice does not undo Charlie's measurement until a time $T$ has passed, that Bob does not undo Debbie's measurement until a time $2T$ has passed, and that Alice and Bob do their respective measurements immediately after they implement their respective undoing operations. By choosing $T$ to be large enough, one can ensure that the correlations $p(a,d)$, $p(a,b)$, and $p(c,d)$ are in principle accessible. Take the example of the joint relative frequency for $p(c,d)$: when $T$ is large enough (and when all runs of the experiment are performed at a parallel), Charlie and Debbie could {\em in principle} have gotten together to combine their experimental records to compute the relative frequencies of the various outcomes, to publish papers about these results, and so on. 
Similarly, the relative frequency for $p(a,b)$ is accessible to Alice and Bob, and the relative frequency for $p(a,d)$ is accessible to Alice, Debbie, and Charlie.
The joint frequency on $(c,b)$, however, is different, in that all records of the outcome of $c$ are erased before the outcome $b$ is even generated. Even in principle, then, there is no observer who can access $p(c,b)$. 

If one instead arranges the symmetric setup where one exchanges the time delays $T$ and $2T$, one then finds that the joint frequency for $p(a,d)$ is not accessible to any observer, while each of the other three joint frequencies is accessible to at least two observers. So we see that the difference between a joint frequency being accessible and being inaccessible comes down to a shift in the relative timing of the procedures. 

Since we are assuming a trivial Hamiltonian, nothing in the quantum formalism depends on this timing information. Now, {\em if} one believes that quantum theory is \emph{Complete} in the sense that there is no deeper theory or deeper set of facts that one can use to reason about the world, one might expect that the joint frequency of observer's outcomes cannot depend on this timing information. We will call this the {\em Timing Irrelevance} assumption, as it will be useful also in later sections.
To reject this would presumably entail imagining some mechanism that communicates to Debbie's system whether the reversal of Charlie's measurement has happened yet or not. In a hidden variable model, one must posit such a mechanism, which would necessarily be nonlocal and contextual, but one would not do so in an interpretation that endorses Completeness.

Assuming Timing Irrelevance, each of the three distributions ($p(c,d)$, $p(c,b)$, and $p(a,d)$) must be invariant under any choice of timing. 
Since each distribution presumably must obey the Born rule when the timing is such that it is actually accessible, it consequently must also obey the Born rule when the timing is such that it is inaccessible. Thus, if one grants the assumption of Completeness, one can (at least loosely) motivate the assumption of Timing Irrelevance, and consequently the assumption of Born Inaccessible Correlations.

For interpretations that violate Completeness, however, we are not certain how one could motivate the Timing Irrelevance assumption.  Bohmian mechanics is an example of a theory which rejects Completeness and explicitly violates Timing Irrelevance (and consequently Born Inaccessible Correlations), and in this manner evades the no-go argument. In Bohmian mechanics, the underlying particle positions determine the outcomes in a manner that violates the Born rule for at least one of $p(c,b)$ and $p(a,d)$, depending on the time ordering of the processes in the preferred foliation (or reference frame) of spacetime stipulated by the theory. For example, if Alice's measurement is earlier than Debbie's measurement in the preferred foliation, such as in Figure~\ref{PM_Loophole}(b), then it is $p(c,b)$ that violates the Born rule prediction. This cannot be empirically witnessed, however, because $p(c,b)$ is inaccessible, as mentioned above. (Of course, this explanation invokes nonlocal causal influences, but if one has an issue with this, one is better served by the Local Friendliness argument, which explicitly makes a locality assumption.) 

In interpretations (most notably including certain Copenhagenish ones) that reject arguments that combine reasoning about incompatible measurement contexts, there is another reason to challenge the attempted justification for Born Inaccessible Correlations using Timing Irrelevance.
Note that any experiment where any of the joint frequencies $p(a,d)$, $p(b,c)$, and $p(c,d)$ are  {\em actually observed} would be a different experiment from the ones depicted in Fig~\ref{PM_Loophole}. To actually observe $p(c,d)$, for instance, Charlie and Debbie must communicate their outcomes with each other, in which case the circuit structure is no longer that shown in Figure~\ref{PM_Loophole}(b). Consequently, Alice and Bob will not generally be able to undo Charlie and Debbie's measurements via local actions on their respective friends; rather, they will need to first undo the interactions between Charlie and Debbie (although provided that Alice and Bob have the requisite level of control, it is still possible in principle for them to undo any interaction between Charlie and Debbie, and other systems the two might interact with). In order to argue that the unverified (though in principle verifiable) joint frequencies in the Pusey-Masanes argument must be identical to those in this alternative experiment, one must appeal to the claim that the joint frequencies are the same in these two different experimental contexts. 

The argument also invokes multiple distinct experimental arrangements if one appeals to different choices of timing to motivate Born Inaccessible Correlations, as discussed above. However, one could reasonably hold that combining these particular distinct measurement contexts is not problematic even in a Copenhagenish view, because the choices in question do not refer to which measurement one chooses from some set of incompatible measurements, but rather simply refer to how long each agent waits before acting (and recall that the Hamiltonian is assumed to be trivial). 

Thus, if one uses Timing Irrelevance to justify Born Inaccessible Correlations, the Pusey-Masanes argument would implicitly appeal to multiple experimental arrangements, despite appearances to the contrary, and so it would not be that different from the Local Friendliness argument in this regard.

On the other hand, one could argue that Born Inaccessible Correlations is not obviously less well motivated than, for example, the universality of unitarity, which is also needed for any EWF argument. In both cases, the motivation seems to largely be the lack of empirical evidence against them and the lack of an obvious alternative law for how these thought experiments should turn out. (And in both cases, there are explicit alternative laws in certain interpretations, so this is not a good argument. Bohmian Mechanics gives an explicit law to supplant Born Inaccessible Correlations, and collapse theories, for example, give an explicit law to supplant unitarity.)
The former is intrinsically a metaphysical assumption, in that no experimental evidence could ever be gathered to falsify it, even in principle, while there are no known principles forbidding one from testing unitarity on the scale of an observer.  Thus, the latter is more of a testable empirical hypothesis rather than a metaphysical principle. But this does not give the latter any more plausibility in regimes wildly far from those in which it has been tested.  Perhaps one could argue that it is more difficult to modify the empirically observable structure of quantum theory in a consistent manner than it is to modify the inaccessible correlations in the theory, but this is quite speculative at present. 

To physicists with empiricist leanings, perhaps it is easier to give up on Born Inaccessible Correlations than it is to give up on something like universality of unitarity. To give up the latter would require that quantum theory be modified on some scales, whereas giving up the former would not require an empiricist to give up anything---one could simply claim that it is not meaningful to reason about such inaccessible quantities. But to a realist, such a resolution feels quite unsatisfactory. 

So in the end, the status of this assumption in the context of the Pusey-Masanes argument remains unclear. (In the context of the Local Friendliness argument, where it can be motivated by Local Agency, we consider it well-motivated.)

\subsection{Related arguments}
\label{sec_relatedPM}

Next, we briefly discuss the EWF arguments in Refs.~\cite{Healey_2018,leegwaterWhen2022,ormrod2022no}, as these are all similar to the Pusey-Masanes argument (and to each other) but require an assumption regarding relativistic frame invariance.

In particular, they assume that the predictions made by some physical theory for all correlations (accessible or inaccessible) are the same in all reference frames. One should take care to recognize that the inaccessible joint frequencies do not {\em have} to be the same in different inertial reference frames on pain of relativity theory being observably false, since the joint frequencies are not accessible, and only observable quantities are {\em necessarily} frame-independent. (Note that a similar comment applies when one motivates Local Agency for Observed Events from relativity theory.) But it would seem odd if the joint frequency {\em did} depend on the reference frame, because nothing in relativity theory suggests such a dependence. This rough argument is quite analogous to the rough argument in Section~\ref{sec_BIC} from Completeness of quantum theory to Timing Irrelevance. Once again, one could reject the argument, but only by accepting an uncomfortable kind of dependence of the inaccessible joint frequencies on seemingly irrelevant facts. 
Moreover, a realist might be uncomfortable postulating that relativity theory only constrains {\em what can be observed} as opposed to constraining {\em what exists}. (A similar debate arises in the sphere of Bell nonclassicality---whether or not relativity merely imposes no-signaling, or a stronger constraint like local causality~\cite{Bell1987Speakable}.)  

On top of that, these arguments assume that one can use the Born rule to predict the outcomes of measurements performed at the same time (in some particular reference frame). 
For example, since Charlie's measurement and Debbie's measurement are space-like-separated in these arguments, there exists an inertial reference frame where the two measurements happen at the same time; then, according to their assumption, the correlation between the two measurement outcomes is predicted by the Born rule. But this correlation is inaccessible, because Alice's and Bob's measurements (and consequently also their undoing operations) are also presumed to be spacelike separated.  Thus, one derives Born Inaccessible Correlations for correlations between space-like separated measurements.

Consequently, one obtains the implication in each row of Table~\ref{tab_PM}. Given that Absoluteness of Observed Events is also assumed in these arguments, the implications across all four rows of Table~\ref{tab_PM} can be chained together, leading to the same contradiction we saw in the Pusey-Masanes argument.

EWF arguments which appeal to relativistic frame invariance are sometimes put forward as illuminating a tension between special relativity and (certain interpretations of) quantum theory. However, it is not clear to us that these arguments actually use relativity theory in any interesting sense. For example, in an argument of the above form, the only role of relativistic frame invariance is to upgrade an explicitly reference frame-dependent assumption (essentially Born Inaccessible Correlations, but {\em only} in one special reference frame) to one which is not reference frame-dependent (namely full Born Inaccessible Correlations). But, absent any motivation for why Born Inaccessible Correlations would hold in the special reference frame {\em which would not also motivate full Born Inaccessible Correlations}, there is no reason to not assume full Born Inaccessible Correlations to begin with.

Finally, let us note Ref.~\cite{ormrod2023theories}, which generalizes the argument to theories other than quantum theory within a framework that explicitly assumes Timing Irrelevance for processes embedded in spacetime.
The argument in Ref.~\cite{ormrod2023theories} also invokes a kind of locality assumption, which the authors term Local Dynamics. This assumption demands that a bipartite transformation whose local input-output pairs are space-like separated must decompose into a common-cause circuit. Both the Pusey-Masanes argument and the Frauchiger-Renner argument also assume the quantum circuit representation of the respective setups is given by a common-cause circuit, and so it seems that this kind of locality assumption is implicitly used in those arguments as well. However, the status of this assumption (in any of these three arguments) is not clear to us. If one takes this common-cause circuit to represent the actual causal structure of the situation, then this is a substantive physical assumption, but if one takes it simply as a calculational tool used to compute the quantum predictions for a given setup, then the substance of this assumption is less clear.

\section{The Frauchiger-Renner no-go theorem} 
\label{sec_FR}

Likely the most well-known extended Wigner's friend argument was introduced by Frauchiger and Renner~\cite{Frauchiger2018}. 
However, it is more complicated and subtle than the other EWF arguments despite having quite similar implications. The complications largely arise because its key assumptions are nuanced epistemological ones rather than more straightforward metaphysical ones (such as, say, Absoluteness of Observed Events), and also because the argument involves many more detailed steps to check.

\begin{figure}[htbp]
    \centering
    \begin{subfigure}[t]{0.32\textwidth}
        \centering
        \includegraphics[width=\textwidth]{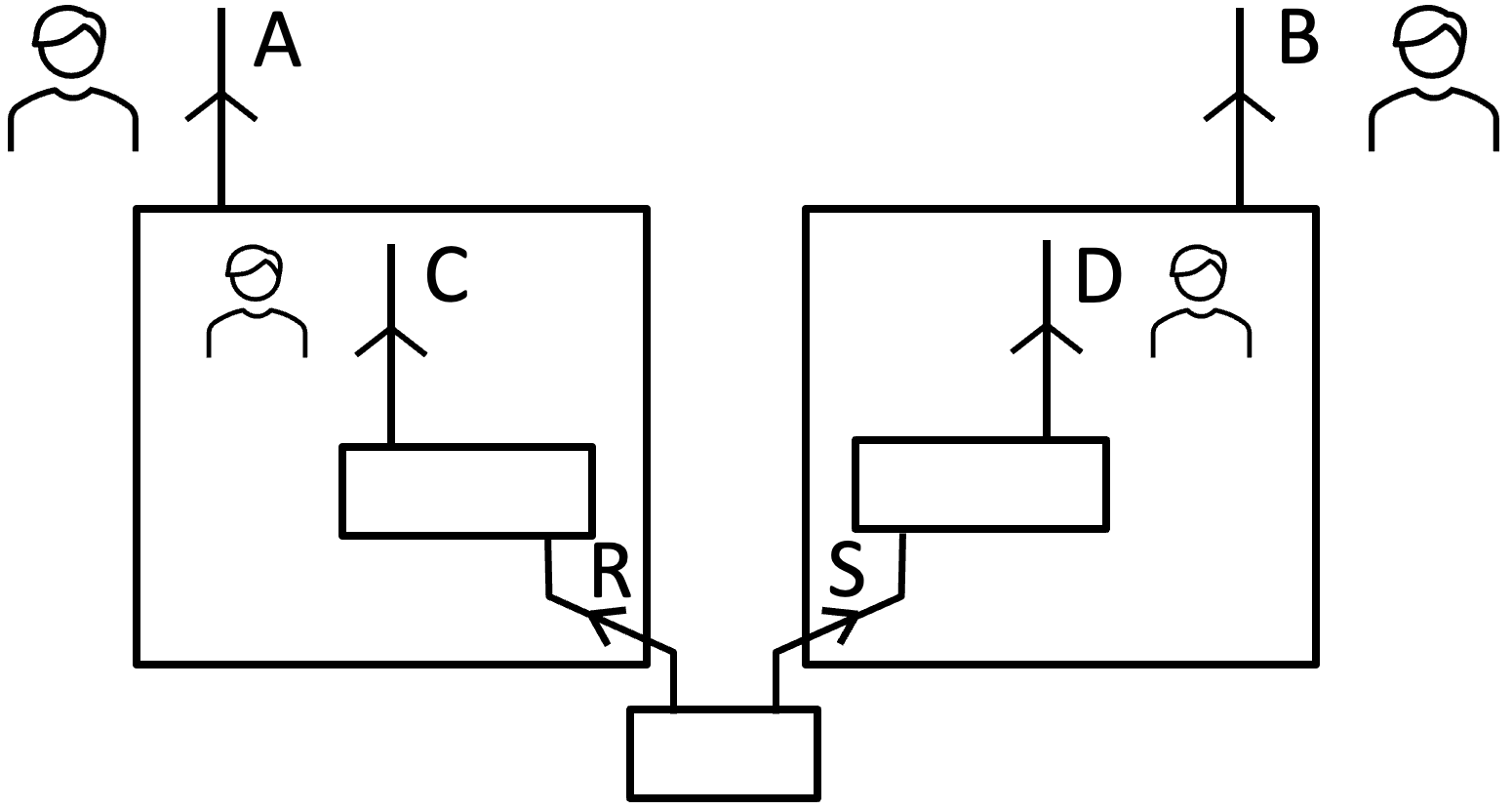}
        \caption{}
    \end{subfigure}
    \hfill
    \begin{subfigure}[t]{0.15\textwidth}
        \centering
        \includegraphics[width=\textwidth]{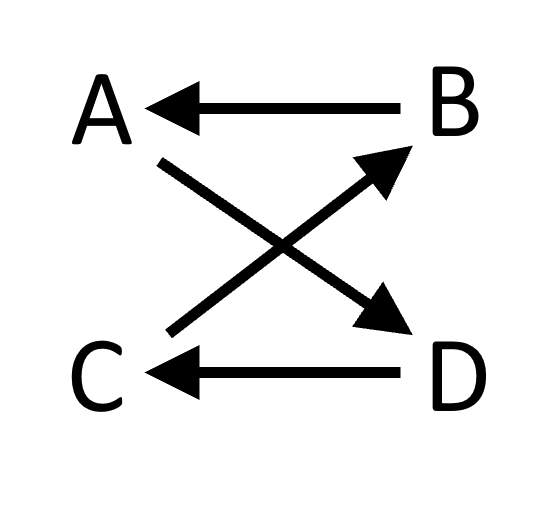}
        \caption{}
    \end{subfigure}
    \caption{a) The Frauchiger-Renner extended Wigner's friend setup. (Exact same as the Pusey-Masanes setup).
 b) The pattern of reasoning in the Frauchiger-Renner argument. For example, $B\rightarrow A$ means that Bob is reasoning about the correlation between his own outcome and Alice's outcome. }
    \label{fig_FRfullfig}
\end{figure}

The Frauchiger-Renner setup is exactly the same as that of the Pusey-Masanes argument, depicted again in Fig.~\ref{fig_FRfullfig}(a). (As already mentioned, the latter originated as a simplified presentation of the former.) It begins with two friends, Charlie and Debbie, sharing a pair of qubits in the Hardy state $
\ket{\psi_{\rm Hardy}}_{\rm RS}$. Charlie and Debbie each measure their system (${\rm R}$ and ${\rm S}$, respectively) in the computational basis. Once again, two superobservers, Alice and Bob, have perfect control over Charlie and Debbie's labs, respectively. Alice undoes the measurement performed by Charlie by performing the inverse unitary, while Bob undoes the measurement performed by Debbie. Then, each of Alice and Bob measures the qubit in their lab (${\rm R}$ and ${\rm S}$, respectively) in the complementary $\pm$ basis.

Each of the four agents then applies quantum theory to reason about part of the experiment. In particular, Alice reasons about Debbie, who reasons about Charlie, who reasons about Bob, who reasons about Alice. This pattern of reasoning is depicted in Figure~\ref{fig_FRfullfig}(b). 

The Frauchiger-Renner argument involves the reasoning of multiple different agents, and relies on the fact that each agent makes a different choice about which systems in the experiment to model as quantum, and which to model as classical. This can be captured by the notion of the movable Heisenberg cut, or \enquote{shifty split}~\cite{Bell_1990}, describing which systems each agent treats as quantum and which as classical.
Most interpretations of quantum theory (with Copenhagenish interpretations as the notable exceptions) do not consider the Heisenberg cut to be a fundamental notion. For example, in each of Many Worlds, Bohmian Mechanics, and collapse interpretations, there is an objective fact of the matter about whether a given system is classical or quantum. Consequently, from their perspective, one might be inclined to present the argument in a manner that does not rely explicitly on the question of which agents view which systems as classical or quantum. This is essentially what the Pusey-Masanes argument achieves. 
However, such modified arguments require slightly strengthened assumptions, as we discuss in Footnote~\ref{foot_con} and Sec.~\ref{sec_shiftyAOE}. Although these stronger assumptions are from most perspectives motivated at least as well as those used in our presentation of the Frauchiger-Renner's argument, Renner and some of his coauthors~\cite{Frauchiger2018,Rio2023,Rio2023a} consider the use of the weaker assumptions to be significant. We will evaluate this in Section~\ref{sec_shiftyAOE}. 

In the argument, each agent reasons according to a particular choice of Heisenberg cut. For example, Alice reasons about Debbie's outcome, treating only the agent Charlie and systems ${\rm R}$ and ${\rm S}$ as quantum systems. In Alice's view, Charlie's measurement is simply a unitary interaction, which she undoes. Consequently, as in the previous arguments, she can compute the distribution over her and Debbie's outcomes by applying the Born rule directly on the initial (Hardy) state, obtaining
\begin{align}\label{eq_-0}
    p(a=-,d=0)=0.
\end{align}
Hence, Alice reasons that whenever her outcome is $a=-$, Debbie will see the outcome $d=1$. 

Similarly, Debbie reasons about Charlie's outcome, treating only systems ${\rm R}$ and ${\rm S}$ as quantum systems. By applying the Born rule to the initial Hardy state, Debbie concludes that 
\begin{equation}\label{eq_11}
    p(c=1,d=1)=0.
\end{equation}
Hence, she reasons that, whenever her outcome $d=1$, Charlie will see the outcome $c=0$.

Similarly, Charlie reasons about Bob, treating only Debbie and systems ${\rm R}$ and ${\rm S}$ as quantum systems, concluding that 
\begin{equation}\label{eq_0-}
    p(c=0,b=-)=0.
\end{equation}
Hence, Charlie reasons that when his outcome $c=0$, Bob will see outcome $b=+$. 

Finally, we consider the perspective of Bob, who views both Charlie and Debbie (as well as qubits ${\rm R}$ and ${\rm S}$) as quantum systems. Because Bob views the measurements of Charlie and Debbie as unitary interactions that are undone by Alice and Bob, respectively, Bob also applies the Born rule directly on the initial Hardy state, and concludes that 
\begin{equation}\label{eq_--}
    p(a=-,b=-)>0.
\end{equation}
Hence, Bob reasons that when he sees the outcome $b=-$, Alice will sometimes see outcome $a=-$. Noting that outcome $b=-$ occurs with nonzero probability, the argument will consider postselecting on those runs where $a=b=-$.

We summarize these quantum predictions in Table~\ref{tab_FR}. 
This table, once again, is analogous to that for Hardy nonlocality (see Table~\ref{tab_HardyNL}), and is exactly the same to that for the Pusey-Masanes argument (see Table~\ref{tab_PM}).

\begin{table}[h]
\setlength\tabcolsep{0.5mm}
\begin{tabular}{|c|c|c|}
    \hline
    \rowcolor{gray!15}
    \textbf{agents} & \textbf{quantum prediction} & \textbf{implication} \\ \hline
    Charlie, Debbie & $p(c=1,d=1)=0$ & $d = 1 \ \implies \ c = 0$  \\ \hline
    Charlie, Bob & $p(c=0,b=-)=0$ & $c = 0 \ \implies \ b = +$  \\ \hline
    Alice, Debbie & $p(a=-,d=0)=0$ & $a =- \ \implies \ d = 1$  \\ \hline
    Alice, Bob & $p(a=-,b=-) \ > \ 0$ & $a = - \ \centernot\implies \ b = +$  \\ \hline
\end{tabular}
\caption{The quantum predictions and implications about the correlations between the pair of outcomes observed by the respective pair of agents in the Frauchiger-Renner argument.}
\label{tab_FR}
\end{table}

Frauchiger and Renner's argument relies on an assumption they term Consistency: that agent ${\rm W}$ can adopt agent ${\rm U}$'s knowledge about the outcome of an agent ${\rm V}$, provided that agent ${\rm W}$ views both agent ${\rm U}$ and ${\rm V}$ as classical systems rather than quantum systems (provided that all agents reason with the same theories and background information.)\footnote{\label{foot_con}When formulating the Frauchiger-Renner argument in a way that does not have the shifty-split, a stronger version of the Consistency assumption is needed---namely, one which allows every agent to adopt any other agent's knowledge, regardless of what systems each agent views as classical. } This last stipulation is included because without it, one would immediately be led to reason about properties of a system that one viewed as quantum, which Frauchiger and Renner take to be a contentious thing to do\footnote{ \label{footFR}They say {\em \enquote{In the general context of quantum theory, the rules for such nested reasoning may be ambiguous, for the information held by one agent can, from the viewpoint of another agent, be in a superposition of different \enquote{classical} states. Crucially, however, in the argument presented here, the agents' conclusions are all restricted to supposedly unproblematic \enquote{classical} cases.}}~\cite{Frauchiger2018} These sentiments are endorsed, for example, in many Copenhagenish interpretations.}. Because of this last stipulation, Bob cannot {\em directly} adopt the reasoning made by Charlie or Debbie (who he views as quantum systems). However, as we will show explicitly in the next section, {\em sequential} applications of the Consistency assumption allow Bob to ultimately chain together the four logical implications in the right column of Table~\ref{tab_FR} by adopting Alice's reasoning, who in turn adopts Debbie's reasoning, who in turn adopts Charlie's reasoning. But this leads Bob to a contradictory chain of implications.

Before elaborating, let us note some differences between our and Frauchiger-Renner's presentation of the argument. Just as in Brukner's presentation, Frauchiger and Renner take the superobservers to measure their respective Friend-and-Friend's-system in an entangled basis, rather than having them undo the Friend's measurement and measure just the system in the $\pm$ basis. The equivalence of these was proved in Section~\ref{usefulfact}.
Moreover, in Frauchiger and Renner's presentation, Charlie and Debbie do not share a Hardy state, but rather implement a particular measure-prepare protocol which ultimately leads them to prepare a shared Hardy state. (The Hardy state version of the argument has been used in some subsequent presentations~\cite{Nurgalieva_2019,Vilasini2019}.)

In addition, Frauchiger and Renner's presentation of the argument adds a good deal of additional complexity by taking care to specify the timing of various laboratory procedures. The timing choices are made in a manner which ensures that in each step of the argument, no agent directly adopts the reasoning of another agent
whose measurement procedure has already been reversed by a superobserver. 
The idea behind this is to circumvent a supposed loophole~\cite{aaronson2018} that when a superobserver applies a quantum transformation on a Friend, the Friend's reasoning becomes unreliable, and so should not be taken on board by another observer. Frauchiger and Renner's choices for the timing of the various measurements ensure that there will be no such loophole in their argument, 
in the sense that each agent only {\em directly} adopts the reasoning of another agent prior to the second agent being acted on by a superobserver. (If one considers {\em indirect} adoption of reasoning, then these timing choices do not avoid the supposed loophole; for example, Bob indirectly adopts Charlie's reasoning after Charlie's measurement has been reversed, by directly adopting Alice's reasoning about Charlie's reasoning.) In our view, however, this supposed loophole should not be taken seriously regardless of the choice of timings or matter of direct versus indirect adoption of reasoning: it is reasonable to accept a logical statement that was made yesterday, even if the person who first made the argument has had their mind erased.
As long as the person reasons using the correct theory and begins with correct information---such as taking into account that their mind will be erased tomorrow, if this is necessary for making the right prediction---their conclusions will be reliable. (For example, in Bohmian mechanics, Alice's undoing {\em is} relevant for predicting the correlation between Charlie's and Bob's outcome, given Frauchiger and Renner's choice of timing. If Charlie is Bohmian, then---knowing this---he will reject the reasoning that Frauchiger and Renner ascribe to him, and will instead make the Bohmian prediction.)
 Consequently, we consider the timing details to be a red herring  (except insofar as one is interested in the question of which exact correlations are accessible or not).

\subsection{Reasoning using the Consistency assumption}
\label{sec_usingCons}

In contrast, details about which observers view which other observers as quantum systems are critical for correctly applying the Consistency assumption, and consequently are critical for the overall argument. So, let us now derive Frauchiger-Renner's contradiction explicitly, being careful to note how the Consistency axiom is used at each step.
Our presentation is inspired by that of Refs.~\cite{Nurgalieva_2019,fraser2020fitch,Rio2023}. In particular, we will keep track of each agent's knowledge by writing the statement \enquote{Alice knows $z$} as ${\cal K}_{\rm Alice}[z]$. The two sorts of instances of the Consistency assumption used in the argument are as follows: if agent ${\rm U}$ and agent ${\rm V}$ (whose outcomes are labeled by $u$ and $v$, respectively) are viewed as classical by agent ${\rm W}$ and ${\rm U}$, then ${{\cal K}_{\rm W}[{\cal K}_{\rm U}[v=v_0]]\implies {\cal K}_{\rm W}[v=v_0]}$; moreover, ${{\cal K}_{\rm W}[{\cal K}_{\rm U}[u=u_0 \implies v=v_0]]\implies {\cal K}_{\rm U}[u=u_0 \implies v=v_0]}$. 

For reference, Table~\ref{tablecuts} describes the placement of the cut made by each of the four agents.

\begin{table}[htb!]
\centering 
\begin{tabular}{|c|c|} 
\hline  \rowcolor{gray!15}
Agent & Who they view as quantum systems \\
\hline
Alice & Charlie \\
\hline
Bob & Charlie, Debbie \\
\hline
Charlie & Debbie \\
\hline
Debbie & no one \\
\hline 
\end{tabular}
\caption{The agents being viewed as quantum systems by each agent in the Frauchiger-Renner argument.} \label{tablecuts}
\end{table}

From the facts that Charlie predicts Eq.~\eqref{eq_0-}, Debbie predicts Eq.~\eqref{eq_11}, and Alice predicts Eq.~\eqref{eq_-0}, one has
\begin{align}
&{\cal K}_{\rm C}[c=0 \implies b=+], \label{eq_Charlieknowl1}\\
&{\cal K}_{\rm D}[d=1 \implies c=0], \label{eq_Debknowl1}\\
&{\cal K}_{\rm A}[a=- \implies d=1]. \label{eq_Aliceknowl1}
\end{align}
Moreover, the fact that Alice and Bob post-select on their outcome $a=-$ and $b=-$ implies that
\begin{align}
&{\cal K}_{\rm A}[a = -], \label{eq_Aliceknowl7}\\
&{\cal K}_{\rm B} [b = -]. \label{eq_Bobknowl5}
\end{align}

Next, the argument goes, each agent can predict the reasoning of the other agents, because all agents know the experimental protocol (including knowing where the other agents will place their Heisenberg cuts) in advance, and they all reason using the same quantum theory.

{\em First}, Debbie can compute Bob's inference expressed in Eq.~\eqref{eq_Charlieknowl1}, so we have
\begin{align} \label{DebbieKnowsCharlieKnows}
&{\cal K}_{\rm D} [{\cal K}_{\rm C}[c=0 \implies b=+]].
\end{align}
It follows from Eq.~\eqref{DebbieKnowsCharlieKnows} and the Consistency assumption that 
\begin{equation} \label{eq_Debbieknowl2}
{\cal K}_{\rm D}[c=0 \implies b=+].
\end{equation}
Note here that the Consistency axiom allows Debbie to assimilate Charlie's knowledge {\em about outcomes $c$ and $b$} because she views Charlie and outcomes $c$ and $b$ as classical systems.
Debbie then can combine this with her reasoning that $d=1$ implies $c=0$, Eq.~\eqref{eq_Debknowl1}, so we have
\begin{align} \label{eq_Debbieknowl3}
    &{\cal K}_{\rm D}[(d=1 \implies c=0) \land (c=0 \implies b=+)] \nonumber\\
    \implies &{\cal K}_{\rm D}[d=1 \implies b=+].
\end{align}

{\em Second}, Alice can compute Debbie's inference expressed in Eq.~\eqref{eq_Debbieknowl3}, so
\begin{align} \label{AliceKnowsDebbieknows}
&{\cal K}_{\rm A} [{\cal K}_{\rm D}[d=1 \implies b=+]]. 
\end{align}
It follows from Eq.~\eqref{eq_Aliceknowl2} and the Consistency assumption that 
\begin{align} \label{eq_Aliceknowl2}
{\cal K}_{\rm A}[d=1 \implies b=+].
\end{align}
Note in particular that the Consistency axiom allows Alice to assimilate Debbie's knowledge {\em about outcomes $d$ and $b$} because she views Debbie and outcomes $d$ and $b$ as classical systems.
Alice can combine this with her reasoning that $a=-$ implies $d=1$, Eq.~\eqref{eq_Aliceknowl1}, so we have
\begin{align} \label{eq_Aliceknowl3}
    &{\cal K}_{\rm A}[(a=- \implies d=1) \land (d=1 \implies b=+)]  \nonumber\\
    \implies &{\cal K}_{\rm A}[a=- \implies b=+].
\end{align}

{\em Third}, Bob can predict Alice's inference expressed in Eq.~\eqref{eq_Aliceknowl3}, so 
\begin{align} \label{eq_Bobknowl2}
&{\cal K}_{\rm B} [{\cal K}_{\rm A}[a=- \implies b=+]].
\end{align}
It follows from Eq.~\eqref{eq_Bobknowl2} and the Consistency assumption that 
\begin{equation} \label{eq_Bobknowl3}
{\cal K}_{\rm B}[a=- \implies b=+].
\end{equation}
Note in particular that the Consistency axiom allows Bob to assimilate Alice's knowledge {\em about outcomes $a$ and $b$} because he views Alice and outcomes $a$ and $b$ as classical systems.

{\em Fourth}, Bob knows that in the subset of experimental runs that they are post-selecting on, Alice's outcome is $-$, so 
\begin{equation} \label{eq_Bobknowl4}
 {\cal K}_{\rm B} [{\cal K}_{\rm A}[a = -]]. 
\end{equation}

Finally, from Eq.~\eqref{eq_Bobknowl4} and the Consistency assumption, we have that 
\begin{equation}\label{eq_Bobknowl6}
{\cal K}_{\rm B} [a=-].
\end{equation}
 Bob can combine this with his reasoning in Eq.~\eqref{eq_Bobknowl3}, so we have
\begin{align}
&{\cal K}_{\rm B}[(a=-) \land (a=- \implies b=+)]  \nonumber\\
    \implies &{\cal K}_{\rm B} [b=+],
\end{align}
in contradiction with Eq.~\eqref{eq_Bobknowl5}.

\subsection[Evaluating Born Inaccessible Correlations]{Evaluating the Born Inaccessible Correlations assumption}

The Frauchiger-Renner argument requires the assumption of Born Inaccessible Correlations, for exactly the same reasons as in the Pusey-Masanes argument as explained in Section~\ref{sec_BIC}. In particular, at least one of the probability distributions in Eq.~\eqref{eq_-0},  Eq.~\eqref{eq_11}, and  Eq.~\eqref{eq_0-} is not accessible even in principle, and yet the argument assumes that they are constrained by the Born rule. 

Frauchiger and Renner recognize this assumption (it is part of what they call Assumption \enquote{Q}). However, they imply that rejecting this assumption involves rejecting quantum theory itself---for instance, they say that \enquote{Assumption (Q) is that any agent A \enquote{uses quantum theory}}, and that \enquote{... [in] standard quantum mechanics...the time order in which agents...carry out their measurements is irrelevant, because they act on separate systems.} 

But as we argued in the context of the Pusey-Masanes argument, this assumption can be rejected without rejecting operational quantum theory, as it is very much interpretation-dependent. 
The challenges we faced in trying to motivate this assumption in that context arise in the present context as well. In particular, Frauchiger and Renner motivate it by a Timing Irrelevance argument like the one we presented in Section~\ref{sec_BIC} (see again the second quote just above). But as we discussed therein, Timing Irrelevance itself demands justification, and it is not clear if even the strong assumption of Completeness of quantum theory would justify the assumption.

This assumption of the argument is usually overlooked in favor of focusing on the Consistency assumption, but in our view, it is similarly important, and similarly tricky to motivate (unless one introduces nontrivial measurement settings and assumpitons such as Local Agency).

\subsection{Evaluating the Consistency assumption}
\label{sec_evaCons}

In addition to assuming Born Inaccessible Correlations, the Frauchiger-Renner argument also requires the assumption of Consistency. Unlike the former assumption, which in our view deserves more attention than it has as yet been given, the latter has already been the subject of a great deal of contention (as just two examples, see Refs.~\cite{renes2021consistency,vilasini2022general}). 

Consider again Eq.~\eqref{DebbieKnowsCharlieKnows}, namely ${{\cal K}_{\rm D} [{\cal K}_{\rm C}[c=0 \implies b=+]]}$. This equation itself is hard to deny, since it simply expresses Debbie's understanding of how Charlie will reason in this scenario.

However, the question of whether it is reasonable for Debbie to {\em herself} know that $c=0 \implies b=+$ is subtle, since this implication rests on a computation (done by Charlie) wherein Debbie is treated quantumly. As discussed in Section~\ref{sec_stalkee}, some interpretations (such as Copenhagenish ones) may reject the idea that an agent can accept any reasoning that treats themself quantumly. For if one is willing to assign a quantum state to oneself, then such interpretations run into trouble even in the simpler Wigner's stalkee experiment.

To put it another way, the Consistency axiom allows Debbie to accept a conclusion that was reached by Charlie using an argument that Debbie herself is not necessarily warranted in accepting. But can a rational agent accept the conclusions of another agent {\em if they do not accept the reasoning which led to that conclusion}?
Without answering this question in the affirmative, it is unclear whether one can reasonably maintain that ${\cal K}_{\rm D}[c=0 \implies b=+]$ follows from ${\cal K}_{\rm D}[ {\cal K}_{\rm C}[c=0 \implies b=+]]$, as the Consistency axiom would allow.

Similarly, consider Eq.~\eqref{AliceKnowsDebbieknows}: ${{\cal K}_{\rm A} [{\cal K}_{\rm D}[d=1 \implies b=+]]}$. This statement is again hard to deny, but does it imply that Alice should herself know that $d=1 \implies b=+$? Recall that the claim that $d=1 \implies b=+$ was derived (by Debbie) using the intermediate step ${d=1 \implies c=0}$, which follows from treating Charlie as a classical system and reasoning about his outcomes. But in some interpretations (like Copenhagenish interpretations, and apparently in Frauchiger and Renner's own view---see again Footnote~\ref{footFR}), Alice cannot safely reason about Charlie's measurement outcomes, since she views Charlie (and his laboratory) as a quantum system. 
So once again, the Consistency assumption allows an agent to accept a conclusion even though that agent may not accept the reasoning which led to the conclusion.

The story is much the same for the application of Consistency to Eq.~\eqref{eq_Bobknowl2}. 

The application of Consistency to Eq.~\eqref{eq_Bobknowl4}, in contrast, seems unproblematic, since all the agents in question view Alice classically.

In many realist interpretations, the Consistency axiom seems to have a priori plausibility:  if all agents are rational, reason using the same theory, begin with the exact same information, and trust each other's means of gathering new information, then it is not clear whether any one of them would ever have cause to deny {\em any} of the logical steps of any of the other agents. Thus, many realist interpretations are likely to reject other assumptions such as Born Inaccessible Correlations instead of Consistency to resolve the no-go theorem.  But in frameworks (primarily Copenhagenish interpretations) where agents are allowed to make different modeling choices (such as where to place the Heisenberg cut) or where there are explicit moratoriums on certain types of reasoning (such as being forbidden from reasoning about the properties of quantum systems), it is natural to question whether one agent can really accept the reasoning of another agent.

Frauchiger-Renner's argument is framed in this latter way, with different agents reasoning according to their own preferred Heisenberg cuts. It is arguably unsurprising, then, that the Consistency assumption could lead to a contradiction in this context.  The Frauchiger-Renner argument provides a formalization of such an argument (if one grants the other assumptions of the argument).\footnote{The Wigner's stalkee argument in Section~\ref{sec_stalkee} also shows that the predictions of measurement outcomes made by different agents with different placements of Heisenberg cuts can be contradictory. Nevertheless, there are three shortcomings of the Wigner's stalkee argument compared to the Frauchiger-Renner argument. First, the stalkee argument involves the projection postulate on the evolution of a system (the Friend) that is elsewhere in the argument treated as a unitary, whereas the Frauchiger-Renner argument does not. 
Second, the stalkee argument explicitly requires the Friend to treat herself quantumly (which will be rejected immediately by certain Copenhagenish interpretations), whereas the Frauchiger-Renner argument does not require such reasoning directly (only if Debbie questions the origin of the implication of $c=0\implies b=+$ does she need to treat herself quantumly). Third, in the stalkee argument, there is still an overlap between the predictions from the two perspectives, while in Frauchiger-Renner, there is a logical contradiction.}

One {\em can} preserve Consistency within such frameworks, however, if one replaces every statement of fact (such as `outcome $\alpha=1$') with a statement expressing a fact {\em relative to a particular choice of cut} (such as `outcome $\alpha=1$ given that system $\beta$ is on the quantum side of the Heisenberg cut). This is shown explicitly in Ref.~\cite{vilasini2022general}, and demonstrates (as a response to a challenge often posed by Renner) that one can write down a consistent set of rules for reasoning in multi-agent scenarios.

\subsection{A weaker notion of Absoluteness of Observed Events?}
\label{sec_shiftyAOE}

A claimed advantage of the Frauchiger-Renner argument relative to other known EWF arguments is that it does not assume that observed outcomes are objective and observer-independent. For instance, the authors state that~\cite{Frauchiger2018}
{\em\enquote{Another distinction is that the framework used here treats all statements about observations as subjective, i.e., they are always defined relative to an agent. This avoids the a priori assumption that measurement outcomes obtained by different agents simultaneously have definite values.}}
In our view, it is not obvious that this is meaningfully the case.
It seems that perspectival interpretations can avoid the Frauchiger-Renner contradiction in a similar manner to how they avoid the other EWF no-go results, and without sacrificing any of their core principles. 
For instance, if one gives up on Absoluteness of Observed Events entirely, so that the measurement outcomes of each different agent exist in distinct realities, then each agent is blocked from making inferences about another agent's outcome based on their own outcome. One sometimes hears it said that perspectival interpretations evade the no-go theorem by rejecting the Consistency assumption~\cite{Frauchiger2018}, but note that perspectival interpretations typically reject much more than just Consistency; for instance, they may reject the meaningfulness of Eqs.~\eqref{eq_-0}-\eqref{eq_0-} when the agents in question have not communicated their results to each other.

There is a sense in which it is true that the Frauchiger-Renner argument does not require the full strength of Absoluteness of Observed Events. The argument can in fact be made using a weaker assumption that we will refer to as {\em Shifty Absoluteness of Observed Events}. This assumption states that only outcomes of observers {\em who are modeled as classical} are necessarily single and absolute (the outcomes of observers who are modeled as quantum may or may not be absolute). 

In interpretations where there is an objectively correct place at which all observers must put their Heisenberg cuts, this is a plausible assumption: some observers are fundamentally classical, and their outcomes are absolute, while others are fundamentally quantum, and their outcomes may or may not be absolute. 

But more generally (including in Frauchiger and Renner's framework and in, e.g., most Copenhagenish interpretations), whether an observer is modeled classically or not is a subjective choice, encoded in their personal placement of the Heisenberg cut (the shifty split). In this case, it is hard to see how one could accept Shifty Absoluteness of Observed Events while rejecting the full Absoluteness of Observed Events. 
Whether or not a given event is absolute is not a subjective matter, and so cannot in our view be a function of the cut.  To put it another way: if any single agent models a given observer classically, then Shifty Absoluteness of Observed Events guarantees that that observer's outcome is absolute. But if it is absolute for one observer, then (by the very meaning of the word) it is absolute for all observers, and so one recovers the full assumption of Absoluteness of Observed Events.

Still, this is our best attempt to capture the sense in which Frauchiger-Renner's assumptions are more permissive to a perspectival view, as they claim. {\em If} a number of agents try to reason  consistently using an interpretation that rejects Absoluteness of Observed Events but endorses Shifty Absoluteness of Observed Events (as well as the other background assumptions), Frauchiger and Renner's argument shows they will reach a contradiction. Consequently, one is forced to reject {\em even} the idea that observers one views as classical have absolute outcomes. In contrast, the other EWF arguments we have seen do not force one to this conclusion, since they can be evaded by stipulating that {\em only} one of the observers who is being viewed as quantum in the argument has a non-absolute measurement outcome. 

For instance, one could evade the Pusey-Masanes argument by holding that Charlie and Debbie's outcomes are perspectival rather than absolute. Since both agents are viewed\footnote{\label{foot_view}As mentioned in the next subsection, one can think of the Pusey-Masanes reasoning as being done by you, me, Alice, Bob, etc.} quantumly throughout the entire argument, one could do so while maintaining the position that Alice and Bob's outcomes are absolute, and hence could maintain Shifty Absoluteness of Observed Events. But in Frauchiger-Renner's argument, every agent only directly reasons about outcomes of observers whom they view as classical. That is, no equation in Section~\ref{sec_usingCons} involves any agent making statements about an outcome of an observer whom they view as quantum. Therefore, allowing only outcomes of quantum observers to be perspectival is not enough to resolve the no-go theorem.

If a perspectival interpretation were to allow for violations of Shifty Absoluteness of Observed Events, then it could resolve the Frauchiger-Renner no-go result while maintaining Consistency (and also Born Inaccessible Correlations). Imagine that Debbie knows that Charlie knows the value of an outcome $z$ is $0$. If $z$ only has meaning relative to Charlie, Debbie cannot simply adopt Charlie's reasoning to herself know that $z=0$ as well (even if she views $z$ to be observed by a classical observer, which we will take to be the case). However, it is not necessary to reject Consistency if one instead considers the property that is \enquote{$z$ relative to Charlie}, which we will denote $z^{\rm C}$. As such, it makes sense for Debbie to herself know that $z^{\rm C}=0$; first, Debbie knows that Charlie knows $0$ to be the value of an outcome that is relative to him, and then Debbie uses Consistency so that she herself knows that $z^{\rm C}=0$, namely, that $0$ is the value of the outcome {\em relative to Charlie}. That is, once Shifty-Absoluteness of Observed Events is rejected, there is no longer a need to reject Consistency. 
 
\subsection{Comparison with Pusey-Masanes and Local Friendliness}\label{comparisonsec}

Frauchiger and Renner's argument has a distinctly epistemological rather than metaphysical flavor: it appeals explicitly to the reasoning of agents and to the inferential rules they follow. Indeed, works building on this argument have often focused on what inferential rules are reasonable in light of the argument~\cite{Nurgalieva_2019,fraser2020fitch,renes2021consistency,Rio2023}. Meanwhile, the other EWF arguments in this work are more about metaphysics, and refer primarily to facts about the world rather than anyone's reasoning about the world. For example, as mentioned in Footnote~\ref{foot_view}, one could either view the Pusey-Masanes argument as being made from the perspective of a single agent (perhaps Alice, or you the reader, or an abstract unspecified agent), or as being made from no particular perspective. Either way, the reasoning is about {\em what actually happened} rather than what another agent knows.

Both the Pusey-Masanes and the Frauchiger-Renner arguments require the assumption of Born Inaccessible Correlations---that even empirically inaccessible correlations obey the Born rule. 
However, the Pusey-Masanes argument does not explicitly refer to the reasoning of agents, and so it does not require the additional Consistency assumption that Frauchiger and Renner require. This avoids the nuanced and contentious considerations regarding if and when one agent can adopt the reasoning of another agent.

In our view, this is a notable advantage of the Pusey-Masanes argument.
The idea that even inaccessible correlations must be constrained by quantum theory is in our view the core of both arguments, and by focusing on this fact, one greatly simplifies the argument, while removing an extra controversial assumption.  

Still, unless one can find some stronger motivation for the assumption of Born Inaccessible Correlations (without appealing to different measurement settings and Local Agency for Observed Events), both arguments remain inconclusive, in our view.

Furthermore, just like the Pusey-Masanes argument, the Frauchiger-Renner argument is not empirically testable. In contrast, as we mentioned in Section~\ref{sec_LF}, the Local Friendliness argument provides empirically testable inequalities based on theory-independent assumptions, much like Bell's theorem. 
 
Finally, we note that only in the Frauchiger-Renner argument does the question of whether an observer can assign a quantum state to themself (with all the associated complications; see Section~\ref{sec_stalkee}) rears its head. In the Frauchiger-Renner theorem, one observer (Debbie) arguably must assign a quantum state to herself in order to reason about Charlie's prediction for $p(c,b)$. On the contrary, in Pusey-Masanes or Local Friendliness, the entire argument can be viewed as the reasoning of a superobserver (or an external agent), and thus at no point does an observer have to assign a quantum state to themself. 
 
\section{Gao's argument}
\label{sec_Gao}

 The EWF arguments discussed so far are all fairly similar, and are constructed without any assumptions about the nature of correlations between measurements done in sequence on a single system. In the present section and the one after it, we will see two rather different EWF arguments which make assumptions of that kind. (One can also find a third such argument in Ref.~\cite{walleghem2024extended}, which appeared after our review.) 

In Ref.~\cite{Gao2019}, Gao introduces an argument claiming to establish that quantum theory and special relativity are incompatible. While we do not agree with his conclusions, the thought experiment he proposes warrants careful study.

\begin{figure}[htbp]
    \centering
    \begin{subfigure}[t]{0.22\textwidth}
        \centering
        \includegraphics[width=\textwidth]{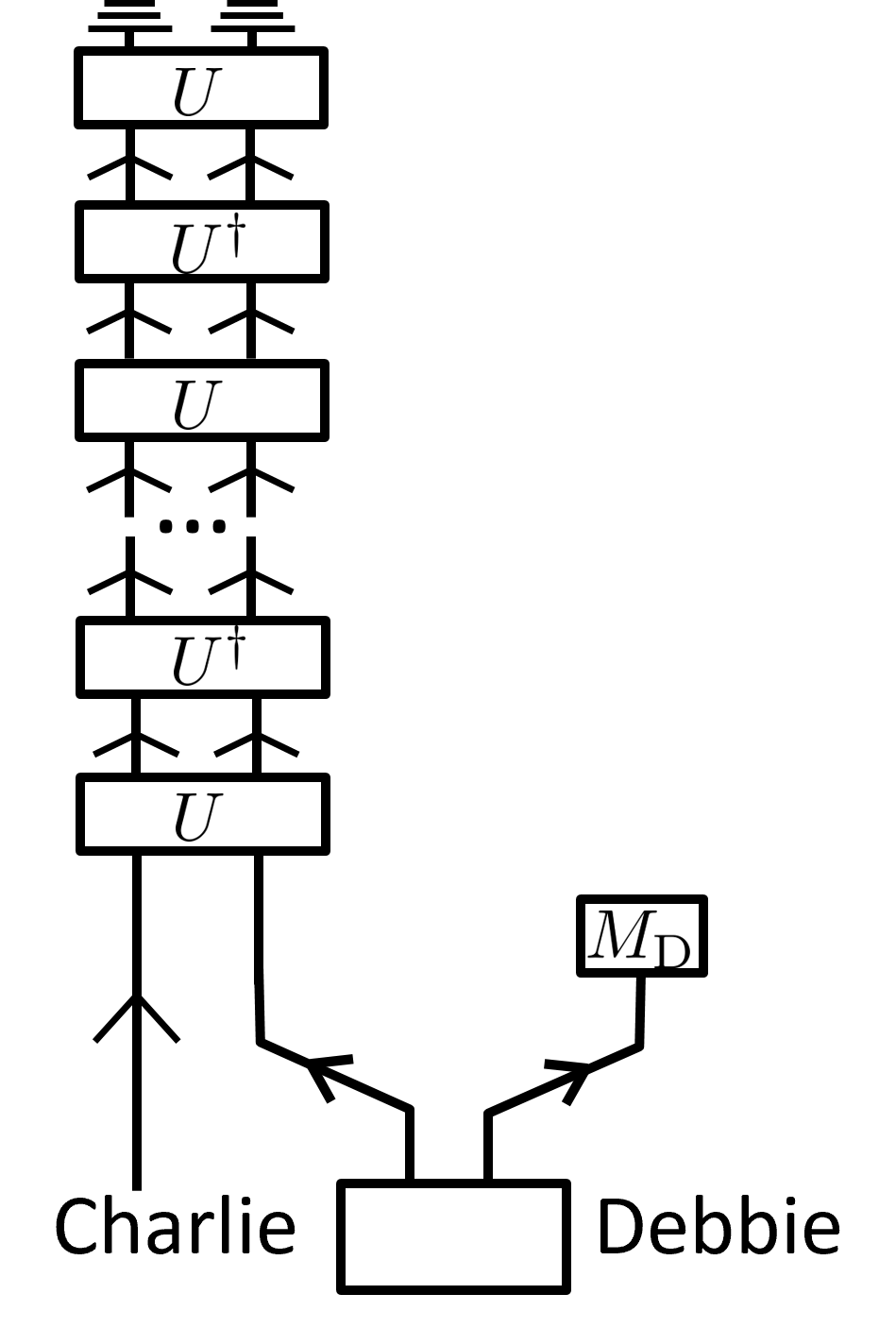}
        \caption{}
        \label{Gao1}
    \end{subfigure}
    \hfill
    \begin{subfigure}[t]{0.22\textwidth}
        \centering
        \includegraphics[width=\textwidth]{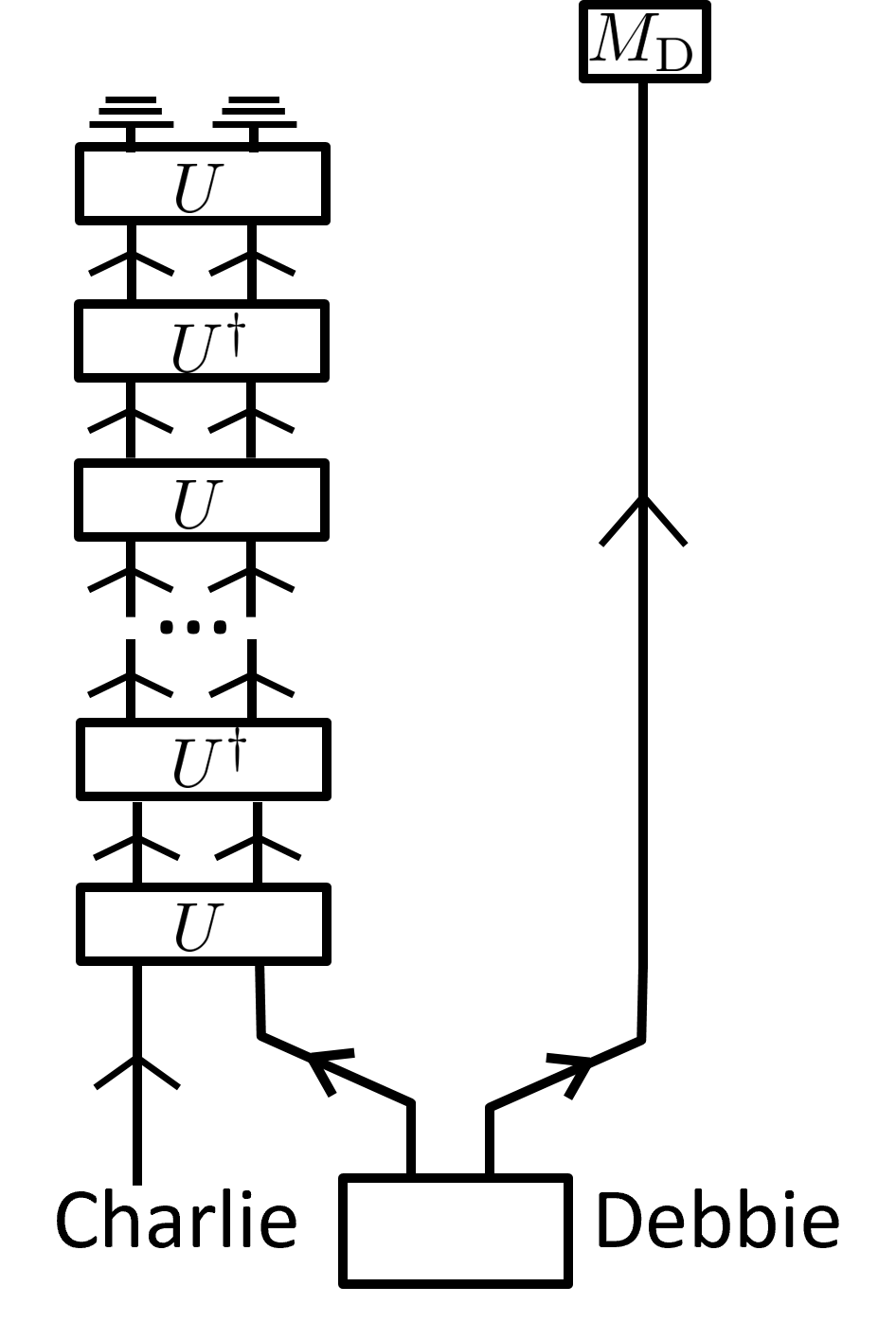}
        \caption{}
        \label{Gao2}
    \end{subfigure}
    \caption{Circuit representation of Gao's setup. 
    a) In one reference frame, all Charlie's measurements happen after Debbie's measurement. b) In another reference frame, all Charlie's measurements happen before Debbie's measurement. }
    \label{fig_Gaofullfig}
\end{figure}

Consider the singlet state $\frac{1}{\sqrt{2}}\left(\Ket{01}-\Ket{10}\right)$ shared between Charlie and Debbie. Charlie measures his part of the entangled system in the computational basis. Charlie's measurement process is described by the unitary $U$. After Charlie measures the qubit, the superobserver undoes Charlie's measurement by applying $U^{\dagger}$ on the joint system of Charlie and the qubit, and then Charlie measures his system again in the same basis, and the superobserver undoes the measurement again. This cycle repeats until Charlie has performed the measurement $k$ times. Debbie measures her half of the entangled system also in the computational basis. No superobserver undoes Debbie's measurement, so Debbie only measures once.

Additionally, it is assumed that all of Charlie's measurements are space-like separated from Debbie's measurement. Consequently, there exists a reference frame in which all of Charlie's measurements occur after Debbie's measurement, and another in which they all take place before Debbie's measurement. The setup, viewed from two such reference frames, is illustrated in Fig.~\ref{fig_Gaofullfig}(a) and Fig.~\ref{fig_Gaofullfig}(b).

If we grant Absoluteness of Observed Events, then each measurement by Charlie yields an absolute outcome, and we denote these as $c_1, c_2,...,c_k$. Likewise, Debbie's measurement produces an absolute outcome $d$. 

Gao argues that, according to quantum theory, in the reference frame where Debbie measures first, the outcomes must satisfy $c_1=c_2=...=c_k$ since the state measured by Charlie is always $\ket{0}$ if Debbie observed $d=1$, or is always $\ket{1}$ if Debbie observed $d=0$. Gao further argues that, in the reference frame where Debbie measures last, Charlie's outcomes do not have to be the same, since the state measured by Charlie is always the maximally mixed state. 
Thus, Gao concludes that the predictions made by quantum theory are not invariant when one changes inertial reference frames, and so he concludes that quantum theory is incompatible with relativity. 

Gao's conclusion rests on the particular pair of claims above about the frequencies with which Charlie's outcomes must arise, depending on the relative timing between Debbie and Charlie's measurements. But both of these claims are far from self-evident.\footnote{Even if the claims are true, the correlations which supposedly contradict relativity theory are not observable to anyone. As we mentioned in Section~\ref{sec_relatedPM}, the only quantities which {\em must} be reference frame invariant on pain of relativity theory being observably false are those which are empirically testable. However, we consider this kind of resolution unsatisfactory; at the level of the theories themselves, this does not in our view alleviate the inconsistency.} (Gao himself merely states the claims; we could not identify any {\em argument} for them in his writing.) Certainly, one cannot justify these claims based on empirical data or by appeal to operational quantum theory, since the relative frequency in question is not accessible to any observer (even in principle). Because all the undoing operations on Charlie's side are space-like separated from Debbie's measurements, Charlie's first $k-1$ outcomes are all erased by the superobserver before the relative frequency over $(c_1, c_2, ..., c_k)$ could be empirically verified by anyone. (Similarly, the correlations among Charlie's outcomes, namely the relative frequency of outcomes $(c_i, d)$ for $i=1,2,...,k-1$ cannot be empirically verified.)  

Much like the assumption of Born Inaccessible Correlations appearing in the Pusey-Masanes and Frauchiger-Renner arguments, Gao seems to us to be basing these particular claims on some kind of implicit assumption of Completeness---that there is nothing beyond the quantum formalism that can be appealed to to make predictions, even when referring to inaccessible correlations. But even if one is willing to grant Completeness (which many interpretations will not do), it does not seem to us that Gao's claims follow.

In the following, we will give two different arguments, motivated (loosely) by Completeness. Regardless of when Debbie's measurement is done (and consequently, also regardless of the reference frame), one argument suggests that all of Charlie's outcomes must be the same, while the other suggests that all of Charlie's outcomes must be random. At least one of these two arguments must be incorrect, but without further assumptions, it is not clear to us which. 

Consider the case when Debbie performs her measurement in the {\em past lightcone} of all of Charlie's measurements (rather than considering the case of spacelike separation). In this case, Debbie's measurement leads one to update Charlie's state\footnote{Note that, although the projection postulate is used for Debbie's measurement here, it is not used on a system that is also modeled unitarily in the argument since there is no superobserver acting on Debbie.}, and it is an uncontestable prediction of quantum theory that all of Charlie's outcomes must be identical, since in this case the distribution $p(c_i,d)$ is {\em actually observable} for every value of $i$, and so one could actually verify that $c_i = \lnot d$ for all $i$. For instance, if Debbie observes outcome $1$, then Charlie's state will be updated to $\ket{0}$, and all of Charlie's outcomes will be $0$.

Next, consider the case when Debbie performs her measurement in the {\em future lightcone} of all of Charlie's measurements. 
In this case, Charlie's system is effectively prepared in a maximally mixed state according to operational quantum theory, as an improper mixture (with no preferred decomposition into any particular basis).  Each of Charlie's measurements is performed on this same improper mixture, since after each measurement, the superobserver reverts the state to what it was prior to his measurement. Consequently, a commitment to Completeness suggests that each of Charlie's outcomes is random, for if there is no further information (such as an underlying ontic state) to break the symmetry, then there is no reason why the same outcome would occur for all of Charlie's measurements. So when Debbie measures in the future lightcone of all of Charlie's actions, it seems that Charlie's outcomes should be random. 

Now, as noted in Section~\ref{sec_BIC}, the assumption of Completeness motivates (at least loosely) the Timing Irrelevance assumption. That is, Completeness motivates the idea that (when the Hamiltonian is trivial) the probability of measurement outcomes for two measurements done in parallel should not depend on the time at which the measurements in question were performed (even if the outcomes for both measurements are not jointly accessible to any observer). 
In particular, to imagine that one of Charlie's outcomes is related to another of Charlie's outcomes would presumably entail imagining some mechanism that transmits information about his earlier outcomes to his later measurements. In a hidden variable model, one could certainly posit such a mechanism, but not in an interpretation that is committed to Completeness.

But by Timing Irrelevance, Charlie's outcomes cannot be different depending on whether Alice's measurement is in the future or past lightcone of Charlie's measurements. So the two arguments just given are in contradiction. So it is simply unclear to us what can be deduced in this context, if anything, from the assumption of Completeness.\footnote{To be clear, Gao presumably would reject Timing Irrelevance, since Timing Irrelevance would immediately say that the two situations in Fig.~\ref{fig_Gaofullfig}(a) and (b) are identical, whereas Gao claims that they are not. But as mentioned previously, Gao does not justify his claims for why the outcome frequencies are as he states in his paper.}

Within interpretations of quantum theory that reject Completeness and posit some underlying states of reality, Gao's scenario is completely trivial to resolve. In such interpretations, it is perfectly reasonable that the outcomes of all of Charlie's measurements would be identical, regardless of what Debbie does, because these outcomes are ultimately determined by the underlying state of reality of the system, which may well be identical for all of Charlie's measurements. Indeed, this is necessarily the case in any interpretation that respects Leibniz's methodological principle for theory construction~\cite{Spe05,SpekLeibniz19,schmid2021unscrambling}, since doing and undoing a measurement must be represented as the ontic identity (because such a procedure is operationally equivalent to doing nothing at all). 

So Gao's argument does not establish any incompatibility between relativity theory and quantum theory. But Gao's scenario does illustrate the fact that quantum theory is underspecified to such a degree that it does not make unambiguous predictions in all setups.

\section{Gu{\'e}rin et. al.'s argument}

In Ref.~\cite{allard2021no}, Gu{\'e}rin et. al. introduce an extended Wigner's friend argument that (unlike the other arguments in this paper) does not involve any Bell-like elements. The authors claim to prove that, under an assumption of linearity, records of an observer's measurement outcome cannot generally persist after a superobserver manipulates that observer. Although this setup leads to a number of interesting questions, we will ultimately argue that the linearity assumption is in fact unmotivated, and so one does not reach the intended conclusion. 

We begin by presenting our modified version of the argument. We explain the modifications and their purpose in Section~\ref{modifiedsetup}.

\begin{figure}[htb!]
    \centering
    \includegraphics[width=0.25\textwidth]{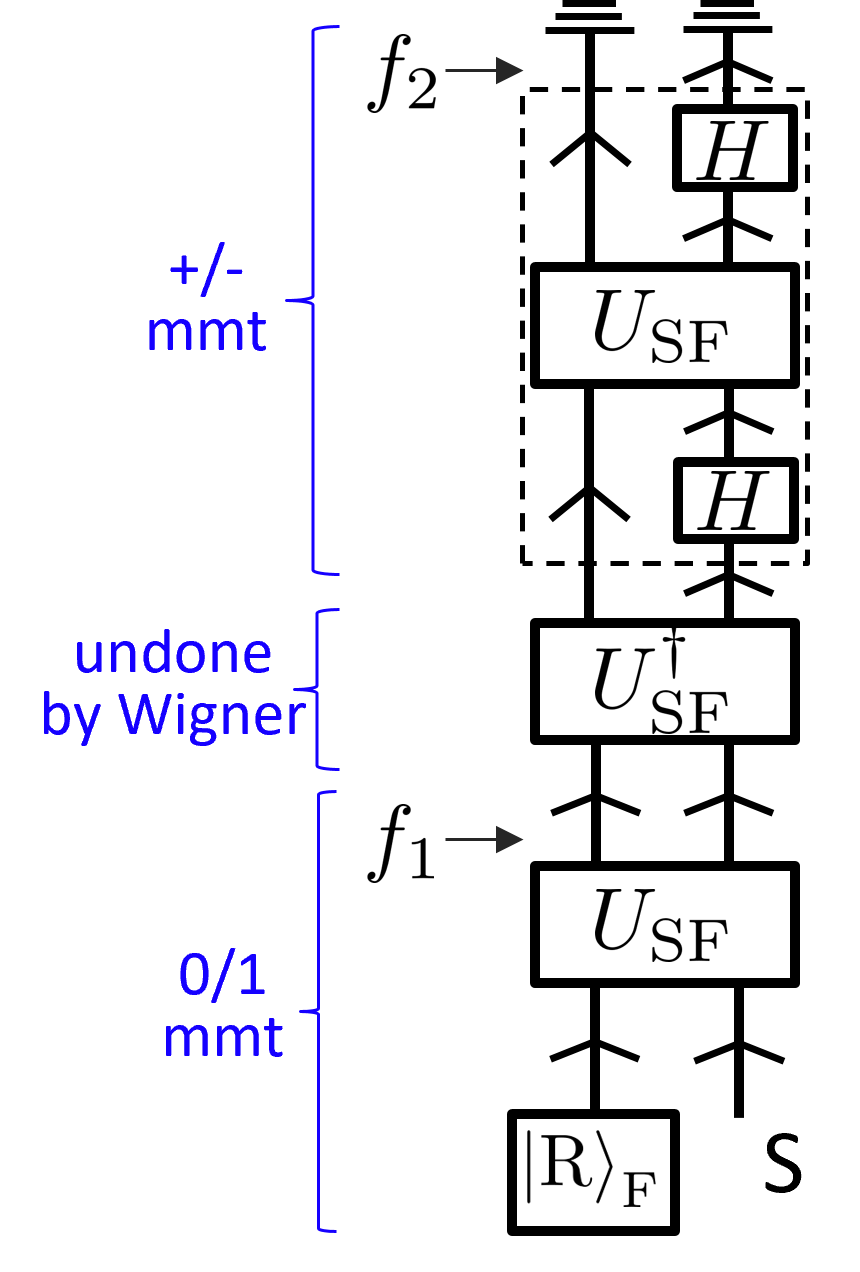}
    \caption{The setup illustrating (our version of) the no-go theorem for the persistent reality of Wigner's friend's perception. The friend observes measurement outcomes at the points indicated by $f_1$ and $f_2$.}
    \label{fig_persistent1}
\end{figure}

In our version of the argument, Wigner's friend, ${\rm F}$, measures a system ${\rm S}$ twice: first in the computational basis, and then in the complementary $\pm$ basis. In between these two measurements, Wigner undoes the Friend's first measurement. As in the previous EWF arguments, we model the Friend as a system in Hilbert space and the Friend's measurements as unitary interactions with ${\rm S}$, leading to the circuit depicted in Fig.~\ref{fig_persistent1}. 

The unitary description of the Friend's measurement in the computational basis is given by the Friend beginning in the ready state $\ket{R}_{\rm F}=\ket{0}_{\rm F}$, followed by the CNOT gate $U_{\rm SF}$. 
This measurement is then undone by Wigner's unitary operation $U^{\dagger}_{\rm SF}$, returning the Friend to the ready state and system ${\rm S}$ to the quantum state it was in prior to the Friend's measurement.

The unitary description of the Friend's complementary $\pm$ basis is given by the Friend beginning in the ready state, followed by the Hadamard gate $H$ applied to the system, then the CNOT gate $U_{\rm SF}$, and then the Hadamard gate $H$ again on the system. (Given that the Hadamard gate rotates between the computational and $\pm$ bases, one can see that this is just like the unitary description of the computational basis measurement, but with the appropriate \enquote{coordinate transformation}.)

Note that Wigner's undoing operation $U_{\rm SF}^{\dagger}$ is critical for the latter portion of the circuit to be interpreted as a measurement by the Friend on the system, since it is only with this operation that the Friend is returned to the requisite Ready state.

Assuming Absoluteness of Observed Events, the Friend will perceive an outcome $f_1$ due to their first measurement, and another outcome $f_2$ due to their second measurement, as indicated in the figure at the appropriate points in the circuit. The frequencies $p(f_1)$ and $p(f_2)$ with which these measurement outcomes are observed are given by the Born rule,
namely 
\begin{equation} \label{born1}
p(f_1=0) = {\rm Tr} ( \ket{0}_{\rm S}\bra{0}\rho_{\rm S} )
\end{equation} 
and 
\begin{equation} \label{born2}
p(f_2=0) = {\rm Tr} ( \ket{+}_{\rm S}\bra{+}\rho_{\rm S}  ).
\end{equation} 
 Now, assuming that both outcomes $f_1$ and $f_2$ are objective, there exists a joint distribution $p(f_1,f_2)$ describing the frequencies at which these outcomes occur. By the arguments just given, this means that the marginal $p(f_1)$ of $p(f_1,f_2)$ must obey the Born rule probabilities for the computational basis measurement, while the marginal $p(f_2)$ of $p(f_1,f_2)$ must obey the Born rule probabilities for the $\pm$ basis measurement. 

If one further assumes {\em Linearity} in the sense that the overall process mapping the state $\rho_{\rm S}$ of system ${\rm S}$ into the (real-valued) probability $p(f_1,f_2)$ is linear, then the Reisz representation theorem~\cite{riesz1914demonstration} states that this process can be represented by
\begin{equation}
{\rm Tr}(E_{f_1,f_2} \rho_{\rm S})
\end{equation}
for some operator $E_{f_1,f_2}$. Since $p(f_1,f_2)$ must moreover be positive for all $\rho_{\rm S}$ (and since the set of $\rho_{\rm S}$ span the space of operators on the Hilbert space associated with ${\rm S}$), it follows that $E_{f_1,f_2}$ is a positive operator. Similarly, the fact that $\sum_{f_1,f_2}p(f_1,f_2)=1$ implies that the set $\{E_{f_1,f_2}\}_{f_1,f_2}$ defined by ranging over all $f_1$ and all $f_2$ satisfies $\sum_{f_1,f_2}E_{f_1,f_2}=\mathbb{I}$, so this set $\{E_{f_1,f_2}\}_{f_1,f_2}$ is a positive operator-valued measure (POVM). 

But every POVM is associated with a valid measurement in quantum theory, and so it seems we have derived the existence of a single quantum measurement from which one can compute the statistics of two incompatible measurements on system ${\rm S}$, namely, the computational basis measurement and the $\pm$ basis measurement. But this is impossible, since (by definition~\cite{incompatibilityReview}) two incompatible quantum measurements cannot be simulated by any single quantum measurement. So we have a contradiction. 

If one grants the assumption of Linearity (which we shall argue one should not), then it seems that perspectival interpretations which aim to circumvent the argument would be forced to a more radical position than is required to evade other EWF arguments, because $f_1$ and $f_2$ are observed by a {\em single} observer. In particular, one would need to imagine that since the Friend cannot at any point in time have access to (or memory of) both outcomes simultaneously, $f_1$ is only defined relative to {\em the Friend prior to Wigner's undoing }, and $f_2$ is only defined relative to {\em the Friend after Wigner's undoing }.
This would mean that not only do different observers generically live in fragmented, incompatible realities, but so too do the versions of a single observer that exist at different points in time.

However, we do not believe the argument supports this conclusion, because we do not think the requisite Linearity assumption is reasonable.

\subsection{The problem with the Linearity assumption}

Let us now analyze the assumption of Linearity: that the process mapping the density operator $\rho_{\rm S}$ of system ${\rm S}$ into the probability $p(f_1,f_2)$ is linear. Ref.~\cite{allard2021no} attempts to motivate this assumption by appealing to a well-known\footnote{Versions of this argument are often used to justify the linearity of transformations in a generalized probabilistic theory~\cite{Hardy,barrett2007,chiribella2010probabilistic} and to justify the linearity of ontological representation maps~\cite{schmid2023addressing}.} argument using classical probability theory. 

The argument---which is itself in our view incontestable---is as follows. 
Imagine two possible preparation procedures on a system, $P_0$ and $P_1$, and denote the probability assignments to the joint outcome $(f_1,f_2)$ made by $P_0$ by $p_0(f_1,f_2)$, and denote the probability assignment made by $P_1$ by $p_1(f_1,f_2)$. Now imagine one has uncertainty about the preparation done on the system, believing with probability $\alpha$ that the preparation procedure is $P_0$ and with probability $1-\alpha$ that the state is $P_1$. As a direct consequence of the law of total probability, it follows that one's probability assignment should be $p_{\rm mixed} = \alpha p_0(f_1,f_2) + (1-\alpha)p_1(f_1,f_2)$. 

However, this argument does not establish linearity on the space of density operators (which is what Ref.~\cite{allard2021no} requires), but rather linearity in the space of preparation procedures. Typically, of course, the space of preparations in quantum theory is exactly the space of density operators, but for the purpose of making predictions about inaccessible correlations, this association is questionable at best. 

The density operator representation arises when one takes quantum theory as an unquotiented operational theory and quotients relative to operational equivalence (see for example Refs.~\cite{chiribella2010probabilistic,schmid2020structure}). That is, for any way of preparing a quantum system, the only details that are relevant for making predictions about future measurements on the system are those which determine the density operator. All other details---such as if the density operator arose as a mixture of one set of pure states or another---are not relevant, and such information is termed the {\em context} of a preparation.

In setups where a superobserver Wigner undoes a Friend's measurements, however,  the context of a preparation could be relevant for computing inaccessible probability distributions describing pairs of outcomes that cannot be simultaneously known to any single observer. For instance, consider preparing the maximally mixed state via mixing the states $\ket{0}$ and $\ket{1}$, versus preparing the maximally mixed state via mixing the states $\ket{+}$ and $\ket{-}$; imagine also that the Friend measures the system in the computational basis many times in a row, but with Wigner undoing the measurements in between. In the case where the preparation procedure is the mixture of computational basis states, the outcomes of the Friend's measurements must all be identical (since the state of the system really {\em is} either $\ket{0}$ or $\ket{1}$). In the case where the preparation procedure is the mixture of $\pm$ basis states, it is unclear what the Friend's measurement outcomes will be, and in some interpretations, they may all be different.
So what is usually considered the context of a preparation---namely, which set of pure states was mixed together to generate the state---might well be relevant for the inaccessible predictions in question.

In other words, the mathematical object which forms the basis for making predictions about such inaccessible correlations cannot be assumed to be the density operator. Ideally, one would {\em derive} the relevant object by considering quotienting relative to not only the usual set of quantum measurements, but also relative to all probing schemes involving inaccessible measurement outcomes. But this could only be done if one had a precise theoretical prescription for the frequency with which such outcomes occur, which would in turn depend on one's interpretational commitments. 

Alternatively, one might take the relevant object to be one's state of knowledge about which pure state was performed.\footnote{The importance of this more refined mathematical object was previously pointed out in other contexts in Ref.~\cite{schmid2021unscrambling}.} That is, for the process mapping system ${\rm S}$ into the probability $p(f_1,f_2)$, one takes the input not as the space of density operators, but as the space of probability distributions over pure states. 
This process must be linear on this domain, by the argument at the start of this section. But it is not even a map on density operators, much less a linear map on density operators.

\subsection{Why our modifications to the argument are useful} \label{modifiedsetup}

While our argument considers the question of the existence of joint outcomes for two measurements of a single observer across time, Ref.~\cite{allard2021no} aims to study the persistence of an observer's record of a single measurement outcome.  

\begin{figure}[htb!]
    \centering
    \includegraphics[width=0.2\textwidth]{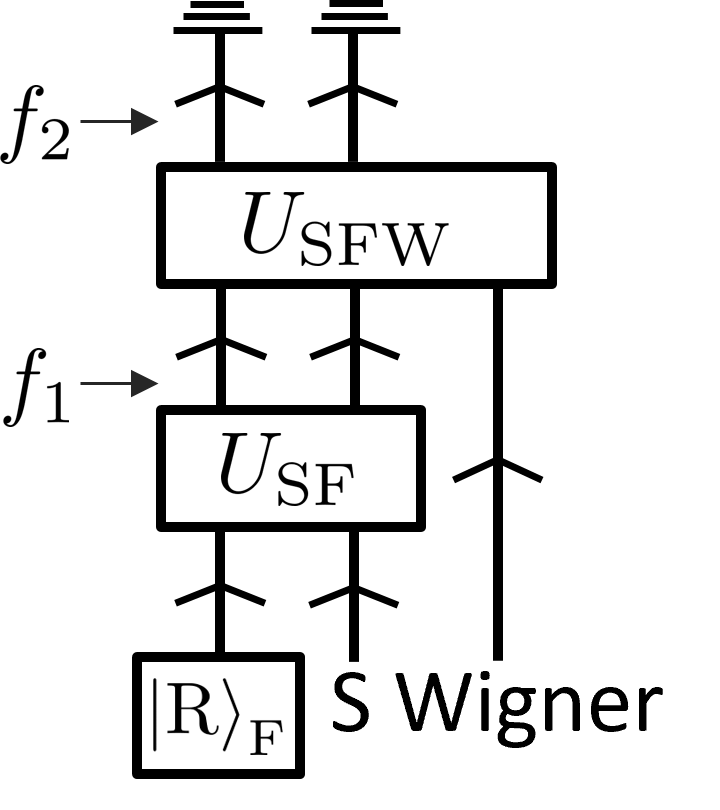} 
    \caption{ The scenario from the original argument in Ref.~\cite{allard2021no}. The Friend is presumed to have some perception of the measurement records at the points indicated by $f_1$ and $f_2$.}
    \label{fig_persistent5}
\end{figure}

The original argument given in Ref.~\cite{allard2021no} is based on the circuit shown in Figure~\ref{fig_persistent5}, rather than our modified circuit in Fig.~\ref{fig_persistent1}. Here, Wigner's friend makes a measurement in the computational basis, after which Wigner performs a joint measurement of the Friend and system ${\rm S}$ (here depicted as a unitary $U_{FSW}$) in a basis which Ref.~\cite{allard2021no} takes to be arbitrary. 

The proof then runs much the same as our proof, but where one assumes that there are events $f_1$ and $f_2$ (just before and just after Wigner's measurement, as depicted in Fig.~\ref{fig_persistent5}) \enquote{corresponding to the perceived measurement records of the Friend}. These perception events are presumed to occur at frequencies governed by the Born rule frequencies for a computational basis measurement on the state of $F$.

Now, for the case of $f_1$, this is clear: the circuit up until this point explicitly depicts the Friend making a measurement of system ${\rm S}$ in the computational basis, and so by Absoluteness of Observed Events there will be an event corresponding to the Friend perceiving an outcome. Then, one can further assume that the frequency with which the outcome is perceived obeys the Born rule. 

For the case of $f_2$, however, it is less clear to us what assumptions are being made governing the system ${\rm F}$ and the perceptions of the Friend. Certainly one cannot simply appeal to Absoluteness of Observed Events to argue that a measurement outcome will be perceived, since the observer in question does not begin in a \enquote{ready state} of the sort we discussed earlier, nor does the observer then interact with the system in a particular manner befitting a measurement (e.g., one which couples some degree of freedom of the system to an amplification process that ultimately generates a macroscopic signature of the measurement outcome for the observer to collect). Rather, Ref.~\cite{allard2021no} seems to assume some novel kind of correspondence between computational basis measurements on qubit ${\rm F}$ (which, recall, is a coarse-graining of a vast collection of Hilbert spaces for all the particles associated with the Friend and the Friend's immediate environment) and the perceptions of the Friend. While this may be reasonable, we find it somewhat difficult to reason about, and it is not necessary to make the point.

By considering two sequential measurement processes carried out by the Friend, our version of the argument need only appeal to Absoluteness of Observed Events in its usual form, guaranteeing that there are observed measurement outcomes at the points labeled by $f_1$ and $f_2$. Moreover, recalling from Section~\ref{sec_OGwigner} that our choice of unitary representation of the Friend's measurement process encodes the measurement outcome in the computational basis of system ${\rm F}$, one can explicitly check that accessing this record---that is, measuring system ${\rm F}$ in that basis---leads to the expected Born rule frequencies in Eq.~\eqref{born1} and Eq.~\eqref{born2}, namely
\small
\begin{align}
p(f_1=0) \nonumber \\ 
= &{\rm Tr} \left( (\mathbb{I}_{\rm S} \otimes  \ket{0}_{\rm F}\bra{0}) U_{\rm SF} \bigl(\rho_{\rm S}\otimes\ket{0}_{\rm F}\bra{0}\bigr)  U_{\rm SF}^{\dagger} \right)  \nonumber \\ \nonumber
= & {\rm Tr} \left( U_{\rm SF}^{\dagger}\bigl((\ket{0}_{\rm S}\bra{0}+\ket{1}_{\rm S}\bra{1})\otimes \ket{0}_{\rm F}\bra{0}\bigr) U_{\rm SF} \bigl(\rho_{\rm S}  \otimes \ket{0}_{\rm F}\bra{0}\bigr) \right)\\ \nonumber
= & {\rm Tr} \Bigl( \bigl(\ket{00}_{\rm SF}\bra{00}+\ket{11}_{\rm SF}\bra{11}\bigr) \bigl(\rho_{\rm S} \otimes \ket{0}_{\rm F}\bra{0}\bigr) \Bigr)\\ 
= & {\rm Tr} \Bigl( \ket{0}_{\rm S}\bra{0}\rho_{\rm S}  \Bigr),
\end{align}
and
\begin{align}
    &p(f_2=0) \nonumber \\ \nonumber
    = &{\rm Tr} \left( \bigl(\mathbb{I}_{\rm S} \otimes  \ket{0}_{\rm F}\bra{0}\bigr) \bigl( H_{\rm S} \otimes \mathbb{I}_{\rm F} \bigr) U_{\rm SF} \bigl( H_{\rm S} \otimes \mathbb{I}_{\rm F} \bigr) U_{\rm SF}^{\dagger} U_{\rm SF} \right.\\ \nonumber
    &\left.\bigl(\rho_{\rm S}\otimes\ket{0}_{\rm F}\bra{0}\bigr)  U_{\rm SF}^{\dagger} U_{\rm SF} \bigl(H_{\rm S}^{\dagger}\otimes \mathbb{I} \bigr) U_{\rm SF}^{\dagger} \bigl(H_{\rm S}^{\dagger} \otimes \mathbb{I} \bigr) \right) \\ \nonumber
    = &{\rm Tr} \left( (\mathbb{I}_{\rm S} \otimes  \ket{0}_{\rm F}\bra{0}) \bigl(H_{\rm S} \otimes \mathbb{I}_{\rm F} \bigr) U_{\rm SF} \bigl( H_{\rm S} \otimes \mathbb{I}_{\rm F} \bigr)  \bigl(\rho_{\rm S}\otimes\ket{0}_{\rm F}\bra{0}\bigr) \bigl(H_{\rm S}^{\dagger} \otimes \mathbb{I} \bigr) \right.\\ \nonumber
    & \left. U_{\rm SF}^{\dagger} \bigl(H_{\rm S}^{\dagger} \otimes \mathbb{I} \bigr) \right) \\ \nonumber
    = & {\rm Tr} \left( U_{\rm SF}^{\dagger}\bigl(\mathbb{I}_{\rm S} \otimes \ket{0}_{\rm F}\bra{0}\bigr) U_{\rm SF} \bigl( H_{\rm S} \otimes \mathbb{I}_{\rm F} \bigr)  \bigl(\rho_{\rm S}\otimes\ket{0}_{\rm F}\bra{0}\bigr) \bigl(H_{\rm S}^{\dagger} \otimes \mathbb{I} \bigr)\right)\\ \nonumber
    = & {\rm Tr} \Bigl( \bigl(\ket{00}_{\rm SF}\bra{00}+\ket{11}_{\rm SF}\bra{11}\bigr) \bigl(H_{\rm S} \otimes \mathbb{I} \bigr) \bigl(\rho_{\rm S} \otimes \ket{0}_{\rm F}\bra{0}\bigr) \bigl(H_{\rm S}^{\dagger} \otimes \mathbb{I} \bigr)\Bigr)\\ \nonumber
    = & {\rm Tr} \Bigl( \bigl(\ket{+0}_{\rm SF}\bra{+0}+\ket{-1}_{\rm SF}\bra{-1}\bigr) \bigl(\rho_{\rm S} \otimes \ket{0}_{\rm F}\bra{0}\bigr) \Bigr)\\ 
    = & {\rm Tr} \Bigl( \ket{+}_{\rm S}\bra{+}\rho_{\rm S}  \Bigr).
        \end{align}
\normalsize

\section{Conclusions}

Extended Wigner's friend arguments have the potential to provide deep insights into how one should understand quantum theory. Most notably, these arguments do not appeal to the classical notion of realism typically codified in the framework of ontological models or classical causal models, and so the usual responses to previous no-go theorems like Bell's theorem or noncontextuality no-go theorems are not sufficient. And unlike the original Wigner's friend thought experiment, these arguments need not assume the projection postulate or any sort of collapse of the wavefunction.

We have introduced and critically analyzed a number of extended Wigner's friend arguments from the literature. Ultimately, we argued that Brukner's argument, Gao's argument, and the argument of  Gu{\'e}rin et. al. do not support their intended conclusions, as they make assumptions that are stronger than intended (in Brukner's case) or that are not motivated (in the latter two cases). 
 Meanwhile, the Pusey-Masanes argument requires the assumption of Born Inaccessible Correlations, which we consider to be plausible, but not obviously backed up by any deep physical principles in the context of the Pusey-Masanes scenario. The Frauchiger-Renner argument requires both this assumption and the assumption of Consistency, which we also consider to be plausible in the context of many realist interpretations, although less obviously reasonable in the context of Copenhagenish interpretations (in which the Frauchiger-Renner argument can indeed be taken to undermine it). Ultimately, we consider the assumptions of the Local Friendliness argument to have the clearest motivations among the EWF arguments to date. The most notable two assumptions in this no-go theorem are that outcomes are single and absolute, and a weak notion of locality---that any freely-chosen setting must be uncorrelated with any set of relevant observed events outside its future light cone. This assumption can be motivated from relativity theory---for example, by the requirement that correlations arising at spacelike separation have a common cause explanation (possibly using a nonclassical notion of common cause). 
As a consequence, the Local Friendliness argument has immediate implications for the research program of quantum causal models, which aims to give a common-cause explanation to all spacelike separated correlations using a nonclassical generalization of the causal modeling framework. In order to achieve such an explanation, then, it seems likely that one must allow for the radical possibility that measurement outcomes are perspectival.

Roughly speaking, EWF arguments always involve three types of assumptions: i) Absoluteness of Observed Events; ii) an assumption about inaccessible-but-actually-existing correlations (such as Born Inaccessible Correlations or Local Agency for Observed Events), and iii) the in principle possibility of a superobserver that can rewind a measurement process. 
Correspondingly, there are three obvious kinds of responses: perspectival interpretations, which relax Absoluteness of Observed Events, interpretations (like Bohmian mechanics) that provide prescriptions other than the Born rule for governing the correlations between measurement outcomes that are inaccessible, and theories (like collapse theories) that reject the universality of unitary quantum dynamics.

There are many interesting questions for future work.

As we have seen, most prior works focused on scenarios with measurements performed in parallel on a bipartite system, and relied on assumptions regarding inaccessible correlations for measurements done in parallel. The two notable exceptions to this rule rested on questionable assumptions. After the appearance of this review, a new argument involving sequential measurements on a single system was constructed~\cite{walleghem2024extended}. The argument relies on an assumption of Commutation Irrelevance which is an extension of the assumption of Timing Irrelevance needed for the Pusey-Masanes (and Frauchiger-Renner) argument. The considerations in Section~\ref{sec_BIC} apply to it as well, and so the argument in Ref.~\cite{walleghem2024extended} is essentially on the same footing as Pusey-Masanes's. This constitutes the most solid EWF argument to date that does not involve a bipartite scenario.

One notable question is to clarify the role of locality assumptions used in the Frauchiger-Renner, Pusey-Masanes, and Gao arguments. Ref.~\cite{ormrod2023theories} goes some way towards addressing this question in the context of the Pusey-Masanes argument, by identifying some relevant notions of locality and compositionality, but the conceptual status of these is not yet clear, as we mentioned in Section~\ref{sec_relatedPM}. Analyses of these arguments within Bohmian mechanics might also be useful in this regard. In a similar vein, it would be interesting to determine if any of these arguments have notable consequences for the program of reproducing quantum theory via nonclassical causal modeling.

Another natural question is to study the relationship between the different causal and inferential assumptions going into the various EWF arguments. The Frauchiger-Renner argument focuses on inferential rules such as Consistency, while the other EWF arguments focus more on physical or metaphysical assumptions (such as Absoluteness of Observed Events or notions of locality). Although it is clear that causal assumptions constrain inferential assumptions and vice versa~\cite{schmid2021unscrambling,pearl2009causality}, the precise logical relationships between these assumptions are very far from clear.

A final question is to develop adequate and compelling perspectival interpretations that explain these EWF arguments. To this end, it behooves existing perspectival interpretations to analyze these arguments carefully;  doing so may shed light on these interpretations, and also may provide an arena in which these interpretations can win new converts by providing compelling explanations. Some very preliminary work of this sort has been attempted in regard to the Frauchiger-Renner argument in Refs.~\cite{bub2017bohr,cavalcanti2023consistency,Healey_2018,stacey2019qbism,debrota2020respecting} and to the Local-Friendliness argument in Refs.~\cite{Cavalcanti2021,Cavalcanti2021bubble,dibiagioStable2021}.

Obviously, these thought experiments are quite far-fetched and completely impractical with modern technology.  Pure operationalism works fine for practical purposes, as we are rarely in doubt about what constitutes an observation in a practical physical experiment.
However, these thought experiments provide insights into the interpretation of quantum theory, which in turn guide us (whether implicitly or explicitly) in how we apply the theory in practice, especially in novel contexts, whether these be causal modeling, biological systems, or artificial intelligence.

Moreover, supposed thought experiments have a creeping habit of becoming real experiments.  The Einstein-Podolsky-Rosesn argument \cite{einsteinCan1935} was originally a thought experiment, but eventually became a real experiment, and one that is a cornerstone of quantum technologies today.  Although extended Wigner's friend arguments are considerably more farfetched than Einstein-Podolsky-Rosen experiments, we do not think one should ever rule out an experiment as unperformable unless there is a strong physical argument against it.\footnote{It might superficially appear that the Pusey-Masanes and Frauchiger-Renner arguments are not experimentally testable, but rather are mere no-go theorems. However, they can be tested experimentally, since they ultimately constrain the (empirically accessible) correlation between the superobservers' outcomes. In the Pusey-Masanes arguments, for example, the assumptions in the arguments allow one to derive Eqs.~\eqref{eq_PM1},~\eqref{eq_PM2} and~\eqref{eq_PM3}, which imply that $a=-\Rightarrow b=+$ as explained in Sec.~\ref{PMsec}; if one finds in the lab that sometimes $a=-$ and $b=-$, then one has experimentally ruled out the assumptions going into the argument. (The fact that Eqs.~\eqref{eq_PM1},~\eqref{eq_PM2} and~\eqref{eq_PM3} are not themselves experimentally falsifiable is not relevant to this point; one may view them simply as facts used in establishing the argument---facts which are motivated by quantum theory and formulated in a theory-independent way.\label{ft:testable})}

There are no thought experiments.  Only experiments that we have not figured out how to do yet.

\section{Acknowledgements}
David and Y\`{i}l\`{e} are co-first authors of this paper. 
David and Y\`{i}l\`{e} thank L\'idia del Rio for hosting us at ETH and for stimulating discussions, especially regarding the Frauchiger-Renner argument. Y\`{i}l\`{e} thanks Eric Cavalcanti and Howard Wiseman for hosting her at Griffith, and we thank Eric and Howard also for stimulating discussions, especially regarding the Local Friendliness argument. Y\`{i}l\`{e} further thanks Howard for insightful discussions on Footnote~\ref{ft:testable} and on the Pusey-Masanes argument, especially regarding Timing Irrelevance. Y\`{i}l\`{e} also thanks Marwan Haddara for explaining Hardy paradox and its relation to the possibilistic Local Friendliness argument. We thank Veronika Baumann, \v{C}aslav Brukner, Ladina Hausmann, Markus Muller, Nuriya Nurgalieva, Nicholas Ormrod, Matt Pusey, Renato Renner, John Selby, Indrajit Sen, Rob Spekkens,  Vilasini Venkatesh, Laurens Walleghem and Marek Zukowski for useful discussions. DS was supported by the Foundation for Polish Science (IRAP project, ICTQT, contract no. MAB/2018/5, co-financed by EU within Smart Growth Operational Programme). YY was supported by Perimeter Institute for Theoretical Physics. Research at Perimeter Institute is supported in part by the Government of Canada through the Department of Innovation, Science and Economic Development and by the Province of Ontario through the Ministry of Colleges and Universities. YY was also supported by the Natural Sciences and Engineering Research Council of Canada (Grant No. RGPIN-2017-04383). ML was supported in part by the Fetzer Franklin Fund of the John E. Fetzer Memorial Trust and by grant number FQXi-RFP-IPW-1905 from the Foundational Questions Institute and Fetzer Franklin Fund, a donor advised fund of Silicon Alley Community Foundation.


\bibliographystyle{apsrev4-2-wolfe}
\setlength{\bibsep}{3pt plus 3pt minus 2pt}
\bibliography{bib.bib}

\end{document}